\documentclass[12pt]{article}

\usepackage[margin=1in]{geometry}
\usepackage{bbm, bm}
\usepackage{amsmath, amssymb, amsthm}
\usepackage{graphicx}
\usepackage{float}  
\usepackage{pbox}  
\usepackage{color}
\usepackage{natbib}
\usepackage[pdftex,colorlinks=true,linkcolor=blue,citecolor=blue,urlcolor=blue,bookmarks=false]{hyperref}
\usepackage{enumitem}
\usepackage{setspace}
\usepackage{fancyhdr}
\usepackage[normalem]{ulem}
\allowdisplaybreaks

\newtheorem{theorem}{Theorem}
\newtheorem{assumption}[theorem]{Assumption}
\newtheorem{proposition}[theorem]{Proposition}
\newtheorem{corollary}[theorem]{Corollary}
\newtheorem{example}[theorem]{Example}
\newtheorem{definition}[theorem]{Definition}
\newtheorem{lemma}[theorem]{Lemma}

\newlist{thmlist}{enumerate}{1}
\setlist[thmlist]{label=(\alph*), ref=\thetheorem(\alph*), noitemsep}

\DeclareMathOperator*{\argmin}{arg\,min}

\newcommand{\calC}{\mathcal{C}}
\newcommand{\E}{\mathbbm{E}}

\newcommand{\N}{\mathcal{N}}
\newcommand{\proba}{\mathbbm{P}}

\newcommand{\calS}{\mathcal{S}}

\newcommand{\var}{\operatorname{Var}}

\newcommand{\Y}{\mathbf{Y}}
\newcommand{\Z}{\mathbf{Z}}
\newcommand*{\defeq}{\mathrel{\vcenter{\baselineskip0.5ex \lineskiplimit0pt
                           \hbox{\scriptsize.}\hbox{\scriptsize.}}}%
                                                =}

\begin{document}

\title{Optimizing Randomized and Deterministic Saturation Designs under Interference}

\author{
\begin{minipage}{10em}
\small
\centering
Chencheng Cai\\
Temple University\\
chencheng.cai@temple.edu
\end{minipage}
\begin{minipage}{10em}
\small
\centering
Jean Pouget-Abadie\\
Google Research NYC\\
jeanpa@google.com
\end{minipage}
\begin{minipage}{10em}
\small
\centering
Edoardo M. Airoldi\\
Temple University\\
airoldi@temple.edu
\end{minipage}
\let\thefootnote\relax\footnote{Chencheng Cai is a Post-doctoral Fellow Research Associate at Department of Statistical Science, Fox School of Business, Temple University. 
Jean Pouget-Abadie is Research Scientist at Google Research, New York. 
Edoardo M. Airoldi is the Millard E. Gladfelter Professor of Statistics, Operations, and Data Science at the Fox School of Business at Temple University.
 This work was supported, in part, by NSF awards CAREER IIS-1149662, and by ONR awards YIP N00014-14-1-0485 and N00014-17-1-2131.
}}
\date{}

\maketitle
\thispagestyle{empty}
\begin{abstract}


Randomized saturation designs are a family of designs which assign a possibly different treatment proportion to each cluster of a population at random. As a result, they generalize the well-known (stratified) completely randomized designs and the cluster-based randomized designs, which are included as special cases. We show that, under the stable unit treatment value assumption, either the cluster-based or the stratified completely randomized design are in fact optimal for the bias and variance of the difference-in-means estimator among randomized saturation designs. However, this is no longer the case when interference is present. We provide the closed form of the bias and variance of the difference-in-means estimator under a linear model of interference and investigate the optimization of each of these objectives.
In addition to the randomized saturation designs, we propose a deterministic saturation design, where the treatment proportion for clusters are fixed, rather than randomized, in order to further improve the estimator under correct model specification. Through simulations, we illustrate the merits of optimizing randomized saturation designs to the graph and potential outcome structure, as well as showcasing the additional improvements yielded by well-chosen deterministic saturation designs. 

\vspace{3em}

\noindent\textbf{Keywords:} Violations of SUTVA, Causal Inference, Potential Outcomes, Saturation Designs
\end{abstract}

\newpage
\tableofcontents

\newpage
\pagestyle{fancy}
\setcounter{page}{1}
\section{Introduction}

In many randomized experiments, the population of interest can be organized into groups (clusters) of units. 
In certain instances, the clustering of units is artificial. For instance, units are grouped according to their distance to the discontinuity point in a regression discontinuity design \citep{owen2020optimizing}, or units can be grouped into subsets of data from the perspective of data fusion \citep{rosenman2021designing}.
A more common incentive for exploring the cluster structure of population is to discover interference between units \citep{toulis2013estimation, tchetgen2012causal}. As a violation of the stable unit treatment value assumption (SUTVA) \citep{imbens2015causal}, units within the same cluster are often assumed to have interference, that is the outcome of one unit can be affected by the treatment status of its group-mates. In certain cases, interference can also occur across clusters.
Examples of such interference clusters include a class of students in educational studies \citep{rosenbaum2007interference}, a group of people with a financial relationship \citep{banerjee2013diffusion}, a social network group \citep{phan2015natural}, or a block of crop field \citep{zaller2004effects}.
In a two-sided market, interference can occur across both customer-side and listing-side~\citep{johari2020experimental}, which provide a natural clustering.
When the clusters representing interference are not immediately clear, \citet{ugander2013graph} explores algorithmic clustering solutions.

Given a clustered population, three group-level experimental designs are commonly used: the stratified completely randomized design, the
cluster-based randomized design and the randomized saturation design. The stratified---sometimes referred to as `blocked'---design extends the standard completely randomized design \citep{rubin1974estimating} to groups of units such that in each cluster, an equal proportion of units is treated \citep{owen2020optimizing, rosenman2021designing}. In cluster-based designs, all units within the same cluster receive treatment or control \citep{eckles2017design}.  
When interference is present, compared to stratified design, it is
generally believed that a cluster-based randomized design will be less
biased~\citep{eckles2017design, ugander2013balanced}, but will have higher variance than a
completely-randomized design that assigns the same proportion of units to
treatment. The complexity of finding balanced partitioning of a large set of
experimental units~\citep{andreev2006balanced} is another aspect to take into
consideration when choosing between both of these standard designs. An optimal cluster-based randomized design is tractable under monotonicity \citep{pouget2018optimizing}.

The randomized saturation design, proposed in \citet{hudgens2008toward} as a compromise between the two previous designs, is a two-step procedure, where clusters are first assigned with treatment proportions, and then units within each group are assigned to treatment and control at random according to the assigned treatment proportion. 
Randomized saturation designs are often used in the
context of interference because they allow the experimenter to infer a unit's
reaction to varying levels of treatment~\citep{banerjee2012can,
sinclair2012detecting, crepon2013labor}. This is especially appropriate if we are
willing to make an anonymous interference
assumption~\citep{manski2013identification} or an assumption of no
peer-effect-heterogeneity~\citep{athey2015exact}. For an excellent reference on
randomized saturation designs, we refer the reader to \citet{baird2016optimal}.

Randomized saturation designs offer an interesting interpolation between
stratified and cluster-based randomized designs. Both can be
conceptualized as a randomized saturation design: the stratified completely randomized
design corresponds to a randomized saturation design with identical treatment
proportions across all clusters; the cluster-based randomized design
corresponds to a randomized saturation design with full treatment or full
control proportions. 

Randomized saturation designs are an example of \emph{model-assisted}
designs~\citep{basse2015model}. Indeed, the distribution of the treatment proportions 
can be chosen to optimize a particular objective under a set of model
assumptions, without sacrificing the validity of the estimation procedure if our model of potential outcomes is misspecified. With
high confidence in our modelling assumptions, we can further optimize the
assignment of each treatment proportions within each cluster of experimental units. We
refer to these designs as \emph{deterministic} saturation designs and show that they yield
additional improvements over their randomized saturation design counterparts
under certain assumptions. Unlike general randomized saturation designs, which randomly assign treatment proportions to clusters in a first stage, deterministic saturation designs predetermine the treatment proportion for each cluster and forgo the initial randomization. Both of them randomly assign treatment within each cluster in the second stage; ``deterministic'' only refers to the first stage. 

\subsection*{Our contribution}
We conduct a complete analysis of the bias and variance of the difference-in-means estimator under any randomized saturation design. Furthermore, we provide general guidance for finding the optimal randomized saturation design in terms of bias, variance, or mean-squared error, particularly when a realistic linear interference model holds.

We start by assuming the stable unit treatment value assumption (SUTVA), where interference is absent. We show that, under SUTVA, all randomized saturation designs are unbiased and at least one of the stratified design and the cluster-based design has the minimum variance for the difference-in-means estimator among all randomized saturation designs. 

When interference is present, we assume a linear interference model where interference occurs both within and across clusters and units can receive heterogeneous levels of interference depending on their local neighbors. This assumption of interference is more realistic than that of the previous study \citep{baird2016optimal}, which assumed isolated clusters and homogeneous interference. 
We show that the closed form of the bias, the variance, and the mean-squared error of the difference-in-means estimator can be optimized analytically within a symmetric proportion family of randomized saturation design. Under this interference structure, we find that the optimal randomized saturation design is not necessarily the cluster-based design or the stratified design, unlike in the SUTVA case.

In addition, we propose the optimal deterministic saturation design, which can further reduce the variance/mean-squared error of the difference-in-means estimator when interference is present. Finding the optimal deterministic saturation design requires more knowledge of certain population statistics than finding the optimal randomized saturation design does. Using an optimal deterministic saturation design takes advantage of additional information when available to better design the experiment.

The manuscript is organized as follows. In Section~\ref{sec:randomized_designs}, we formally introduce randomized saturation designs and explore the bias and variance of the standard difference-in-means estimator under the stable unit treatment value (SUTVA) assumption, as well as under a heterogeneous linear model of interference. These results can be extended to random graph model setting and other model-assisted estimator as discussed in Section~\ref{sec:extensions}. In Section~\ref{sec:optimized_designs}, we introduce and define optimal deterministic saturation
designs and show that they can yield additional improvements over randomized
saturation designs, even optimal randomized ones. The benefits from optimizing the randomized saturation design, as well as the additional improvement obtained from optimizing the deterministic saturation design, is demonstrated in Section~\ref{sec:simulations} with simulations. We conclude this paper with practical considerations in Section~\ref{sec:conclusion}.

\section{Randomized Saturation Designs}
\label{sec:randomized_designs}
In this section, we formally define randomized saturation designs, and study the bias and variance of the standard difference-in-means estimator under various potential outcome models.

\subsection{Definitions}
\label{sec:randomized_definitions}

A randomized saturation design is any two-stage design that first assigns
clusters of experimental units at random to treatment proportions, and then
assigns the units within each cluster to treatment and control, respecting the
assigned treatment proportion for each cluster. Let $N$ be the number of experimental
units, let $\Y$ be their outcome vector, and let $\Z \in \{0,1\}^N$ be the
assignment vector stating whether each unit $i$ is in treatment ($Z_i = 1$) or
control ($Z_i = 0$). Let $M$ be the number of clusters of the experimental
units; they partition the experimental cohort such that each unit $i \in [1,N]$ belongs to exactly one cluster $\calC \in \{\calC_j\}_{j \in [1,M]}$. There are many possible kinds of randomization saturation designs. We
list two below, and show that they are equivalent when the number of clusters is large.

\begin{definition}
  The \emph{independently-sampled randomized saturation design} is a two-stage
  design defined by a probability distribution $\mathcal{D}$ on $[0,~1]$ and the
  following procedure: for each cluster $\calC_j$, sample $\pi_j \sim
  \mathcal{D}$ and assign $n_j = \lfloor \pi_j \cdot N_j \rfloor$
  randomly-chosen units of cluster $\calC_j$  to treatment and the remainder
  $N_j - n_j$ units of cluster $\calC_j$ to control.
\end{definition}

The independently-sampled randomized saturation design is entirely characterized
by its distribution $\mathcal{D}$. The total number of treated units
is a random variable, given by $n_t = \sum_{j = 1}^M \lfloor \pi_j \cdot N_j
\rfloor$. Assuming the size of each cluster is large ($N_j >> 1$), the expected
number of treated units over the sampling of $\bm{\pi} \sim D^M$ is the
expectation of $\mathcal{D}$ times the total number of experimental units $N$: $\E_{\bm{\pi} \sim D^M} \left[ n_t \right] \approx N \cdot \E_{\pi \sim D}[\pi]$.

\begin{definition}
The \emph{permutation-based randomized saturation design} is a two-stage
  randomized design defined by a fixed vector $\bm{\pi} \in [0,1]^M$ of length
  $M$ and the following procedure: sample a random permutation $P$ of $[1, M]$,
  letting $\bm{P}$ be the corresponding permutation matrix of $P$. For each
  block $\calC_j$, assign $n_j = \lfloor (\bm{P} \bm{\pi})_j N_j \rfloor$
  randomly-chosen units of $\calC_j$ to treatment, and the remainder $N_j - n_j$
  units of $\calC_j$ to control, where $(\bm{P \pi})_j$ is the $j^{th}$
  coordinate of the permuted vector $\bm{P \pi}$.
\end{definition}

The permutation-based design is entirely characterized by its vector $\bm{\pi}$.
The total number of treated units is fixed when the clusters are of
equal size: $n_t = \sum_{j =1}^M \left\lfloor \pi_j \frac{N}{M}\right\rfloor$. For this reason, we will always refer to the second implementation of randomized
saturation designs, unless stated otherwise. To simplify the analysis, we assume the clusters are of equal size throughout this paper so that the total number of treated units $n_t$ is fixed. This is the case when $M$ clusters of equal size are drawn from a super-population. A similar equal-sized cluster assumption was made in \citet{baird2016optimal}.
For further ease of exposition, we will assume that the number of units in each
cluster is large enough to ignore the flooring function.

The treatment-proportions vector $\bm{\pi}$ can be chosen explicitly by the
experimenter or be the result of an optimization program; it can also be
randomly sampled from a probability distribution.  In the latter case, the
independently-sampled and permutation-based randomized saturation designs are
equivalent when the number of clusters is large.  Assuming that the treatment
proportions vector is sampled from a probability distribution $\bm{\pi} \sim
\mathcal{D}^M$,  the $k^{th}$ moment of the number of units assigned to
treatment in each the permutation-based design is equal asymptotically to its
$k^{th}$ moment under the independently-sampled design by the law of large
numbers:
$
  \forall k \in \mathbb{N},~\E_P \left[(\bm{P \pi})_j^k \right] = \sum_{j=1}^M
  \frac{\pi_j^k}{M} \xrightarrow[M \rightarrow + \infty]{} \E_{\pi_j \sim
  \mathcal{D}}[\pi_j^k]
$
, where $\E_P \left[(\bm{P \pi})_j^k \right]$ is the $k^{th}$
moment of the $j^{th}$ coordinate of $(\bm{P \pi})$, and is shown to be
equivalent to the $k^{th}$ moment of the $j^{th}$ coordinate of $\pi$ sampled
according to $\mathcal{D}$, $\E_{\pi_j \sim \mathcal{D}}[\pi_j^k]$, when the
number of clusters is large.

Finally, both cluster-based randomized designs and stratified completely randomized designs are in fact
instantiations of randomized saturation designs.  The cluster-based randomized
design is an example of a randomized saturation design where $\bm{\pi} \in \{0,
1\}^M$, assigning either all units in a cluster to treatment or to control, whereas a stratified completely randomized
assignment, which assigns the same proportion of units to treatment in each cluster, corresponds to a
randomized saturation design with constant vector $\bm{\pi} =
\left(\frac{n_t}{N}\right)^M$.

In the subsequent sections, we adopt the following notational convention. 
The plain letter with subscript, $Y_i(\bm Z)$, stands for the potential outcome of unit $i$ under treatment assignment $\bm Z$. The plain letter with superscript, $Y^{(j)}(\bm Z)$, is the cluster-level potential outcome for clusters $j$ such that $Y^{(j)}(\bm Z):=\sum_{i\in\mathcal C_j} Y_i(\bm Z)$. The bolded letters, $\bm Y(\bm Z)$ and $\bm Y^{(j)}(\bm Z)$, denote the vector of all unit-level potential outcomes and the sub-vector restricted to cluster $j$, such that $\bm Y(\bm Z):=\{Y_i(\bm Z)\}_{i=1}^N$ and $\bm Y^{(j)}(\bm Z):= \{Y_i(\bm Z)\}_{i\in\mathcal C_j}$. We define $\bm Y^+(\bm Z):= (Y^{(1)}(\bm Z), \cdots, Y^{(M)}(\bm Z))$ to be the vector of all cluster-level potential outcomes under treatment assignment $\bm Z$. Finally, for any vector $\bm a\in\mathbb R^n$, $\overline{a}:=n^{-1}\sum_{i=1}^N a_i$ denotes its average, and
for any two vectors $\bm a, \bm b\in\mathbb R^n$, we define the sample covariance operator $\mathbb S(\cdot, \cdot)$ such that

\begin{equation}
\mathbb S(\bm a, \bm b) := \frac{1}{n-1}\sum_{i=1}^n \left(a_i - \overline{a}\right)\left(b_i - \overline{b}\right).
\label{eq:sample-covariance-operator}
\end{equation}

We adopt the convention $\mathbb S(\bm a):=\mathbb S(\bm a, \bm a)$ for the sample variance of vector $\bm a$. For example, $\mathbb S(\bm Y^{(j)}(\bm 1), \bm Y^{(j)}(\bm 0))$ is the sample within-cluster covariance between the potential outcomes with every unit treated and the potential outcomes with every unit untreated; $\mathbb S(\bm Y^+(\bm Z))$ is the sample variance of cluster-level potential outcomes under treatment $\bm Z$. 


\subsection{Bias and variance under SUTVA}

A starting point to understanding any class of designs is to consider the resulting bias and variance of the commonly used difference-in-means estimator under the Stable Unit Treatment Value Assumption (SUTVA)~\citep{imbens2015causal}. Let $\hat \tau$ denote the difference-in-means estimator, defined by
$$\hat \tau \defeq \frac{1}{n_t} \sum_{i= 1}^N Z_i Y_i(\Z) - \frac{1}{n_c} \sum_{i= 1}^N (1 - Z_i) Y_i(\Z),$$
where $n_t$ is the total number of treated units and $n_c \defeq N - n_t$ is the total number of control units. 

We have defined the difference-in-means estimator $\hat\tau$ under the assumption of equal cluster size. Without assuming equal cluster size, \citet{hudgens2008toward} defined the difference-in-means estimator at the cluster-level such that
$$\hat\tau = \frac{\sum_{j=1}^M\hat\tau_j(1) \mathbb I\{\pi_j > 0\}}{\sum_{j=1}^M \mathbb I\{\pi_j > 0\}} - \frac{\sum_{j=1}^M\hat\tau_j(0) \mathbb I\{\pi_j < 1\}}{\sum_{j=1}^M \mathbb I\{\pi_j < 1\}},$$
where $\hat\tau_j(1) = (N_j\pi_j)^{-1}\sum_{i\in\mathcal C_j}Y_i(1)$, $\hat\tau_j(0) = (N_j(1-\pi_j))^{-1}\sum_{i\in\mathcal C_j}Y_i(0)$ are the sample mean estimators for cluster $j$. This definition of $\hat\tau$ is identical to ours under a equal cluster size assumption, which we will make throughout the rest of this paper.


By the law of iterated expectations,
the shorthand $\E_\Z[\hat \tau]$ denotes $\E_{\bm{\pi}}\left[ \E_{\Z}
\left[ \hat{\tau} \middle| \bm{\pi} \right] \right]$, i.e. the expectation taken
with respect to the permutation of the treatment proportions assignment $\bm{\pi}$ to
clusters, and with respect to the assignment of units to treatment
and control $\Z$, conditioned on the assignment of $\bm{\pi}$. We first show that, when SUTVA holds, the difference-in-means
estimator $\hat\tau$ is unbiased under a randomized saturation design for the
total treatment effect $TTE \defeq \overline{\Y(1)} - \overline{\Y(0)}$.


\begin{proposition}
  \label{prop:bias_sutva}
  IF SUTVA holds,
  \begin{align*}
      \E_{\Z}\left[\hat{\tau} | \bm{\pi}\right] & =\frac{1}{n_t} \sum_{j=1}^M \pi_j
    Y^{(j)}(1) - \frac{1}{n_c} \sum_{j=1}^M (1 - \pi_j) Y^{(j)}(0) \\ 
    \E_{\Z}\left[\hat{\tau} \right] & = TTE
  \end{align*}
  where $Y^{(j)}\defeq \sum_{i \in \calC_j} Y_j$ is the cluster-level outcome of
  cluster $\calC_j$. 
\end{proposition}

%
A proof is included in the supplementary materials. From
Proposition~\ref{prop:bias_sutva}, the difference-in-means estimator is
\emph{not} guaranteed to be unbiased if we condition on a specific assignment of
clusters to treatment proportions. Only by randomizing over the assignment
of treatment proportion do we guarantee unbiasedness.

We can also give a concise formula of the variance of the difference-in-means
estimator under SUTVA and a randomized
saturation design. 


%
\begin{proposition}
  \label{prop:variance_sutva}
  When SUTVA holds, 
  the variance of the difference-in-means estimator under a randomized
  saturation design is
  \begin{align}
    \mathrm{Var}_{\bm Z}[\hat\tau]&=\frac{1}{n_tn_c}\sum_{j=1}^MN_j\mathbb S[\bm W^{(j)}] + \frac{N^2}{n_t^2n_c^2}\left[M\mathbb S[\bm W^+] - \sum_{j=1}^M N_j \mathbb S[\bm W^{(j)}]\right]\mathrm{Var}[\bm\pi],
  \label{eq:global_variance_sutva}
  \end{align}
  where $W_i:= \frac{n_t}{N}Y_i(0) + \frac{n_c}{N}Y_i(1)$ is a weighted average of the potential outcomes and $\bm W^{(j)}$, $\bm W^+$ denote the vector of $W_i$'s in cluster $j$ and the vector of all cluster-level $W$'s correspondingly.
\end{proposition}

A proof can be found in the supplementary materials.  The important
takeaway from Equation~\ref{eq:global_variance_sutva} is that the variance of the difference-in-means estimator for a randomized saturation design under SUTVA is linear in the empirical variance of the
treatment-proportions vector $\bm{\pi}$. Optimizing the variance of the difference-in-means
estimator under SUTVA, and holding the number of treated units $n_t$ constant,
can be reduced to choosing the optimal variance of the treatment proportions vector
$\bm{\pi}$. This leads to the following simple characterization for which randomized saturation design will lead to the lowest variance of the difference-in-means estimator under SUTVA.
\begin{corollary}
\label{cor:variance_regimes_sutva}
Assuming SUTVA and holding the number of treated units $n_t$ fixed, if 
    $\mathbb S[\bm W^+]\geq 1/M \sum_{j=1}^MN_j\mathbb S[\bm W^{(j)}]$
    , then $\min_{\bm{\pi}} \var_{\Z} \left[ \hat{\tau} \right]$ is attained when $\var[\bm{\pi}] =0$, corresponding to a stratified completely randomized assignment with $\bm{\pi} = \{\frac{n_t}{N}\}^M$. Otherwise, $\min_{\bm{\pi}} \var_{\Z} \left[ \hat{\tau} \right]$ is attained when $\var[\bm{\pi}] = \frac{n_t n_c}{N^2}$, corresponding to a cluster-based randomized assignment with $\bm{\pi} = \{0,1\}^M$
\end{corollary}
In other words, if the variance of the cluster-level aggregate outcomes is higher than the average of the intra-cluster outcome variances, then a cluster-based randomized assignment will only exacerbate the variance of the difference-in-means estimator. Without any further assumptions, a cluster-based randomized assignment is appropriate only when the variance of the cluster-level aggregate outcomes is lower than the average of the intra-cluster outcome variances. Furthermore, only a stratified completely randomized design or a cluster-based randomized designs can be the variance-minimizing design in the class of randomized saturation designs for the difference-in-means estimator under SUTVA, unless 
    $\mathbb S[\bm W^+]= M^{-1}\sum_{j=1}^MN_j\mathbb S[\bm W^{(j)}]$, in which case, all randomized saturation designs will lead to the same variance and mean-squared error. A proof of Corollary~\ref{cor:variance_regimes_sutva} is included in the supplementary materials.

\subsection{Bias under a linear interference model}

In the previous section, we explored the bias and variance of the
difference-in-means estimator under SUTVA.
In this section, we seek to extend these results to a setting where interference
is present. For the sake of exposition, we will focus on a commonly-used linear model of
interference. Consider a network over the units of experimentation, such that an edge between
two units indicates they are likely to interfere with one another. Let the
neighborhood $\N_i$ of unit $i$ be the set of all units linked by a direct edge
to unit $i$ and let $(\alpha_i, \beta_i, \gamma_i) \in \mathbb{R}^3$, such that the
outcome of unit $i$ can be expressed as
\begin{equation}
  \label{eq:dirg:ch3}
  Y_i(\Z) = \alpha_i + \beta_i Z_i + \gamma_i \rho_i
\end{equation}
where $\rho_i = \frac{1}{|\N_i|} \sum_{j \in \N_i} Z_j$ is the proportion of
$i$'s neighborhood that is treated.
The $\beta_i$ coefficient can be interpreted as a
direct effect parameter, while the $\gamma_i$ coefficient can be interpreted as
an interference parameter: if $\forall i, \gamma_i = 0$, then SUTVA holds. The linear model of interference in Equation~\ref{eq:dirg:ch3} is an example of an anonymous interaction model~\citep{manski2013identification} for which randomized saturation designs are appropriate. For any two assignment vectors $\Z$ and $\Z'$, such that
the treatment status of unit $i$ and the number of its treated neighbors is identical, unit $i$'s outcome is held constant:
$Z_i=Z'_i\text{ and }\sum_{j \in \N_i} Z_j = \sum_{j \in \N_i} Z'_j \implies Y_i(\Z) = Y_i(\Z')$. See \citet{eckles2017design} and \citet{forastiere2021identification} for more details on the linear interference model.

We adopt the same notational conventions for $\alpha_i, \beta_i, \gamma_i$ as we did for $Y_i(\bm Z)$ in Section~\ref{sec:randomized_definitions} such that $\alpha_i$ , $\alpha^{(j)}$, $\bm\alpha^{(j)}$, $\bm\alpha$ and $\bm\alpha^+$ correspond to the unit-level value, the cluster-level value, the vector of unit-level values in cluster $j$, the vector of all unit-level values, and the vector of cluster-level values of the parameter $\alpha$ correspondingly. 
We begin by quantifying the total treatment effect for this linear model of
interference, for which a proof is provided in the supplementary materials.

\begin{proposition}
  \label{prop:tte:ch3}
Under the model of interference in Equation~\ref{eq:dirg:ch3}, the total
  treatment effect is the sum of the average direct effect and the average interference effect: $TTE = \bar{\bm{\beta}} + \bar{\bm{\gamma}}$.
\end{proposition}

To write the bias of the classic difference-in-means estimator in closed-form,
we introduce the following linear combination of the different components of
$\bm{\gamma}$, where each component is down-weighted by the proportion of
each unit's intra-cluster number of neighbors to its total number of neighbors:
$
\gamma' \defeq \frac{1}{N} \sum_j \sum_{i \in \calC_j}
\gamma_i\frac{|\N_i \cap \calC_j|}{|\N_i|}$.
$\gamma'$ can be interpreted as a measure of clustering quality and is contained in the segment
$[0, \overline{\gamma}]$. For a perfect clustering where no graph edges are cut (i.e. endpoints belong to different clusters), $\gamma' = \bar {\gamma}$. For a
random clustering, $\gamma' \approx \frac{\bar {\gamma}}{M}$. For a clustering
which places no unit in the same cluster as one of its neighbors, $\gamma' = 0$. The expectation, and by extension the bias, of the difference-in-means estimator under the linear interference model defined in Eq.~\ref{eq:dirg:ch3}, can be expressed using $\gamma'$.

\begin{theorem}
  \label{thm:bias_cond_graph}
  Assuming the linear interference model in Eq.~\ref{eq:dirg:ch3},
the expectation of the difference-in-means estimator is given by
  $\E_\Z \left[\hat \tau \right] = \bar \beta + \frac{N^2}{n_t
  n_c} \left(\gamma' - \frac{\bar \gamma - \gamma'}{M-1} \right) \var[\bm{\pi}]$.
\end{theorem}
The important takeaway of Theorem~\ref{thm:bias_cond_graph} is that the
expectation---and therefore bias---of the difference-in-means estimator
under a randomized saturation design is linear in the empirical variance of the
treatment-proportions vector $\bm{\pi}$.  Much like in Proposition~\ref{prop:variance_sutva}, optimizing the bias of a randomized
saturation design under the linear model of interference in Eq.~\ref{eq:dirg:ch3} can be reduced to choosing the optimal variance of the treatment-proportions vector. This leads to the following characterization of which randomized saturation design minimizes the bias of the difference-in-means estimator.

\begin{corollary}
  \label{cor:linear_interference_bias_randomized}
  Assume 
  that the linear
  interference model from Equation~\ref{eq:dirg:ch3} holds. We can distinguish
  three cases. If $\gamma' \geq \frac{\bar \gamma}{M}$, then the bias of the
  difference-in-means estimator under a randomized saturation design is
  minimized for a cluster-based randomized assignment $\bm{\pi} \in \{0,1\}^M$. The bias is then
  equal to
    $|TTE - \E_\Z[\hat \tau]| = \frac{M}{M-1} \left( \bar \gamma - \gamma'\right)$.
  If $\gamma' \leq \frac{\bar\gamma}{M}$, then the bias of the
  difference-in-means estimator under a randomized saturation design is
  minimized for a constant treatment-proportions vector $\bm{\pi} =
  \left(\frac{n_t}{N}\right)_M$ and is equal to $|TTE - \E_\Z[\hat \tau]| = \bar \gamma$.
  If $\gamma' = \frac{\gamma}{M}$, then both results hold: the bias is constant,
  such that every randomized saturation design minimizes the bias.
\end{corollary}

A proof is included in the supplementary materials. The significance of $\frac{\gamma}{M}$ as the cut-off point is intuitive: when
the graph is randomly-clustered, $\gamma' \approx \frac{\gamma}{M}$. Hence, the
first regime corresponds to a ``better-than-random'' clustering of the
experimental units, where cluster-based randomized designs will improve the bias of the difference-in-means estimator, while the second regime corresponds to a
``worse-than-random'' clustering. In conclusion, to optimize the bias of the difference-in-means estimator for a
randomized saturation design under the linear interference model in
Equation~\ref{eq:dirg:ch3}, the optimal randomized saturation design is
either a stratified completely randomized design or a cluster-based randomized
design---the parameter $\gamma'$, an indicator of the quality of the clustering,
being the deciding factor between the two.


\subsection{Variance under a linear interference model}
\label{sec:randomized_variance}

In the previous section, we discussed the bias of the difference-in-means estimator under a linear interference model, and provided a treatment-proportions vector $\bm \pi$ which minimizes this bias. In certain circumstances, we may be more interested in minimizing the variance of the estimator instead of its bias. We explore this in the following section, under the same linear interference model from Eq.~\ref{eq:dirg:ch3}.

Under the asymptotic regime where both the number of clusters and the size of each cluster are large enough, we can express the variance of the difference-in-means estimator in closed-form, in terms of the centralized moments of $\bm\pi$, as shown in the following theorem. 

\begin{theorem}\label{thm:variance-dim}
Suppose $M\rightarrow\infty$ and $\inf_j N_j\rightarrow \infty$. The total variance of the difference-in-means estimator is given by
\begin{equation}
    \mathrm{Var}_{\bm Z}[\hat\tau]= V_0 + V_1\mathrm{Var}[\bm\pi] + V_2\mathrm{Var}^2[\bm\pi]+V_{3}\mu_{3c}[\bm\pi] + V_4(\mu_{4c}[\bm\pi] - \mathrm{Var}^2[\bm\pi]) + o(N^{-1}),\label{eq:variance-dim}
\end{equation}
where $\mathrm{Var}[\bm\pi]$ is the variance of the vector $\bm\pi$, and $\mu_{3c}[\bm\pi]:= M^{-1}\sum_{j=1}^M(\pi_j-\bar{\bm\pi})^3$ and $\mu_{4c}[\bm\pi]:= M^{-1}\sum_{j=1}^M(\pi_j-\bar{\bm\pi})^4$ are the third and fourth central moments of the vector $\bm\pi$. All five coefficients $V_0$ to $V_4$ depend only on the potential outcomes as well as certain statistics of the interference graph.
\end{theorem}


The explicit formulas for $V_0$ to $V_4$ are provided in the supplementary materials so as to not overburden the reader with notation. Unlike previous results, which were linear in the variance of the treatment-proportions vectors $\mathrm{Var}[\bm \pi]$, the total variance of the difference-in-means estimator $\mathrm{Var}_{\bm Z}[\hat\tau]$ depends on up to the fourth central moment of the treatment-proportions vector $\bm \pi$, as expressed in \eqref{eq:variance-dim}. 

\subsubsection{Simplifying assumptions}

Before determining which treatment-proportions vector $\bm\pi$ minimizes the variance of the difference-in-means estimator $\mathrm{Var}_{\bm Z}[\hat\tau]$, we first introduce a set of common assumptions under which the expressions for the coefficients $V_0,\dots, V_4$ can be significantly simplified. 

\begin{assumption}\label{assump:variance-sufficient}
  As $M\rightarrow\infty$ and $\inf_j N_j\rightarrow \infty$, we suppose
  \begin{thmlist}
  \item \label{assump:boundedness}(Boundedness) There exists a constant $\epsilon_1 > 0$ such that $\max_i |\alpha_i| + |\beta_i|+|\gamma_i| < \epsilon_1$.

  \item \label{assump:dense}(Dense Connection) There exists a constant $\epsilon_2>0$ such that $\min_i |\mathcal N_i|\geqslant \epsilon_2N/M$.

  \item \label{assump:edge-prob}(Proxy Edge Probability) There exists a constant $\epsilon_3 > 0$ such that for all $1\leqslant j, l\leqslant M$, we have
  $$p_{jl} - \epsilon_3\sqrt{\frac{p_{jl}}{N_l}\log (NM)}\leqslant \min_{i\in\mathcal C_j}\frac{|\mathcal N_i\cap\mathcal C_l|}{N_l}\leqslant \max_{i\in\mathcal C_j}\frac{|\mathcal N_i\cap\mathcal C_l|}{N_l}\leqslant p_{jl} + \epsilon_3\sqrt{\frac{p_{jl}}{N_l}\log (NM)},$$
  where 
  $$p_{jl}:= \frac{\sum_{i\in\mathcal C_j}|\mathcal N_i\cap\mathcal C_l|}{N_jN_l}=\frac{\text{number of edges between $\mathcal C_j$ and $\mathcal C_l$}}{N_jN_l}$$
  is the observed edge-formation probability between cluster $\mathcal C_j$ and $\mathcal C_l$.


  \item \label{assump:network-unconfoundedness}(Unconfoundedness of Network) We assume the edge formation between units $i$ and $k$ is approximately independent with their potential outcome parameters $(\alpha_i, \beta_i, \gamma_i, \alpha_k, \beta_k, \gamma_k)$ in the sense that, for any fixed bounded function of potential outcome parameters $f(\alpha, \beta, \gamma)$, there exists a constant $\epsilon_f>0$ such that for any unit $i$ and for any cluster $l$ 
  $$\left|\frac{1}{N_j}\sum_{i\in\mathcal N_k\cap\mathcal C_j}f(\alpha_i,\beta_i,\gamma_i)-\frac{|\mathcal N_k\cap \mathcal C_j|}{N_j}\frac{\sum_{i\in\mathcal C_j}f(\alpha_i,\beta_i,\gamma_i)}{N_j}\right| \leqslant \epsilon_f\sqrt{\frac{\log (NM)}{N_j}}.$$
\end{thmlist}
\end{assumption}
Assumption~\ref{assump:boundedness} assumes the potential outcome parameters are uniformly bounded as the size of the network increases to infinity. 
Assumption~\ref{assump:dense} requires that the degree of each unit be at least of the same order as its cluster size $\sim N/M$, such that for unit $i$, the individualistic interference effect $\gamma_i\rho_i$ in \eqref{eq:dirg:ch3} can be well approximated by its expectation $\gamma_in_t/N$ with a negligible deviation of order $O(|\mathcal N_i|^{-1/2})$. 
In Assumption~\ref{assump:edge-prob}, $p_{jl}$ ($=p_{lj}$) is the ratio of the number of observed edges to the maximum possible number of edges between clusters $\mathcal C_j$ and $\mathcal C_l$. $p_{jl}$ serves as a proxy edge-formation probability between clusters $\mathcal C_j$ and $\mathcal C_l$, such that the proportion $|\mathcal N_i\cap\mathcal C_l|/N_l$ of observed edges formed with unit $i$ in cluster $\mathcal C_j$ departs at most $O(N_l^{-1/2})$ from the proxy probability $p_{jl}$. The factor $\log (NM)$ in Assumption~\ref{assump:edge-prob} comes from a union bound over all possible pairs of $(i, l)$.

Finally, assumption~\ref{assump:network-unconfoundedness} impose a bound on the sample covariance between the edge formations $\{\mathbb I\{i\in\mathcal N_k\}\}_{i\in\mathcal C_j}$ and the potential outcomes $\{f(\alpha_i, \beta_i, \gamma_i)\}_{i\in\mathcal C_j}$. 
When this upper-bound is small, the formation of edges with some unit $i$ in cluster $\mathcal C_l$ is approximately independent from the potential outcome parameters of that unit $i$, hence the name ``unconfoundedness of network''.


In Section~\ref{sec:extensions}, we extend our results to graphs which are generated by a known random process. In particular, we will show that if the graph is generated under a stochastic block model, Assumption~\ref{assump:variance-sufficient} is satisfied with high probability. While Assumptions~\ref{assump:boundedness}-\ref{assump:edge-prob} are common assumptions in real applications, Assumption~\ref{assump:network-unconfoundedness} may require further examination. In the stochastic block model example, Assumption~\ref{assump:network-unconfoundedness} holds because the edge forming probability only depends on a predetermined probability matrix; it may fail under other random graph models that incorporate potential outcomes in the graph-generating process (e.g. a graphon model where the edge-forming probability is a bivariate function of the potential outcomes of the two nodes). Handling cases where the graph is confounded with the potential outcomes is out-of-scope for our paper. 

To further simplify Eq.~\eqref{eq:variance-dim}, we can assume that the interference effects are ``block-fixed'', formalized in the following assumption---the full expression for Eq.~\eqref{eq:variance-dim} without this simplifying assumption can be found in the supplementary materials.

\begin{assumption}[Block-fixed Interference Effect]\label{assump:block-fixed-effect}
The interference effect is fixed within each cluster $\mathcal C_j$ such that $\gamma_i = \gamma_{[j]}$, $\forall i\in\mathcal C_j$, where $\gamma_{[j]}$ is the common interference parameter in cluster $\mathcal C_j$.
\end{assumption}

Under Assumption~\ref{assump:block-fixed-effect}, we can rewrite $\gamma'=\frac{1}{M} \sum_j \frac{\gamma_{[j]}}{N_j}\sum_{i \in \calC_j}
\frac{|\N_i \cap \calC_j|}{|\N_i|}$, where the common value of interference parameters $\gamma_{[j]}$ in each cluster $\mathcal C_j$ is down-weighted by the average proportion of intra-cluster edges per cluster.

\subsubsection{Simplified form}

We provide the simplified form of the coefficients in Theorem~\ref{thm:variance-dim} under Assumption~\ref{assump:variance-sufficient} and Assumption~\ref{assump:block-fixed-effect} (block-fixed interference effect) in the following corollary.

\begin{corollary}\label{cor:var-simplified}
  Under Assumption~\ref{assump:variance-sufficient} and Assumption~\ref{assump:block-fixed-effect} , the coefficients in Eq.~\eqref{eq:variance-dim} can be simplified to
\begin{align*}
    V_0&=\frac{2}{\tilde n} \sum_{j=1}^M\frac{N_j}{N} \mathbb S\left(\bm\alpha^{(j)} + \frac{n_c}{N}\bm\beta^{(j)}\right)\\
    V_1&=\frac{4M}{\tilde n^2}\left[\mathbb S\left(\bm\alpha^{+} + \frac{n_c}{N}\bm\beta^{+} + \frac{n_t}{N}\bm\gamma^{+}\right) - \frac{1}{M}\sum_{j=1}^MN_j\mathbb S\left(\bm\alpha^{(j)} + \frac{n_c}{N}\bm\beta^{(j)}\right)\right]\\
    V_2&=\frac{2}{\tilde n^2}\sum_{j=1}^M\mathbb S[\{q_{jl}\gamma^{(j)} + q_{lj}\gamma^{(l)}:l\neq j\}]\\
    V_3&=\frac{8M}{\tilde n^2}\mathbb S\left(\bm\alpha^{+} + \frac{n_c}{N}\bm\beta^{+} + \frac{n_t}{N}\bm\gamma^{+}, \tilde{\bm\gamma}^{+}\right)\\
    V_4&=\frac{4M}{\tilde n^2}\mathbb S\left(\tilde{\bm\gamma}^{+}\right),
\end{align*}
where $\tilde n:= \frac{2n_t n_c}{N}$ is the harmonic mean of $(n_t, n_c)$, $q_{jl}:=p_{jl}/\sum_k p_{jk}$ is the row-normalized transformation of the proxy edge-forming probabilities $p_{jl}$, and $\tilde{\bm\gamma}^{+}$ is the vector of down-weighted cluster-level interference parameter, whose coordinates are $\tilde{\gamma}^{(j)}:= q_{jj}\gamma^{(j)}$. Recall that $\mathbb S(\cdot)$ and $\mathbb S(\cdot, \cdot)$ are the sample variance and sample covariance operators introduced in \eqref{eq:sample-covariance-operator}.
\end{corollary}



It is also possible to compute these coefficients in closed-form under Assumption~\ref{assump:variance-sufficient} without the block-fixed effect assumption (Assumption~\ref{assump:block-fixed-effect}). To ease the notational burden on the reader, we defer this formula to the supplementary material, and present only their simplified form here. 

By definition of the coefficients
$V_0,\dots,V_4$ in Eq.~\ref{eq:variance-dim}, the coefficient $V_0$ corresponds to the variance of the difference-in-means estimator under a stratified completely randomized assignment, where $\forall i~\pi_i = \frac{n_t}{N}$, such that $\var[\pi] = \mu_{3c}[\pi] = 0$ and $\mu_{4c}[\pi] = \var^2[\pi]$. As expected, if $\forall i~\gamma_{i} = 0$, the expression for $V_0$ coincides with the variance of the difference-in-means estimator under SUTVA, found in Equation~\ref{eq:global_variance_sutva} of Proposition~\ref{prop:variance_sutva}, for a stratified completely randomized assignment. Similarly, the expression for $V_1$, the coefficient before the variance of $\bm\pi$, is almost identical to the expression found of the analog parameter in the SUTVA case, in Equation~\ref{eq:global_variance_sutva} of Proposition~\ref{prop:variance_sutva}, except that the between-cluster sample covariance terms $\mathbb S(\bm\alpha^{+} + \frac{n_c}{N} \bm\beta^{+})$ now appear with interference coefficients: $\mathbb S(\bm\alpha^{+} + \frac{n_c}{N} \bm\beta^{+} + \frac{n_t}{N} \bm\gamma^{+})$.

The connection between the expressions for the coefficients $V_0$ and $V_1$ in the linear interference model and in the SUTVA case is salient when each unit has a sufficiently large number of neighbors. In that case, using the law of large numbers, the potential outcomes of unit $i$ can be approximated by 
$$Y_i(\bm Z) = \alpha_i +\beta_i Z_i + \frac{\gamma_i}{|\mathcal N_i|}\sum_{k\in\mathcal N_i}Z_k\approx \alpha_i + \beta_i Z_i + \frac{n_t}{N}\gamma_i=\tilde\alpha_i + \beta_i Z_i,$$
where $\tilde\alpha_i = \alpha_i+ \frac{n_t}{N}\gamma_i$. As a result, in the asymptotic regime where each unit has many neighbors in the graph, the coefficients $V_0$ and $V_1$ have the same expression as the SUTVA case by replacing $\alpha$ with $\alpha + \frac{n_t}{N}\gamma$.

The coefficients $V_2,\dots,V_4$ do not have similar correspondences in the expression of the variance of the estimator under SUTVA.
Recall that $p_{jl}$ corresponds to the average proportion of neighbors, belonging to cluster $l$, of units in cluster $j$; $q_{jj}$ corresponds to the within-cluster proportion of neighbors in cluster $j$. The coefficient $V_2$ can be described as the between-cluster second moment of $\gamma$, weighted by the proportions $q_{jl}$. The coefficient $V_3$ is the between-cluster sample covariance of the potential outcome combination $\alpha^{(j)}+\frac{n_t}{N}\beta^{(j)}+\frac{n_c}{N}\gamma^{(j)}$ and the interference parameter $\gamma^{(j)}$, down-weighted by the intra-cluster edge proportion $q_{jj}$. Similarly, $V_4$ is the sample covariance of interference parameters $\gamma^{(j)}$, down-weighted by $q_{jj}$. In the extreme case when the clusters are isolated, we have $V_2=0$ and $q_{jj}=1$ in both $V_3$ and $V_4$.

\subsubsection{Choosing a variance-minimizing distribution of $\bm{\pi}$ under interference}

Previously, the statistics of interests---like the bias and variance under SUTVA, and the bias under interference---were linear in the variance of $\bm\pi$ such that determining an optimal distribution of $\bm\pi$ was straightforward.
However, finding the optimal variance $\mathrm{Var}_{\bm Z}[\hat\tau]$ under interference over $\bm\pi\in [0, 1]^M$ is more challenging as the variance depends on the first four central moments of the vector $\bm\pi$. It is an especially difficult problem when $M$ is large. In order to minimize $\mathrm{Var}_{\bm Z}[\hat\tau]$ in a tractable way, we consider optimizing it within the following set for the treatment proportions vector $\bm\pi$:
\begin{equation}
    \mathcal F:=\left\{(\pi_1,\dots, \pi_M)\in[0, 1]^M: \pi_1\leqslant \cdots \leqslant \pi_M \text{ and }\pi_j + \pi_{M+1-j} = \frac{2n_t}{N},\ \forall j \in[1,M]\right\}\label{eq:symmetric-dist}
\end{equation}
The set $\mathcal F$ consists of all symmetric M-tuples bounded within $[0, 1]$ and centered at $n_t/N$.  One benefit of considering $\bm\pi\in\mathcal F$ is that it contains only symmetric distributions, which have zero skewness, such that the third moment term in~\eqref{eq:variance-dim} disappears. Another benefit of considering the set of distributions $\mathcal F$ is the ability to optimize the variance with respect to the second moment and the fourth moment separately. To see this, we observe that $\mathcal F$ fixes the mean $\bar{\bm\pi}$ at $n_t/N$ and that the vector $\bm\pi$ can be entirely determined by its first half of coordinates $0\leqslant \pi_1\leqslant\cdots\leqslant \pi_{\lceil (M+1)/2\rceil}\leqslant n_t/N$. Define the square of the distance between each point and the mean as
$\delta_j = \left(\pi_j -\frac{n_t}{N} \right)^2,\quad\forall 1\leqslant j\leqslant \left\lceil \frac{M+1}{2}\right\rceil.$
One can easily verify that $\mathrm{Var}[\bm\pi] = \bar\delta$ and $\mu_{4c}[\bm\pi] - \mathrm{Var}^2[\bm\pi] = \overline{\delta^2} - \overline\delta^2=\mathrm{Var}[\bm\delta]$, which correspond to the mean and variance of $\bm\delta=(\delta_1,\dots, \delta_{\lceil (M+1)/2\rceil})$ respectively. 
To optimize the variance with respect to the second and fourth moments separately, we begin by maximizing (or minimizing) $\mathrm{Var}[\bm\delta]$, depending on the sign of $V_4$. The resulting optimization program is quadratic in $\overline\delta$, which we can set independently  of $\mathrm{Var}[\bm\delta]$ to its optimal value. Finally, we set $\mathrm{Var}[\bm\delta]$ to its optimal value as suggested in the first step of the optimization, consider the case when $n_t \leqslant N/2$ and each element in $\bm\delta$ is bounded between $0$ and $n_t/N$, without loss of generality. For any given $\mathrm{Var}[\bm\pi]$ (or equivalently, a given $\overline{\delta}$), $\mu_{4c}[\bm\pi] - \mathrm{Var}^2[\bm\pi]$ is lower bounded by 0, attained when $\delta_j = \overline\delta,\forall j$, and $\mu_{4c}[\bm\pi] - \mathrm{Var}^2[\bm\pi]$ is upper bounded by $n_t\bar\delta/N - \bar\delta^2$, attained when $N\bar\delta/n_t$-portion of $\bm\delta$ are at $n_t/N$ and the rest stays at $0$. This procedure is formalized in the following proposition.

\begin{proposition}\label{prop:optimizatin-closed-form}
  Consider the optimization such that
  $$\bm\pi^* = \argmin_{\bm\pi\in\mathcal F}\ \mathrm{Var}_{\bm Z}[\hat\tau],$$
  where $\mathrm{Var}_{\bm Z}[\hat\tau]$  and $\mathcal F$ are given by \eqref{eq:variance-dim}  and \eqref{eq:symmetric-dist} correspondingly. Suppose, without loss of generality, $n_t \leqslant N/2$. Then, we have the following optimal values $\bm\pi^*$.
  \begin{enumerate}[label=(\roman*)]
      \item if $V_4 \geqslant 0$, the first half of $\pi_j^*$'s lie at $\frac{n_t}{N}-d$ and the second half of $\pi_j^*$'s lie at $\frac{n_t}{N}+d$, 
      where 
      $$d = \argmin_{x\in[0, n_t/N]}\ V_1x^2 + V_2x^4.$$
      \item if $V_4 < 0$, $\frac{N^2d^2}{2n_t^2}$-portion of $\pi_j^*$'s lie at $0$,
      $\left(1-\frac{N^2d^2}{n_t^2}\right)$-portion of $\pi_j^*$'s lie at $n_t/N$, and $\frac{N^2d^2}{2n_t^2}$-portion of $\pi_j^*$'s lie at $2n_t/N$, where 
      $$d = \argmin_{x\in[0, n_t/N]}\  \left(V_1+\frac{n_t^2}{N^2}V_4\right)x^2 + (V_2-V_4)x^4.$$
  \end{enumerate}
\end{proposition}
We considered optimizing the variance of the difference-in-means estimator under three different settings of clustering quality and interference effect structure as illustrating examples to the reader. For the sake of brevity, we relegated two of these examples to the appendix (cf. Examples~\ref{example:interference2} and Examples~\ref{example:interference3} in Section~\ref{sec:appendix_examples} of the appendix).
\begin{example}[Perfect clustering and block-fixed interference]\label{example:interference1}
Suppose Assumption~\ref{assump:variance-sufficient} holds, interference effects are block-fixed and the graph is perfectly clustered such that
$$q_{jl}=\mathbb I\{j=l\}\quad \forall 1\leqslant j,l\leqslant M.$$ Assume $n_t \leqslant N/2$. \\
(i) If the intra-cluster variance of potential outcomes dominates the inner-cluster one ($V_1 > 0$), the optimal assignment vector is
$\pi_j^* = \frac{n_t}{N}$ for all $j$. The optimal variance is 
$$\mathrm{Var}_{\bm Z}[\hat\tau] = \frac{2}{\tilde n}\sum_{j=1}^M\frac{N_j}{N}\mathbb S\left(\bm\alpha^{(j)}+\frac{n_c}{N}\bm\beta^{(j)}\right).$$
(ii) If the intra-cluster variance of potential outcomes dominates the inter-cluster one ($V_1 < 0$), the optimal assignment vector is
$$\pi_j^* = \begin{cases}
0 &,\ j< M/2,\\
\frac{2n_t}{N}&,\ j>M/2.
\end{cases}$$
The corresponding optimal variance is
$$\mathrm{Var}_{\bm Z}[\hat\tau] = \frac{M}{n_c^2}\mathbb S\left(\bm\alpha^{+}+\frac{n_c}{N}\bm\beta^{+}+\frac{n_t}{N}\bm\gamma^{+}\right)+ \frac{n_c-n_t}{n_c}\frac{2}{\tilde n}\sum_{j=1}^M\frac{N_j}{N}\mathbb S\left(\bm\alpha^{(j)}+\frac{n_c}{N}\bm\beta^{(j)}\right).$$
\end{example}

When $n_t=n_c$, in case (i) and case (ii) of Example~\ref{example:interference1}, the optimal designs are the stratified design and cluster-based design, respectively. In Example~\ref{example:interference3}, relegated to the appendix, we show the same is true even if the interference effects are not block-fixed. However, when the units are randomly clustered--with block-fixed interference effect---the optimal design is no longer one of the two extremes necessarily, as shown in case (ii) of Example~\ref{example:interference2}, which can also be found in the appendix.

\subsection{Extensions}\label{sec:extensions}

There are some important extensions that can be made to the results presented here. For example, we focused thus far on the difference-in-means estimator, but other estimators may also be appropriate. While the difference-in-means estimator is agnostic to any model assumptions or validity of the clustering---and is unbiased if the stable unit treatment value assumption holds---we might benefit from using a stratified estimator if we believe that the clustering of units is representative of the potential outcomes in some way. When the stable unit treatment value assumptions holds, we can choose an appropriate configuration for the stratified estimator
such that it becomes unbiased, conditionally on the assignment of treatment proportions to clusters $\bm\pi$, and in expectation over any randomized saturation assignment. We can also show that its variance depends only on the harmonic mean of $\bm\pi$ and covariances of the potential outcomes. Finally, we can show that under the linear interference model introduced in Eq.~\ref{eq:dirg:ch3}, the expectation of the stratified estimator is a constant function of $\mathbb{\pi}$. To improve the brevity of our paper, we have relegated these initial extensions to the supplementary materials.

Furthermore, many of our results are implicitly conditioned on a fixed observed graph. In certain cases, it may be more appropriate to consider an underlying random graph model and include this randomness as an additional integration step in our results. Consider, for example, one of the simplest and well-studied random graph models: the stochastic block
model~\citep{holland1983stochastic, anderson1992building, wasserman1994social,
goldenberg2010survey}.  It states that the probability that an edge exists
between two units in a graph $G$ depends only on the clusters they belong to. In
other words, two units belonging to clusters $\calC_k$ and $\calC_l$---with $l$
and $k$ possibly equal---are linked by an edge with probability $A_{kl}$. We
define $\mathbf{A} \defeq (A_{kl}) \in \mathbbm{R}^{M^2}$ the
\emph{block-matrix} of the graph $G$, such that $\forall (i,j,k,l),~i \in \calC_k,~j \in \calC_l \implies \proba((i, j) \in G)
  = A_{k,l}$. 
Under such a random graph model, the measure of clustering quality $\gamma'$, introduced in Theorem~\ref{thm:bias_cond_graph}, has expectation $\E_{\bm{A}}[\gamma'] \approx \frac{1}{M} \sum_{j=1}^M \frac{\gamma^{(j)}}{N_j} \frac{A_{jj}}{\sum_k A_{jk} N_k}$, with respect to the stochastic block-model with block-matrix $\mathbf{A}$. Because all other quantities in Theorem~\ref{thm:bias_cond_graph} are constant with respect to the graph, the expectation of the difference-in-means estimator $\E_{\bm{Z}}[\hat \tau]$ can easily be extended to incorporate the random graph model $\E_{\bm{A}, \bm{Z}}[\hat \tau]$ by replacing $\gamma'$ with its expectation $\E_{\bm{A}}[\gamma']$, computed above. Additionally, by considering a stochastic block model, we can ensure with high probability the validity of the assumptions made to simplify the expression of Equation~\ref{eq:variance-dim}.

\begin{proposition}\label{prop:random-graph-assumption}
Suppose that the interference graph is generated according to a stochastic block model with block-matrix $\bm A$, such that $M\ll N/\log N$. Let $\underline{A}=\min_j\ \sum_{l=1}^MA_{jl}$, then Assumptions \ref{assump:dense}-\ref{assump:network-unconfoundedness} are satisfied with high probability. More specifically,
for any constant $\epsilon_2 < \underline{A}$, Assumption~\ref{assump:dense} is satisfied with probability at least $1 - \exp\{-\frac{N}{4M\underline{A}}(\underline{A}-\epsilon_2)^2\}$; for any constant $\epsilon_3 > \sqrt{3}$, Assumption~\ref{assump:edge-prob} is satisfied with probability at least $1-(NM)^{1-\epsilon_3^2/3}$; for any constant $\epsilon_f > 2\sqrt{3}\|f\|_\infty$, Assumption~\ref{assump:network-unconfoundedness} is satisfied with probability at least $1-(NM)^{1-\epsilon_f^2/(12\|f\|_\infty^2)}$.
\end{proposition}
A proof is included in the supplementary materials. A more thorough extension of our results to other random graph models is left for future work.

\section{Deterministic Saturation Designs}
\label{sec:optimized_designs}
In the previous section, we investigated randomized saturation designs, which assign random treatment proportions to clusters of the experimental cohort. We then analysed the bias and variance of the
difference-in-means estimator for this class of randomized
saturation design. From these results, we determined which randomized
saturation design optimized these objectives, under a regime where SUTVA holds and a regime where a linear model of
interference holds. We show that the bias and variance of these estimators can often be
expressed in terms of moment of the treatment-proportions vector
$\bm{\pi}$, therefore reducing the objective of finding  the ``optimal randomized saturation designs'' among all possible vectors $\bm{\pi}$ to
optimizing over the moments of this vector instead.

This optimization is limited by the random assignment of coordinates of
$\bm{\pi}$ to each cluster. \emph{Optimal deterministic saturation designs}, which we
introduce below, go one step further in their optimization by removing the
permutation step and choosing the optimal treatment proportion per cluster.
\begin{definition}
\label{def:locally_randomized}
  Let $f$ be an objective function, taking as input a treatment-proportions
  vector $\bm{\pi}$, a clustering $\calC$ of the experimental units, and a set
  of parameters $\Theta$. Let $\calS$ be an allowable set of
  treatment-proportions vectors. An \emph{optimal deterministic} design selects
  $\bm{\pi}^* \in \calS$ that minimizes $f$
  \begin{equation}
    \bm{\pi}^* \in \arg \min_{\bm{\pi} \in \calS} f(\bm{\pi}, \calC, \Theta)
  \end{equation}
  and, for each cluster
  $\calC_j$, assigns $n_j = \lfloor \pi_j N_j \rfloor$ randomly-chosen units to treatment and the remaining $N_j - n_j$ units to control.
\end{definition}

It is up to the practitioner to choose which objective function $f$ is most relevant. Generally, she will choose $f$ to be the bias, variance, or mean-squared error of her estimator of choice, conditioned on the assignment of a specific treatment proportion to each cluster. In fact, many of these conditional expectations and variances were previously computed in Section~\ref{sec:randomized_designs}, as a step in applying Adam's law (law of iterated expectations) or Eve's law (law of total variance). We list some common examples below, and show how each objective can be optimized in each scenario.

\begin{example}
  \label{ex:sutva_bias_optimized}
Let $f$ be the bias of the difference-in-means estimator $\hat
  \tau$ under the stable unit treatment value assumption. Let $\calS \defeq
  \{\bm{\pi} \in [0,1]^M : \bar{\pi} = \frac{n_t}{N} \}$ be the set of
  treatment proportion vectors with fixed average $\frac{n_t}{N}$, where $n_t
  \in (0,N)$ is some fixed number of treated units.
\begin{equation}
  f : (\bm{\pi}, \calC, \{\Y(0), \Y(1) \})  \mapsto  \left| TTE -
  \E_\Z\left[\hat \tau| \bm{\pi}\right] \right|
\end{equation}
The constant vector $\bm{\pi}^* \defeq \left(\frac{n_t}{N}\right)_M$ minimizes
the objective function $f$ and belongs to $\calS$:
\begin{equation*}
\forall \calC, \Y(0), \Y(1),~\left(\frac{n_t}{N}\right)_M \in \arg
  \min_{\bm{\pi} \in \calS} f(\bm{\pi}, \calC, \{\Y(0), \Y(1) \} )
\end{equation*}
In other words, the stratified completely randomized assignment is an optimal deterministic
  saturation design for $f$, the bias of the difference-in-means
  estimator under the stable unit treatment value assumption.
\end{example}

A proof can be found in Section~\ref{proof:ex:sutva_bias_optimized}. Practitioners may wish to choose an optimal deterministic
saturation design that optimizes not just for the bias of an
estimator, but its variance as well. A common objective is to optimize
them jointly in the form of the mean-squared error, as is done in the following
example.

\begin{example}
\label{ex:sutva_mse_optimized}
Let $f$ be the mean-squared error of the difference-in-means estimator $\hat
  \tau$ under the stable unit treatment value assumption. Let $\calS \defeq
  \{\bm{\pi} \in [0,1]^M : \bar{\bm{\pi}} = \frac{n_t}{N}\}$ be the set of
  treatment proportion vectors with fixed average $\frac{n_t}{N}$, where $n_t$
  is some fixed number of treated units.
  \begin{equation}
    f: (\bm{\pi}, \calC, \{\Y(1), \Y(0)\}) \mapsto MSE_\Z[\hat \tau | \bm{\pi}]
    =  \left(TTE - \E_\Z[\hat \tau | \bm{\pi}] \right)^2 + \var_\Z[\hat \tau |
    \bm{\pi}]
  \end{equation}
  From Propositions~\ref{prop:bias_sutva} and \ref{prop:variance_sutva}, we can express this objective in closed-form: 
  \[
    f(\bm{\pi}, \calC, \{\Y(1), \Y(0)\})=
    \frac{N^2}{n_t^2n_c^2}\left[\bm\pi^T\left(\tilde{\bm W}^+[\tilde{\bm W}^+]^T - \cal S^+\right)\bm\pi + \bm\pi^T\bm S^+\right],
  \]
  where $\tilde {\bm W}^+ = \{W^{(j)} - N_j\overline{W}\}$ with $W^{(j)}$ defined in Proposition~\ref{prop:variance_sutva}, $\bm S^+=\{N_j\mathbb S[\bm W^{(j)}]\}_{j=1}^M$ is a vector of length $M$, and $\cal S^+$ is an $M\times M$ diagonal matrix whose diagonals are $\bm S^+$. Therefore, the optimal proportion vector $\bm\pi$ can be obtained by the following quadratic optimization. 
  \begin{align}
      \operatorname{minimize}&\quad\bm\pi^T\left(\tilde{\bm W}^+[\tilde{\bm W}^+]^T - \cal S^+\right)\bm\pi + \bm\pi^T\bm S^+\label{eq:example-optimize-sutva-obj}\\
      \textnormal{subject to}&\quad \sum_{j=1}^M\pi_j=M\overline{\pi}=\frac{n_tM}{N},\ \forall j=1,\dots, M
      \label{eq:example-optimize-sutva-constr}
      \\
      &\quad 0\leqslant \pi_j\leqslant 1,\ \forall j=1,\dots, M. \nonumber
  \end{align}
  The feasible set $\Omega_{\pi}$ defined by the constraints of \eqref{eq:example-optimize-sutva-obj} is a convex set with vertices belonging to $\{0,1\}^M$, corresponding to the class of cluster-based randomized saturation designs. 
  
  When $M\mathbb S[\bm W^+] < \min_j\ N_j\mathbb S[\bm W^{(j)}]$, the matrix $\tilde{\bm W}^+[\tilde{\bm W}^+]^T - \cal S^+$ is negative semi-definite, resulting in the concavity of the objective function \eqref{eq:example-optimize-sutva-obj}, in which case, the optimal proportion vector $\bm\pi$ must lie on the vertices of the feasible set $\Omega_\pi$, corresponding to a $\{0,1\}$-valued proportion vector. 
  
  For general cases, when ${\cal S}^+\succ 0$ and $\tilde{\bm W}^+[\tilde{\bm W}^+]^T - \cal S^+$ is neither positive or negative semi-definite, the optimal proportion vector $\bm\pi$ must lie on the boundary of $\Omega_\pi$, but not necessarily the vertices. In other words, the stratified randomized saturation design $\bm\pi = (\frac{n_t}{N})_M$ is never optimal. See Appendix~\ref{proof:ex:sutva_mse_optimized} for more details. 
 
\end{example}

Beyond operating under the stable unit treatment value assumption, optimal deterministic saturation design can postulate a parametric model of potential outcomes with interference, like the one in Equation~\ref{eq:dirg:ch3}, which we do in the following final example.

\begin{example}
\label{example:interference_optimize_conditional_varaince}
Let $f$ be the conditional mean squared error of the difference-in-means estimator $\hat
  \tau$ under the linear interference model in \eqref{eq:dirg:ch3}. Let $\calS \defeq
  \{\bm{\pi} \in [0,1]^M : \bar{\bm{\pi}} = \frac{n_t}{N}\}$ be the set of
  treatment proportion vectors with fixed average $\frac{n_t}{N}$, where $n_t$
  is some fixed number of treated units.
  \begin{equation}
    f: (\bm{\pi}, \calC, \{\bm\alpha,\bm\beta,\bm\gamma\}) \mapsto \mathrm{MSE}_{\bm Z}[\hat \tau | \bm{\pi}]=\mathrm{Var}_{\bm Z}[\hat \tau | \bm{\pi}] + |TTE - \mathbb E_{\bm Z}[\hat \tau | \bm{\pi}]|^2
  \end{equation}
  Specifically, assuming Assumptions~\ref{assump:boundedness} and \ref{assump:dense}, if the units are perfectly clustered, we can rewrite the  objective function as follows:
  \begin{multline}
  f=\frac{4}{\tilde n^2}\Bigg\{\left[\sum_{j=1}^M\pi_j^2\gamma^{(j)}+\sum_{j=1}^M\pi_j\left(W^{(j)}-N_j\overline{W}-\frac{n_t}{N}\gamma^{(j)}\right)-\frac{n_tn_c}{N}\overline{\gamma}\right]^2\\ 
  +\sum_{j=1}^MN_j\pi_j(1-\pi_j)\mathbb S\left(\bm W^{(j)} - \frac{n_t}{N}\bm H^{(j)}+\pi_j(\bm\gamma^{(j)}+\bm H^{(j)})\right)\Bigg\} + f_0,\label{eq:objective-conditional-mse}
  \end{multline}
  where $f_0$ is a constant term with respect to $\bm\pi$, $H_i:= \sum_{k\in\mathcal N_i}\frac{\gamma_k}{|\mathcal N_k|}$, and $W_i := \alpha_i + \frac{n_c}{N}\beta_i$. 
  The objective function shown in \eqref{eq:objective-conditional-mse} is quadratic in $\bm\pi$. In practice, one can find the optimal point of $f$ subject to $\overline{\bm\pi}=n_t/N$ through a quadratic programming solver (for example, the Sequential Least Squares Programming (SLSQP) algorithm).
  
  Under certain additional assumptions, we can express the solution in closed-form. First, we assume there is no significant difference in the distributions of outcome parameters across different clusters such that 
  $W^{(j)} = N_j\overline{W},\quad \gamma^{(j)}=\frac{N}{M}\overline{\gamma}\quad \text{and}\quad \mathbb S(\bm W^{(j)})=S_0\quad \forall j=1,\dots, M.$
  Second, we assume that the interference effects are block-fixed and that the number of neighbors is identical for units in the same cluster such that $\gamma_i$ and $H_i$ are constant in any cluster. When these two additional assumptions hold, the objective function \eqref{eq:objective-conditional-mse} is a quadratic function of $\mathrm{Var}[\bm\pi]$:
  \begin{equation}
  f=\frac{4N}{\tilde n^2}\Bigg\{N\overline{\gamma}^2\left[\mathrm{Var}[\bm\pi] - \frac{n_tn_c}{N^2}\right]^2 
  -S_0\mathrm{Var}[\bm\pi]\Bigg\}+f_0',\label{eq:objective-conditional-mse-simplified}
  \end{equation}

The support for $\mathrm{Var}[\bm\pi]$ is $[0, n_tn_c/N^2]$. The minimum value of the objective function \eqref{eq:objective-conditional-mse-simplified} is attained when 
$\mathrm{Var}[\bm\pi]=\frac{n_tn_c}{N}$, i.e. when $\bm\pi\in\{0, 1\}^M$. From Corollary~\ref{cor:linear_interference_bias_randomized}, when units are perfectly clustered such that $\gamma' = \bar\gamma$, the bias is minimized for any $\{0,1\}$-valued proportion vector. Similarly, the conditional variance term in \eqref{eq:objective-conditional-mse} also obtains its minimum ($0$ in this case) when $\pi_j(1-\pi_j)=0$ for all $j$. Therefore, the conditional mean squared error is minimized at any $\bm\pi\in\{0,1\}^M$ with $\sum_{j}\pi_j=n_tM/N$. 

\end{example}
%



\section{Simulations}
\label{sec:simulations}


In order to validate the results of the previous section, we implement a
small-scale simulation study to validate the bias-variance trade-offs available
to randomized saturation designs. We also illustrate the potential upside of using deterministic saturation designs.

\subsection{Optimal Randomized Saturation Designs}\label{sec:simulation-optimal-global}

In Section~\ref{sec:randomized_designs}, we presented several objectives for which, under certain assumptions on the potential outcomes, either the stratified randomized design ($\var[\mathbf{\pi}]$ is minimized) or the cluster-based randomized design ($\var[\mathbf{\pi}]$ is maximized) are optimal. This is the case in Corollary~\ref{cor:variance_regimes_sutva} and Corollary~\ref{cor:linear_interference_bias_randomized} for example. However, these designs are not always optimal, as we show in Section~\ref{sec:randomized_variance}, notably when the variance---and consequently the mean-squared error---of the difference-in-means estimator under a linear model of interference is concerned. In this first simulation, we consider the variance of the difference-in-means estimator under a linear interference model where the stratified randomized design and the cluster-based randomized designs are \emph{not} optimal, and illustrate the benefit of using an optimal randomized saturation design instead.

We consider a population of 2,000 units, grouped into 40 equally-sized clusters which are labeled $1$ through $40$. The interference graph is generated according to a stochastic block model such that the probability of observing an edge between units $i\in\mathcal C_j$ and $k\in\mathcal C_l$ is 
$\mathbb P[i\sim k]=\exp\{-|j-l|/2\}.$ The edge-formation probability is chosen so that different pairs of clusters $(j,l)$ have different forming probabilities in general. In particular, this increases the value of $V_2$ in the variance of the difference-in-means estimator in Theorem~\ref{thm:variance-dim} by increasing the variance terms $\mathbb S[\{D^{(jl)}:l\neq j\}]$ in the closed-form expression of $V_2$, which can be found in the appendix. As a result, the variance of difference-in-means estimator will, in general, have a larger curvature as a function of the second moment of $\mathbf{\pi}$, as we show below. 

The potential outcomes are randomly generated according to the following distributions:
\begin{align*}
     \forall i=1,\dots, N,\quad\alpha_i&\sim N(0, \sigma_\alpha^2)\quad\text{and}\quad\beta_i=\gamma_i=1.
\end{align*}
After sampling these outcome parameters, we transform each $\alpha_i, \forall i\in\mathcal C_j$ to $\alpha_i - \alpha^{(j)}/N_j$ so that $\alpha^{(j)}=0\ \forall j$ after the transformation. The normalization of $\alpha$ fixes the inter-cluster variance of $\alpha$ to $0$, such that the variance coefficient $V_1$ in Theorem~\ref{thm:variance-dim} can be tuned by a single parameter $\sigma_\alpha$. Furthermore, we consider the proportion vector $\bm\pi$ within the family of symmetric Beta distribution Beta$(\lambda, \lambda),\lambda\in[0, \infty]$ such that
$$\pi_j = F^{-1}_{\lambda}\left(\frac{j}{J+1}\right),\ j=1,\dots, J$$
where $\lambda$ is the shape parameter and $F_{\lambda}$ is the c.d.f function of $Beta(\lambda, \lambda)$. Within the symmetric Beta distribution family, we recover our two baseline randomized designs.  When $\lambda = 0$, $\bm\pi\in\{0,1\}^M$ corresponds to the cluster-based randomized design where half of the clusters are assigned to treatment and all others to control. When $\lambda = \infty$, $\bm\pi=(0.5)_M$ corresponds to the stratified completely randomized design where exactly half of the units in each cluster are treated. 

Recall Theorem~\ref{thm:variance-dim}, where the variance of the difference-in-means estimator can be expressed as 
$$\mathrm{Var}[\hat\tau] = V_0 + V_1\mu_{2c}+V_2\mu_{2c}^2,$$
by ignoring the fourth order term $\mu_{4c}-\mu_{2c}^2$ and by observing that $\mu_{3c}=0$ for symmetric beta distributions. In the subsequent simulations, we carefully choose $\sigma_\alpha$ such that $-V_1/(2V_2)\approx 1/(12)$, making $Beta(\lambda = 1, \lambda = 1)$ the optimal distribution for $\bm\pi$ within symmetric Beta distributions.

\begin{figure}[!htp]
    \centering
    \includegraphics[width=0.6\textwidth]{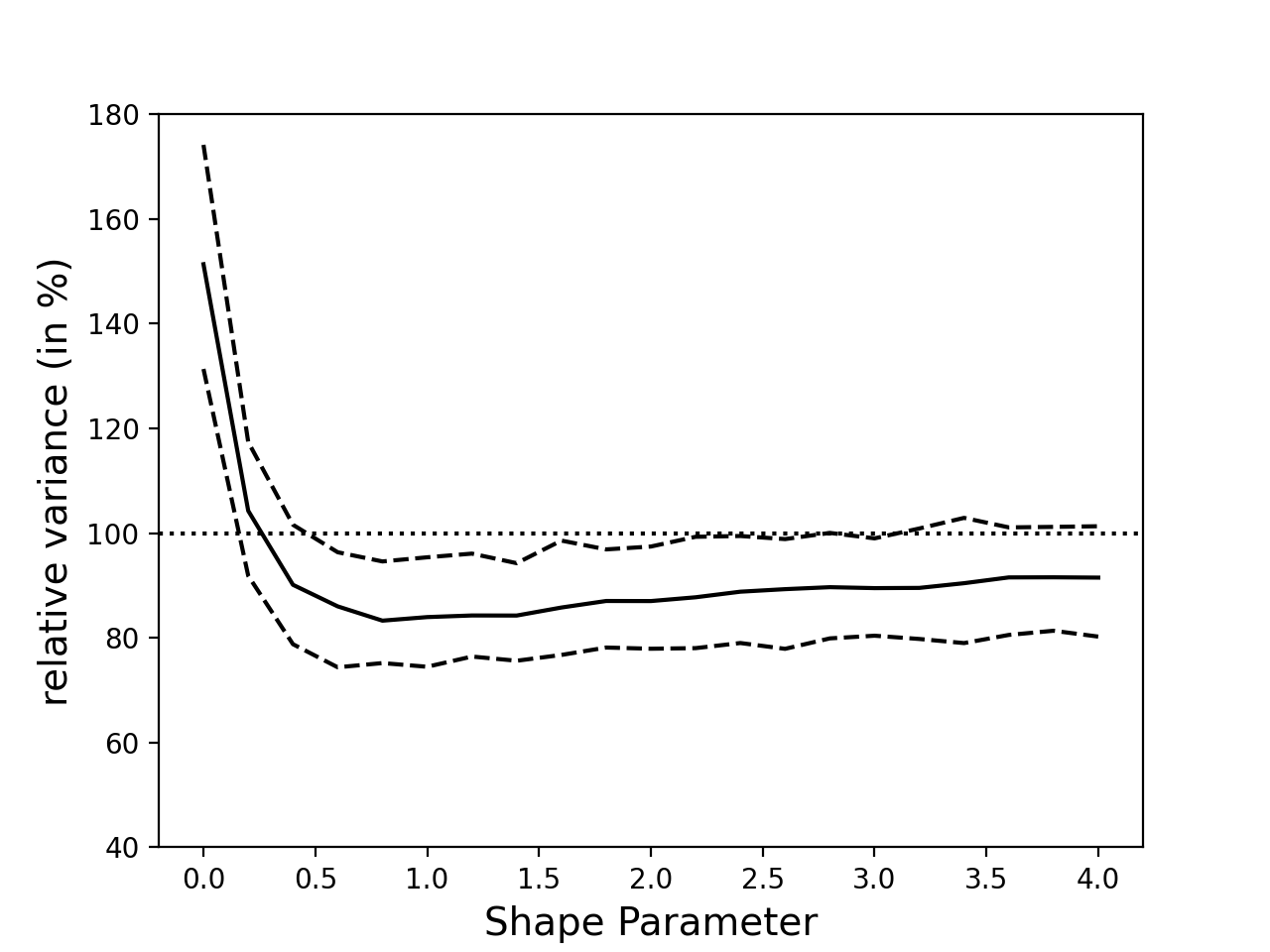}
    \caption{The average relative variance (in \%) over the one of shape $\lambda=\infty$. Lower and upper quantile are plotted in dashed lines. }
    \label{fig:var-shape}
\end{figure}

Finally, we generated 100 realizations of the interference graph and outcome parameters according to the distributions described above. For each realization, we investigated the variance of the difference-in-means estimator under a randomized saturation design for 22 different shape parameters $\lambda\in\{0, 0.02, 0.04,\dots,\allowbreak 0.38, 0.40, \infty\}$. For each shape parameter, the variance is estimated by the sample variance of difference-in-means estimates from 1000 random assignments. We report the relative change in variance over 100 realizations for each value of the shape parameters in Figure~\ref{fig:var-shape}, where the relative change in variance= is calculated with respect to the variance of the estimator under a stratified randomized saturation design ($\lambda = \infty$). The lower 2.5\% quantile and the upper 2.5\% quantile are also reported in dashed lines.

Figure~\ref{fig:var-shape} shows that by fixing $-V_1/(2V_2)$ at $1/12$, neither the cluster-based randomized design ($\lambda = 0$) nor the stratified randomized design $(\lambda=\infty)$ achieve the minimum variance. The minimum average percentage of variance is obtained at about $\lambda = 0.8$, which reduces the variance of the difference-in-means estimator in a stratified randomized design by roughly 17\% in our simulations. 

\subsection{Optimal Deterministic Saturation Designs}\label{sec:simulation-optimal-local}

In Section~\ref{sec:optimized_designs}, we showed that, rather than randomizing over which treatment proportion gets assigned to which cluster---the first step of any randomized saturation design---we could also consider skipping this permutation step and directly optimize each coordinate of the treatment proportions vector $\bm \pi$. We referred to this latter category of designs as a deterministic saturation design in Definition~\ref{def:locally_randomized}. Because a deterministic saturation design can select the optimal treatment proportion for each cluster, as opposed to the optimal \emph{distribution} of treatment proportions, we expect the statistical measure that is being optimized---like the bias and variance of the difference-in-means estimator---to improve when using a well-chosen deterministic saturation design over even the best possible randomized saturation design.

In this section, we demonstrate the benefits of using an optimal deterministic saturation design, as compared to various randomized saturation designs. More specifically, we will compare an optimally-chosen deterministic saturation design over the best possible randomized saturation design---including both the cluster-based and stratified completely randomized designs. Additionally, we will also compare our optimally-chosen deterministic saturation design with fixed treatment proportions vector $\bm{\hat \pi}$ to a randomized design which permutes $\bm{\hat \pi}$. Naturally, we expect this re-randomized design to perform worse than the best possible randomized saturation design on the metric for which we are optimizing for. 

We construct the following simulation using the setting presented in Example~\ref{example:interference_optimize_conditional_varaince}, which considers the mean-squared error of the difference-in-means estimator under a linear model of interference.
We again consider 40 clusters, each containing 50 units. We assume the outcomes follow a linear interference model with the interference graph generated according to a stochastic block model with edge probability:
$$p_{jl}=\begin{cases}
    0.5 &\text{if }j=l\\
    0&\text{if }j\neq l.
    \end{cases}$$
We consider the following distributions for the outcome parameters:
\begin{align*}
    \alpha_{[j]}&\sim \mathrm{Unif}(0, 3)\\
    \alpha_i & = \alpha_{[j]} + \mathcal N(0, 0.01)\quad\forall i\in\mathcal C_j\\
    \beta_i&=1\\
    \gamma_{[j]}&\sim \mathrm{Unif}(0, 1)\\
    \gamma_i&=\gamma_{[j]}\quad\forall i\in\mathcal C_j.
\end{align*}
Recall that three designs will be simulated and compared:
\begin{itemize}
\item The optimal randomized saturation design. Because the clustering is ``perfect'' (no edges are cut), we know from Corollary~\ref{cor:linear_interference_bias_randomized} that the bias of the difference-in-means estimator is minimized for a cluster-based randomized design.
Specifically, the bias is given by $\mathbb E[\hat\tau]-TTE = 2\mathrm{Var}[\bm\pi] - \frac{1}{2}$. On the other hand, from Corollary~\ref{cor:var-simplified}, we have the expected coefficients are $V_1\approx \frac{12}{M}=1.3$, $V_2=0$ and $V_4 >0$. When $V_4 > 0$, the optimal design allocates $\pi_j$ on two points, or equivalently, has zero fourth central moment (see Proposition~\ref{prop:optimizatin-closed-form}). Therefore the mean squared error of $\hat\tau$ can be written in a quadratic form of $\mathrm{Var}[\bm\pi]$ as
$$MSE(\hat\tau) \approx 4[\mathrm{Var}[\bm\pi]]^2 - 1.7\mathrm{Var}[\bm\pi],$$
up to some constant term. The MSE obtains its minimum at $\mathrm{Var}[\bm\pi]=0.2125$, corresponding to $\bm\pi_j^*\in\{0.078, 0.922\}^M$. As $\pi_jN_j$ must be an integer, the closest possible assignment $\bm\pi$ is $\bm\pi\in\{0.02, 0.98\}^M$. We use $\bm\pi\in\{0, 1\}^M$ in this simulation for simplicity.

%
\item The optimal deterministic saturation design. Recall that for this category of designs, we fix the vector $\bm\pi$ in the treatment assignment procedure. The optimal choice of the assignment-proportion $\hat{\bm\pi}$ is obtained by optimizing the conditional MSE in \eqref{eq:objective-conditional-mse} in Example~\ref{example:interference_optimize_conditional_varaince}, subject to $\sum_j\pi_j = M/2$, using a quadratic programming solver. We expect this design to have the lowest mean-squared error, since it optimizes that objective for each coordinate of the treatment proportions vector, rather than optimizing over its distribution.
\item The re-randomized saturation design. Instead of fixing $\bm\pi$ at $\bm{\hat\pi}$ as we did for the optimal deterministic saturation design, we permute $\bm{\hat \pi}$ when assigning treatment. Therefore, the suggested re-randomized saturation design is a special case of a randomized saturation design, which uses $\hat{\bm\pi}$ instead of $\{0, 1\}^M$. While we expect this design to perform the worse of the three suggested designs for the mean-squared error of the difference-in-means estimator, this comparison allows us to showcase that (i) the optimal deterministic randomized design does outperform a non-optimally-chosen randomized saturation design, (ii) the out-performance of the chosen deterministic saturation design is not due to the \emph{distribution} of $\bm \hat \pi$, but its ability to fix and optimize each coordinate of $\bm \hat \pi$.
\end{itemize}

\begin{figure}[!htbp]
    \centering
    \includegraphics[width=0.3\textwidth]{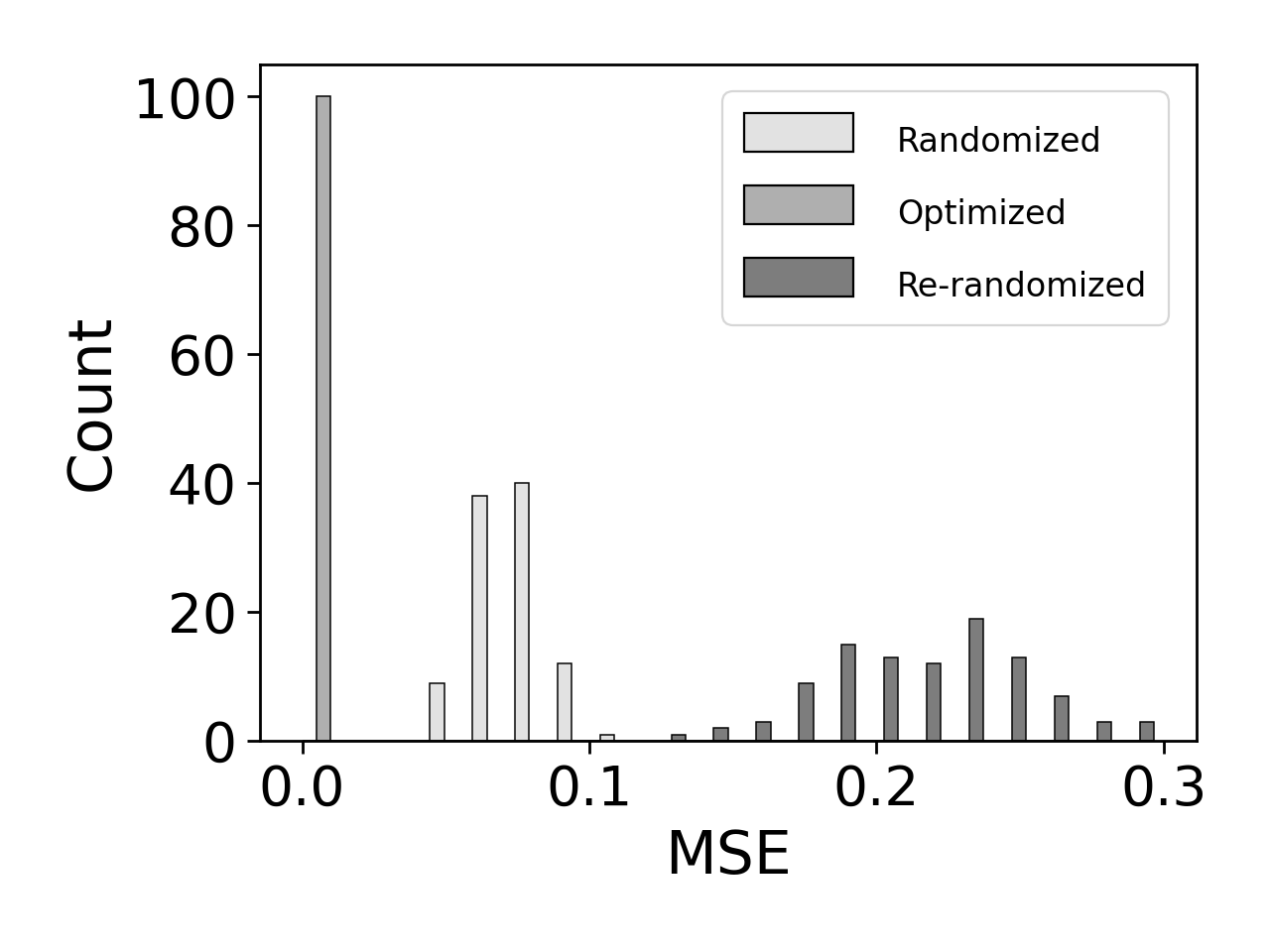}
    \includegraphics[width=0.3\textwidth]{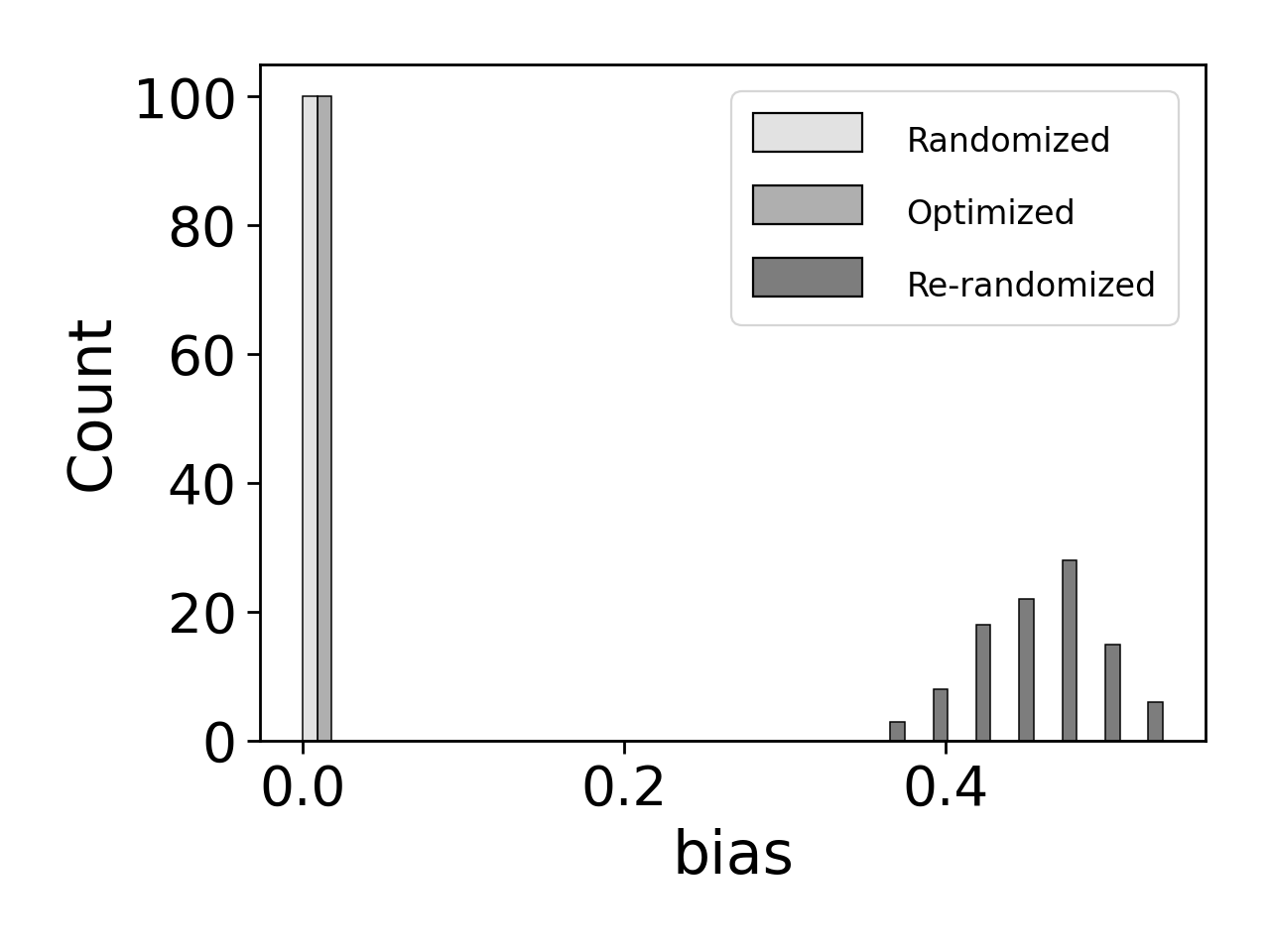}
    \includegraphics[width=0.3\textwidth]{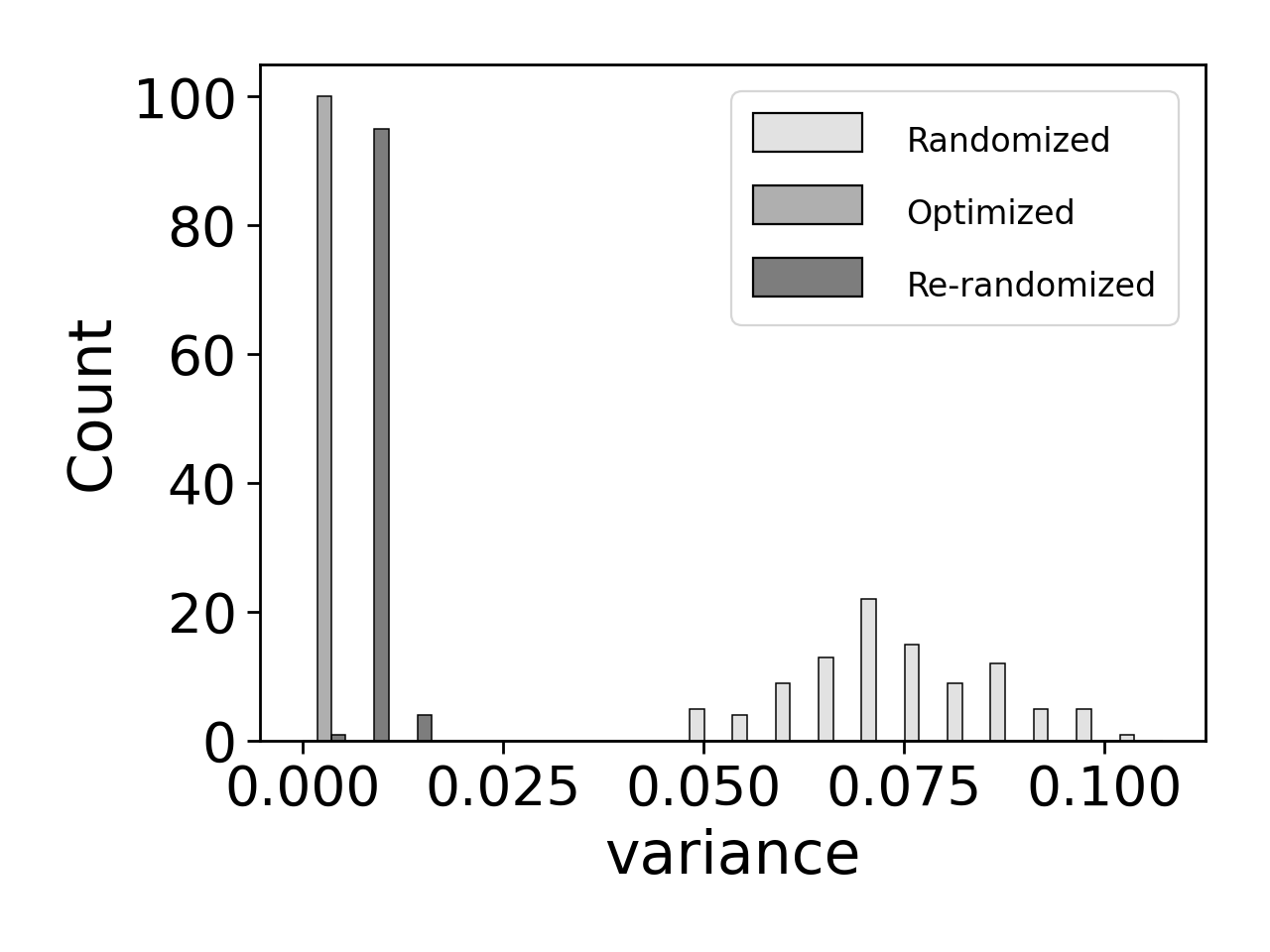}
    \caption{Bias, variance and mean squared error of the orandomized saturation design, the optimal deterministic saturation design and the re-randomized saturation design. }\label{fig:optimized-compare.png}
\end{figure}

We simulated 100 realizations of interference graph and outcome parameters. For each realization, we estimate the bias, variance and mean squared error of the difference-in-means estimator from $1,000$ random assignments from the three suggested designs. The histogram in Figure~\ref{fig:optimized-compare.png} reports the distribution of bias, variance and mean squared error of the three designs. 

As shown in Figure~\ref{fig:optimized-compare.png}, the randomized saturation design optimized for the mean-squared error of the difference-in-means estimator achieves the best (lowest) value for that objective. Expectedly, the optimal deterministic randomized design performs the second best, since the re-randomized design is able to neither fix the optimal coordinate values for $\bm{\pi}$ nor optimize its distribution.  When examining the bias and the variance of each design separately, we notice that both the optimal deterministic saturation design and the optimal randomized saturation design have similar and small bias compared to the one of the re-randomized design, but the variance of the optimal randomized saturation design is significantly greater than the two others. Compared to the optimal randomized and the re-randomized saturation design, the optimal deterministic saturation design reduces the bias and variance simultaneously. 

\begin{figure}[!htbp]
    \centering
    \includegraphics[width=0.6\textwidth]{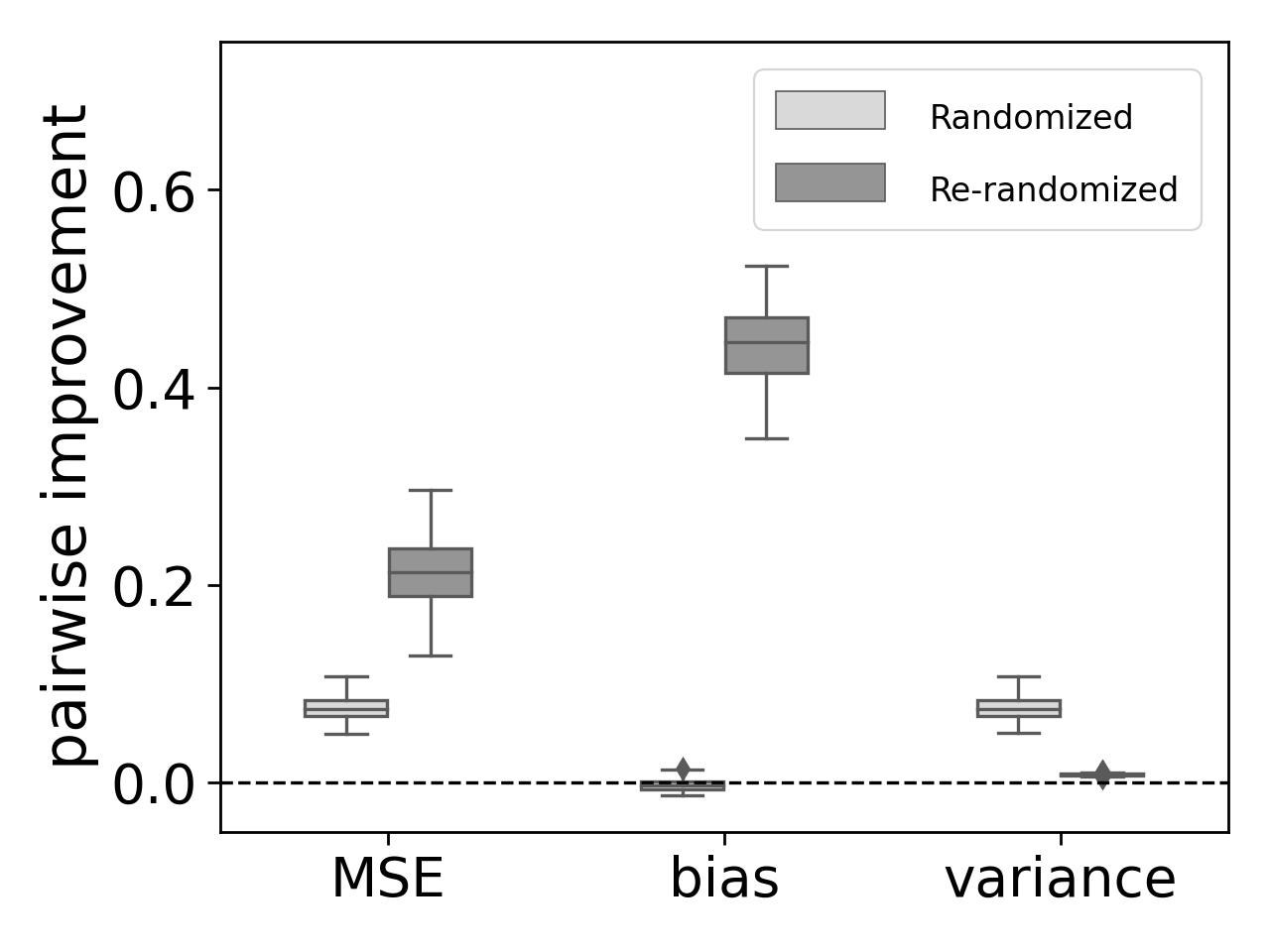}
    \caption{Improvements in Bias, variance and mean squared error of the optimal saturation design over the other two methods. }\label{fig:optimized-improvement.png}
\end{figure}

Figure~\ref{fig:optimized-compare.png} demonstrates the population comparison over 100 repetitions. Although the optimal deterministic saturation design outperforms the optimal randomized one on average, it is still necessary to check if there is any circumstance that the optimal deterministic saturation design is sub-optimal.  A stronger assertion would be whether the optimal deterministic saturation design improves the optimal randomized saturation design on every single realization. Our answer is affirmative. 
Figure~\ref{fig:optimized-improvement.png} demonstrates the improvement in bias, variance, and the mean-squared error of the optimal deterministic saturation design over the other two designs in each of the 100 realizations. One can observe that the optimal deterministic saturation design always improves over the re-randomized saturation design and consistently reduces the variance and MSE from the randomized saturation design. 

\section{Conclusion and Practical Considerations}
\label{sec:conclusion}

This manuscript focuses on the randomized saturation designs, where each cluster of units is first assigned with a treatment proportion and then units within this cluster are randomly assigned to treatment. Depending on whether the treatment proportions assigned to clusters are randomized, we distinguish between two types of randomized saturation designs: the randomized saturation design and the deterministic saturation design. The stratified randomized saturation design, where the treatment proportions are constant, and the cluster-based randomized saturation design, where the treatment proportions are $\{0,1\}$-valued are two well-studied special cases of the randomized saturation design. 

When the potential outcomes satisfy SUTVA, the difference-in-means estimator is unbiased under all randomized saturation designs, and in terms of variance/mean squared error, either the stratified randomized saturation design or the cluster-based one is optimal, depending on the relative values of the inter-cluster variance of outcomes and the intra-cluster variance of outcomes.

When interference is present, we show the bias of the difference-in-means estimator is linear in the variance of the treatment proportion vector for a linear model of interference. In addition, the variance of the difference-in-means estimator is a linear function of the variance, the squared variance, the third central moment and the fourth central moment of the treatment proportion vector. The minimization of such a variance is tractable when we restrict the proportion vector to all symmetric distributed vectors around the fixed mean. It is possible that neither the cluster-based design nor the stratified design is optimal as demonstrated in Section~\ref{sec:simulation-optimal-global}.

The performance of the difference-in-means estimator can be further improved through the optimal deterministic saturation design, when the model is correctly specified. Specifically, the conditional mean squared error of the difference-in-means estimator under SUTVA or under the linear interference model can be optimized by choosing a fixed proportion vector as discussed in Examples~\ref{ex:sutva_mse_optimized} and \ref{example:interference_optimize_conditional_varaince}. The benefits achieved from the optimal deterministic saturation design are demonstrated in the simulation in Section~\ref{sec:simulation-optimal-local}. 


We note that, although our analysis on optimizing randomized saturation designs is based on the exact values of potential outcomes, it does not require full knowledge on all units' potential outcomes to find the optimal proportion vector $\bm\pi^*$. As illustrated in Theorem~\ref{thm:variance-dim} and its simplified form in Corollary~\ref{cor:var-simplified}, the knowledge of certain cluster-level statistics of potential outcomes (e.g., intra-cluster and inter-cluster variances) is sufficient for optimization. Therefore, if an experimenter has certain prior knowledge on these cluster-level statistics---perhaps a pilot experiment has been run on a small scale---an optimal deterministic saturation design can be generated using our approach by plugging in the pre-assumed/estimated potential outcome statistics. The resulting saturation designs are exactly optimal when the `plug-in' parameters are accurate and may still improve over the two extreme designs (cluster-based and stratified) when these `plug-in' parameters are misspecified. 
Furthermore, as one would expect, there is no free lunch on the performance improvement obtained from the optimal deterministic saturation designs.
Optimizing the deterministic saturation design as shown in Definition~\ref{def:locally_randomized} requires more cluster-specific statistics than finding the optimal randomized saturation design, resulting in compromising the robustness under model misspecification. In practice, one would choose the optimal randomized saturation design or the optimal deterministic design based on the feasibility of accurately estimating the necessary population-level statistics of the potential outcomes.
The idea of using plug-in values to design optimal experiments is related to the model-assisted designs \citep{basse2015model}, where an artificial model is assumed to generate designs which are guaranteed to be optimal when properly specified.


\vskip 0.2in
\bibliographystyle{plainnat}
\addcontentsline{toc}{section}{References}
\bibliography{reference}

\newpage
\appendix
\section{Analysis of the Difference-in-means Estimator under SUTVA}




\subsection{Proof of Proposition~\ref{prop:bias_sutva}}
\label{sec:proof:bias_sutva}

The expectation of the difference-in-means estimator conditioned on the
proportion of units assigned to treatment is given by:
\begin{align*}
  \E_\Z \left[ \hat \tau \middle| \bm{\pi} \right] & = \E_\Z \left[\sum_{i=1}^N
  \left(Z_i Y_i(1) + (1 - Z_i) Y_i(0) \right) \frac{(-1)^{1-Z_i}}{n_t^{Z_i}
  n_c^{1 - Z_i}}\right] \\
 &= \E_\Z \left[\sum_{j =1}^M \sum_{i \in \calC_j}\left(Z_i Y_i(1) + (1 - Z_i)
  Y_i(0) \right)  \frac{(-1)^{1 - Z_i}}{n_t^{Z_i} n_c^{1 - Z_i}} \right] \\
  &=  \frac{1}{n_t} \sum_{j =1}^M  \pi_j \sum_{i \in \calC_j} Y_i(1)    -
  \frac{1}{n_c} \sum_{j = 1}^M  \left(1 - \pi_j \right) \sum_{i \in
  \calC_j} Y_i(0)
\end{align*}
We now introduce the permutation matrix $P$.
\begin{align*}
  \E_{\Z,\bm{\pi}} \left[ \hat \tau \right] & = \E_{P} \left[ \frac{1}{n_t}
  \sum_{j, k =1}^M  P_{jk} \pi_k \sum_{i \in \calC_j} Y_i(1)    - \frac{1}{n_c}
  \sum_{j, k = 1}^M P_{jk} \left(1 - \pi_k \right) \sum_{i \in \calC_j} Y_i(0)
  \right] \\
  &= \frac{1}{n_t} \sum_{j, k = 1}^M \frac{\pi_k}{M} \sum_{i \in \calC_j} Y_i(1) -
  \frac{1}{n_c} \sum_{j, k = 1}^M \frac{1 - \pi_k}{M} \sum_{i \in \calC_j} Y_i(0)
  \\
  &=  \frac{1}{n_t} \left(\sum_{k = 1}^M \frac{\pi_k}{M} \right) \left( \sum_j
  \sum_{i \in \calC_j} Y_i(1) \right) -
  \frac{1}{n_c} \left( \sum_{k = 1}^M \frac{1 - \pi_k}{M}\right) \left(\sum_j
  \sum_{i \in \calC_j} Y_i(0)\right) \\
  &= \frac{1}{N} \sum_i Y_i(1) - \frac{1}{N} \sum_i Y_i(0)
\end{align*}

This last quantity corresponds to the total treatment effect, hence the proof that the
difference-in-means estimators is unbiased under the stable unit treatment value
assumption for a randomized
saturation design.


\subsection{Proof of Proposition~\ref{prop:variance_sutva}}
\label{proof:prop:variance_sutva}

Using Eve's law, we have that
\begin{equation*}
  \var_{\Z, \bm{\pi}}\left[\hat \tau\right] = \var_{\bm{\pi}}\left[ \E_\Z\left[
    \hat \tau \mid \bm\pi\right] \right] + \E_{\bm{\pi}} \left[\var_\Z \left[\hat \tau
  \mid \bm\pi\right] \right]
\end{equation*}
We first compute the variance of the estimator conditional on an assignment of
the treatment proportions vector $\bm{\pi}$. We start from 
$$\hat\tau = \sum_{i=1}^NZ_i\left(\frac{Y_i(1)}{n_t} + \frac{Y_i(0)}{n_c}\right) - \frac{1}{n_c}\sum_{i=1}^NY_i(0)=\frac{N}{n_tn_c}\sum_{i=1}^NZ_iW_i - \frac{1}{n_c}\sum_{i=1}^NY_i(0),$$
where $W_i = \frac{n_t}{N}Y_i(0) + \frac{n_c}{N}Y_i(1)$. Its conditional variance is given by
\begin{align*}
    \mathrm{Var}_{\bm Z}[\hat\tau\mid\bm\pi]&=\frac{N^2}{n_t^2n_c^2}\mathrm{Var}_{\bm Z}\left[\sum_i W_iZ_i\right]\\
    &=\frac{N^2}{n_t^2n_c^2}\sum_{j=1}^M \sum_{i,k\in\mathcal C_j}W_iW_k\mathrm{Cov}[Z_i, Z_k]\\
    &=\frac{N^2}{n_t^2n_c^2}\sum_{j=1}^M \left(\sum_{i\in\mathcal C_j}W_i^2\pi_j(1-\pi_j)-\sum_{i\neq k\in\mathcal C_j}W_iW_k\frac{\pi_j(1-\pi_j)}{N_j-1}\right)\\
    &=\frac{N^2}{n_t^2n_c^2}\sum_{j=1}^M\pi_j(1-\pi_j)N_j \mathbb S[\bm W^{(j)}].
\end{align*}
The expectation of the conditional variance is therefore
$$\mathbb E_{\bm\pi}[\mathrm{Var}_{\bm Z}[\hat\tau\mid\bm\pi]=\frac{N^2}{n_t^2n_c^2}\sum_{j=1}^M\mathbb E[\pi_j(1-\pi_j)]N_j \mathbb S[\bm W^{(j)}]=\left(\frac{1}{n_tn_c}-\frac{N^2}{n_t^2n_c^2}\mathrm{Var}[\bm\pi]\right)\sum_{j=1}^MN_j \mathbb S[\bm W^{(j)}].$$
Next, from Proposition~\ref{prop:bias_sutva}, the conditional expectation of $\hat\tau$ is
$$\mathbb E_{\bm Z}[\hat\tau] = \frac{N}{n_tn_c}\sum_{j=1}^M\pi_jW^{(j)}(1) - \frac{1}{n_c}\sum_{j=1}^MY^{(j)}(0),$$
where $W^{(j)}=\sum_{i\in\mathcal C_j}W_i$ is the cluster-level averaged potential outcome. Its variance is
\begin{align*}
    \mathrm{Var}_{\bm\pi}[\mathbb E_{\bm Z}[\hat\tau]]&=\frac{N^2}{n_t^2n_c^2}\mathrm{Var}\left[\sum_{j=1}^M\pi_jW^{(j)}\right]\\
    &=\frac{N^2}{n_t^2n_c^2}\sum_{j,l} W^{(j)}W^{(l)}\mathrm{Cov}[\pi_j, \pi_l]\\
    &=\frac{N^2}{n_t^2n_c^2}\left(\sum_{j=1}^M [W^{(j)}]^2\mathrm{Var}[\bm\pi] - \sum_{j\neq l}W^{(j)}W^{(l)}\frac{\mathrm{Var}[\bm\pi]}{M-1}\right)\\
    &=\frac{N^2}{n_t^2n_c^2}\mathrm{Var}[\bm\pi]M\mathbb S[\bm W^+].
\end{align*}
The total variance is now immediate.

\subsection{Proof of Corollary~\ref{cor:variance_regimes_sutva}}
It suffices to prove that the variance of $\bm\pi$ is maximized at $\{0, 1\}^M$ with a constrained mean. 

We show that the variance of the treatment proportions vector $\bm{\pi}$ is
maximized, constrained to verify $\bar \pi = \frac{n_t}{N}$, only for vectors
$\bm{\pi}^* \in \{0,1\}^M$ assigning either all of a cluster to treatment or
none, assuming that a solution in $\{0,1\}^M$ verifying the equality constraint
exists.  One direction is easy. Let $\bm{\pi^*}$ be any assignment placing all of a cluster
to treatment or none, and verifying the inequality constraint.
\begin{align*}
  \var[\bm{\pi}] & = \frac{1}{M} \sum_{j=1}^M \pi_j^2 -
  \left(\frac{n_t}{N}\right)^2  \\
  & = \frac{1}{M} \sum_{j=1}^M \pi_j - \left(\frac{n_t}{N}\right)^2 \\
   & = \frac{n_t n_c}{n^2}
\end{align*}

We prove the other direction.  Consider $\pi_j < \pi_i$. Consider increasing
$\pi_i$ and decreasing $\pi_j$ by $\epsilon$ such that the total number of
treated units is constant: $\pi'_i = \pi'_i + \epsilon$, $\pi'_j = \pi'_j -
\epsilon$, and $\forall k \notin \{ i,j\}, \pi'_k = \pi_k$, such that:
\begin{align*}
  \var[\bm{\pi}'] = \var[\bm{\pi}] + (\pi_i + \epsilon)^2 + (\pi_j - \epsilon)^2
  = \pi_i^2 + \pi_j^2 + 2 \epsilon^2 + 2 \epsilon (\pi_i - \pi_j)
\end{align*}
Since $\pi_i > \pi_j$, $\var[\bm{\pi}'] \geq \var[\bm{\pi}]$, which concludes the
proof.



\section{Examples of variance-minimizing distributions of $\bm\pi$ under interference}
\label{sec:appendix_examples}

\begin{example}[Random clustering and block-fixed interference]\label{example:interference2}
Suppose Assumption~\ref{assump:variance-sufficient} holds, interference effects are block-fixed and the graph is randomly clustered such that
$$q_{jl}=\frac{1}{M}\quad \forall 1\leqslant j,l\leqslant M.$$ Assume $n_t \leqslant N/2$. Then $V_4 > 0$ and 
$$V_2 =\frac{2}{\tilde n^2 M}\mathbb S[\bm\gamma^+]>0.$$\\
(i) If $V_1 > 0$, the optimal assignment vector is $\pi_j^*=n_t/N$ for all $j$. The optimal variance is $V_0$.\\
(ii) If $V_1 < -\frac{1}{n_c^2M}\mathbb S[\bm\gamma^+]$, the optimal assignment vector is 
$$\pi_j^* = \begin{cases}
0 &,\ j< M/2,\\
\frac{2n_t}{N}&,\ j>M/2.
\end{cases}$$
(iii) If $-\frac{1}{n_c^2M}\mathbb S[\bm\gamma^+]\leqslant V_1\leqslant 0$, the optimal assignment vector is
$$\pi_j^* = \begin{cases}
\frac{n_t}{N}-\sqrt{-\frac{V_1}{2V_2}} &,\ j< M/2,\\
\frac{n_t}{N}+\sqrt{-\frac{V_1}{2V_2}}&,\ j>M/2.
\end{cases}$$

\end{example}

\begin{example}[Perfect clustering and non-block-fixed interference]\label{example:interference3}
Suppose Assumption~\ref{assump:variance-sufficient} holds and the graph is perfectly clustered such that
$$q_{jl}=\mathbb I\{j=l\}\quad \forall 1\leqslant j,l\leqslant M.$$ Assume $n_t \leqslant N/2$. Then 
\begin{align*}
    V_2 &= -\frac{4}{\tilde n^2}\sum_{j=1}^MN_j\mathbb S(\bm\gamma^{(j)})<0,\\
    V_4 &= \frac{4M}{\tilde n^2}\mathbb S(\bm\gamma^{+}) + V_2>V_2. 
\end{align*}\\
(i) If $V_4>0$ and $-\frac{V_1}{V_2}\geqslant \frac{n_t^2}{N^2}$, or $V_4 \leqslant 0$ and $-\frac{V_1+V_4n_t/N}{V_2-V_4}\geqslant \frac{n_t^2}{N^2}$, the optimal assignment vector is $\pi_j^*=n_t/N$ for all $j$. \\
(ii) If $V_4>0$ and $-\frac{V_1}{V_2}< \frac{n_t^2}{N^2}$, or $V_4 \leqslant 0$ and $-\frac{V_1+V_4n_t/N}{V_2-V_4}< \frac{n_t^2}{N^2}$, the optimal assignment vector is 
$$\pi_j^* = \begin{cases}
0 &,\ j< M/2,\\
\frac{2n_t}{N}&,\ j>M/2.
\end{cases}$$

\end{example}

\section{Analysis for Linear Interference Model}
\subsection{Notation for Linear Interference Model}
We define the interference coefficient from unit $k$ to unit $i$ by
$$T_{ik}:=\frac{\gamma_i}{|\mathcal N_i|}\mathbb I\{i\sim k\},$$
such that the potential outcome of unit $i$ under treatment $\bm Z$ in a linear interference model can be expressed as 
\begin{equation}
    Y_i(\bm Z) = \alpha_i + \beta_i Z_k + \sum_{k=1}^NT_{ik}Z_k + \epsilon_i.
\end{equation}
The corresponding difference-in-means estimator is therefore 
\begin{align}
    \hat\tau(\bm Z) &= \frac{1}{n_t}\sum_{i=1}^NY_i(\bm Z)Z_i - \frac{1}{n_c}\sum_{i=1}^NY_i(\bm Z)(1-Z_i)\nonumber\\
    &=-\frac{N\bar\alpha}{n_c} + \frac{N}{n_tn_c}\sum_{i=1}^N\left(W_i-\frac{n_t}{N}H_i\right)Z_i + \frac{N}{n_tn_c}\sum_{i=1}^N\sum_{k=1}^NT_{ik}Z_iZ_k,\label{eq:proof-dim-interference}
\end{align}
where 
$$W_i:=\alpha_i + \frac{n_c}{N}\beta_i\quad\text{and}\quad H_i:=\sum_{k=1}^NT_{ki}=\sum_{k\in\mathcal N_i}\frac{\gamma_k}{|\mathcal N_k|}$$
are the weighted sum of potential outcomes coefficients of unit $i$ and the aggregated interference coefficients of the neighbors of $i$ correspondingly. 

Note that $T_{ik}$ is asymmetric in the sense that $T_{ik}\neq T_{ki}$ for $i\neq k$. One can symmetrize $T_{ik}$ by defining $D_{ik}:= T_{ik}+T_{ki}$. In addition, we define the clustered aggregated interference coefficients by 
$$D_i^{(l)} := \sum_{k\in\mathcal C_l} D_{ik}\quad \text{and}\quad D^{(jl)} := \sum_{i\in\mathcal C_j}\sum_{k\in\mathcal C_l}D_{ik}=\sum_{i\in\mathcal C_j}D_i^{(l)}.$$
The difference-in-means estimator in \eqref{eq:proof-dim-interference} can be rewritten to 
\begin{equation}
    \hat\tau(\bm Z) =-\frac{N\bar\alpha}{n_c} + \frac{N}{n_tn_c}\sum_{i=1}^N\left(W_i-\frac{n_t}{N}H_i\right)Z_i + \frac{1}{2}\frac{N}{n_tn_c}\sum_{i=1}^N\sum_{k=1}^ND_{ik}Z_iZ_k,\label{eq:proof-dim-interference-symm}
\end{equation}
so that the quadratic coefficients of $Z_iZ_k$ is now symmetrized. 

\subsection{Proof of Proposition~\ref{prop:tte:ch3}}
When all units are treated, the observed outcome of unit $i$ is $Y(\bm 1) = \alpha_i + \beta_i + \gamma_i$. In opposite, when all units are untreated, the observed outcome of unit $i$ is $Y(\bm 0) =\alpha_i$. Therefore, the total treatment effect for unit $i$ is $TTE_i=Y(\bm 1)-Y(\bm 0) = \beta_i + \gamma_i$. The average total treatment effect is hence $TTE = \overline{\beta} + \overline{\gamma}$, where $\overline{\beta}$ and $overline{\gamma}$ are population averages. 

\subsection{Proof of Theorem~\ref{thm:bias_cond_graph}}
Consider the expectation of the difference-in-means estimator under an observed treatment proportion vector $\bm\pi$. In additional to the idiosyncratic noise $\epsilon_i$, the randomness also comes from the completely randomization of $\{Z_i:i\in\mathcal C_j\}$ given $\sum_{i\in\mathcal N_j}Z_i=n_j = \lfloor\pi_j N_j\rfloor$ for all $j\in[M]$. \\
According to the notation in \eqref{eq:proof-dim-interference}, the conditional expectation is 
\begin{align}
    \mathbb E_{\bm Z}[\hat\tau\mid\bm\pi]&=
    -\frac{N\bar\alpha}{n_c} + \frac{N}{n_tn_c}\sum_{i=1}^N\left(W_i-\frac{n_t}{N}H_i\right)\mathbb E_{\bm Z}[Z_i\mid\bm\pi] + \frac{N}{n_tn_c}\sum_{i=1}^N\sum_{k=1}^NT_{ik}\mathbb E_{\bm Z}[Z_iZ_k\mid\bm\pi]\nonumber\\
    &\quad+ \frac{N}{n_tn_c}\sum_{i=1}^N\mathbb E\left[\left(Z_i-\frac{n_t}{N}\right)\epsilon_t\mid\bm\pi\right] \nonumber\\
    &=-\frac{N\bar\alpha}{n_c} + \frac{N}{n_tn_c}\sum_{j=1}^M\sum_{i\in\mathcal C_j}\left(W_i-\frac{n_t}{N}H_i\right)\pi_j  + \frac{N}{n_tn_c}\sum_{j=1}^M\sum_{l\neq k}\sum_{i\in\mathcal C_i}\sum_{k\in\mathcal C_l}T_{ik}\pi_j\pi_l\nonumber\\
    &\quad +\frac{N}{n_tn_c}\sum_{j=1}^M\sum_{i\neq k\in\mathcal C_j}T_{ik}\left(\pi_j^2-\frac{\pi_j(1-\pi_j)}{N_j-1}\right) + \frac{N}{n_tn_c}\sum_{j=1}^M\sum_{i\in\mathcal C_j}T_{ii}\pi_j + 0\nonumber\\
    &=-\frac{N\bar\alpha}{n_c} + \frac{N}{n_tn_c}\sum_{j=1}^M\sum_{i\in\mathcal C_j}\left(W_i-\frac{n_t}{N}H_i\right)\pi_j + \frac{N}{n_tn_c}\sum_{j, l}\pi_j\pi_l\sum_{i\in\mathcal C_j}\sum_{k\in\mathcal C_k}T_{ik}\nonumber\\
    &\quad - \frac{N}{n_tn_c}\sum_{j=1}^M\frac{\pi_j(1-\pi_j)}{N_j-1}\sum_{i,k\in\mathcal C_j}T_{ik}.\label{eq:proof-dim-conditional-expectation}
\end{align}
Notice that under the permutation of $\bm \pi$, $\mathbb E_{\bm\pi}[\pi_j] = \overline{\bm\pi} = n_t/N$ for all $j\in[M]$ and 
\begin{align*}
    \mathbb E_{\bm\pi}[\pi_j\pi_l]=
    \begin{cases}
        \overline{\bm\pi}^2 + \mathrm{Var}[\bm\pi] & \text{if } j=l,\\
        \overline{\bm\pi}^2 - \frac{\mathrm{Var}[\bm\pi]}{M-1} &\text{if } j\neq l.
    \end{cases}
\end{align*}
Then we have from \eqref{eq:proof-dim-conditional-expectation},
\begin{align*}
    \mathbb E_{\bm Z, \bm\pi}[\hat\tau]&=\mathbb E_{\bm\pi}[\mathbb E_{\bm Z}[\hat\tau\mid \bm\pi]]\\
    &=-\frac{N\bar\alpha}{n_c} + \frac{N}{n_tn_c}\sum_{j=1}^M\sum_{i\in\mathcal C_j}\left(W_i-\frac{n_t}{N}H_i\right)\mathbb E_{\bm\pi}[\pi_j] + \frac{N}{n_tn_c}\sum_{j, l}\mathbb E_{\bm\pi}[\pi_j\pi_l]\sum_{i\in\mathcal C_j}\sum_{k\in\mathcal C_k}T_{ik}\nonumber\\
    &\quad - \frac{N}{n_tn_c}\sum_{j=1}^M\frac{\mathbb E_{\bm\pi}[\pi_j(1-\pi_j)]}{N_j-1}\sum_{i,k\in\mathcal C_j}T_{ik}\\
    &=-\frac{N\bar\alpha}{n_c} + \frac{1}{n_c}\sum_{j=1}^M\sum_{i\in\mathcal C_j}\left(W_i-\frac{n_t}{N}H_i\right)+\frac{N}{n_tn_c}\sum_{i,k}T_{ik}\left(\frac{n_t^2}{N^2}-\frac{\mathrm{Var}[\bm\pi]}{M-1}\right)\\
    &\quad+\frac{N}{n_tn_c}\sum_{j=1}^M\sum_{i,k\in\mathcal C_j}T_{ik}\left(\frac{M}{M-1}\mathrm{Var}[\bm \pi] + O\left(N_j^{-1}\right)\right)\\
    &=-\frac{N\bar\alpha}{n_c} + \frac{1}{n_c}\left(N\bar\alpha + n_c\bar\beta -n_t\bar\gamma\right) + \frac{N}{n_tn_c}\left(\frac{n_t^2}{N^2}-\frac{\mathrm{Var}[\bm\pi]}{M-1}\right)N\bar\gamma\\
    &\quad+ \frac{N}{n_tn_c}\frac{M}{M-1}\mathrm{Var}[\bm \pi]N\gamma' + O(M/N)\\
    &=\bar\beta + \frac{N^2}{n_tn_c}\left(\gamma' - \frac{\bar\gamma - \gamma'}{M-1}\right)\mathrm{Var}[\bm\pi]+O(M/N),
\end{align*}
where we utilize the fact that 
$$\sum_{i=1}^N H_i = \sum_{i=1}^N\sum_{k=1}^NT_{ki} =\sum_{k=1}^N\sum_{i=1}^NT_{ki}= \sum_{k=1}^N \gamma_k = N\bar\gamma.$$

\subsection{Proof of Corollary~\ref{cor:linear_interference_bias_randomized}}
Theorem~\ref{thm:bias_cond_graph} reveals that the bias of the $\hat\tau$ is $|TTE - \mathbb E_{\bm Z}[\hat\tau]|=\bar\gamma - \frac{N^2}{n_tn_c}(\gamma' - \frac{\bar\gamma-\gamma'}{M-1})\mathrm{Var}[\bm\pi]$. Since the dependence of bias on $\mathrm{Var}[\bm\pi]$ is linear, it suffices to provide lower and upper bound for $\mathrm{Var}[\bm\pi]$ when $\overline{\bm\pi}=n_t/N$ is fixed. On the one hand, the lower bound of $\mathrm{Var}[\bm\pi]$ is obviously zero, attained when $\pi_j = n_t/N$ for all $j$. On the other hand, $\mathrm{Var}[\bm\pi] = M^{-1}\sum_{j=1}^M\pi_j^2 - \overline{\pi}^2$ is bounded above by $\overline{\bm\pi}-\overline{\bm\pi}^2$ because $\pi_j^2 \leqslant \pi_j$ for all $0\leqslant \pi_j\leqslant 1$. The upper bound is attained when $\bm\pi \in\{0, 1\}^M$ since $\pi_j=\pi_j^2$ for all $j$. Depending on the sign of $\gamma' - \frac{\bar\gamma-\gamma'}{M-1}$, the minimal bias is achieved when $\mathrm{Var}[\bm\pi]$ takes the lower bound or upper bound. The corollary now is straightforward.

\subsection{Proof of Theorem~\ref{thm:variance-dim}}\label{sec:proof:variance-dim}
By Eve's law (or the law of total variance), the total variance of the difference-in-means estimator is decomposed to 
$$\mathrm{Var}_{\bm Z, \bm\pi}[\hat\tau] = \mathrm{Var}_{\bm\pi}[\mathbb E_{\bm Z}[\hat\tau\mid\bm\pi]] + \mathbb E_{\bm\pi}[\mathrm{Var}_{\bm Z}[\hat\tau\mid\bm\pi]].$$

The two terms on the right hand side are given by the following two lemmas.

\begin{lemma}\label{lemma:proof-variance-of-expectation}
The variance of the conditional expectation of difference-in-means estimator is given by 
\begin{multline*}
    \mathrm{Var}_{\bm\pi}[\mathbb E_{\bm Z}[\hat\tau\mid\bm\pi]]
    =\frac{N^2M}{n_t^2n_c^2}\Bigg\{ \mu_{2c}\mathbb S[\bm W^++\mu\bm \gamma^+] + \mu_{3c}\mathbb S\left[\bm D^+, \bm W^++\mu\bm \gamma^+\right]\\ +\frac{1}{4}(\mu_{4c}-\mu_{2c}^2)\mathbb S\left[\bm D^+\right]+\frac{1}{2}\mu_{2c}^2\sum_{j}\mathbb S[\{D^{(jl)}:l\neq j\}]\Bigg\},
\end{multline*}
where $\mu = n_t/N = \overline{\bm\pi}$, $\bm D^+=[D^{(jj)}]_{j=1,\dots, M}$ and $\mu_{2c}$, $\mu_{3c}$ and $\mu_{4c}$ are the second, third and fourth central moments of $\bm\pi$.
\end{lemma}

\begin{lemma}\label{lemma:proof-expectation-of-variance}
The conditional variance of the difference-in-means estimator is 
\begin{align*}
    \mathrm{Var}_{\bm Z}[\hat\tau\mid\bm\pi]
    =&\frac{N^2}{n_t^2n_c^2}\Bigg\{\sum_{j=1}^MN_j\pi_j(1-\pi_j)\mathbb S\left[\bm W^{(j)} - \frac{n_t}{N}\bm H^{(j)} + \sum_{l=1}^M\pi_l\bm D^{(l)(j)}\right]\\
    &\qquad +\frac{1}{2}\sum_{j=1}^M\sum_{l=1}^M\pi_j(1-\pi_j)\pi_l(1-\pi_l)\mathbb S_\times [\bm D_{jl}]\Bigg\},
\end{align*}
where $\bm D^{(l)(j)}=[D_i^{(l)}]_{i\in\mathcal C_j}$, $\bm D_{jl}=[D_{ik}]_{i\in\mathcal C_j, k\in\mathcal C_l}$ and 
$$\mathbb S_\times[\bm D_{jl}]:= \frac{1}{(N_j-1)(N_l-1)}\sum_{i\in\mathcal C_j}\sum_{k\in\mathcal C_l}\left(D_{ik} - N_j^{-1}D_k^{(j)} - N_l^{-1}D_i^{(l)} + (N_jN_l)^{-1}D^{(jl)}\right)^2.$$
Its expectation is 
\begin{align*}
    &\mathbb E_{\bm\pi}[\mathrm{Var}_{\bm Z}[\hat\tau\mid \bm\pi]]\\
    =&\frac{N^2}{n_t^2n_c^2}\Bigg\{\left[\mu(1-\mu)-\mu_{2c}\right]\sum_{j=1}^MN_j\mathbb S[\bm W^{(j)}+\mu\bm\gamma^{(j)}]\\
    &+ 2\left[(1-2\mu)\mu_{2c} -\mu_{3c}\right]\sum_{j=1}^MN_j\mathbb S[\bm W^{(j)}+\mu\bm\gamma^{(j)}, \bm {\tilde{D}}^{(j)(j)}]\\
    &+\left[\mu(1-\mu)\mu_{2c}-\mu_{2c}^2\right]\sum_{j=1}^M\sum_{l=1}^MN_j\mathbb S[\tilde{\bm D}^{(l)(j)}]\\
    &+\left[\mu_{2c}^2-\mu_{4c}+(1-2\mu)\mu_{3c}\right]\sum_{j=1}^MN_j\mathbb S[\tilde{\bm D}^{(j)(j)}]\\
    &+ \frac{1}{2}\left[\mu(1-\mu) - \mu_{2c}\right]^2\sum_{j=1}^M\sum_{l=1}^MN_jN_l\mathbb S_\times[\bm D_{jl}]\\ &+\frac{1}{2}\left[(1-2\mu)\mu_{2c}-2(1-2\mu)\mu_{3c}+\mu_{4c}-\mu_{2c}^2\right]]\sum_{j=1}^MN_j^2\mathbb S_\times[\bm D_{jj}]\Bigg\},
\end{align*}
where $\tilde{\bm D}^{(l)(j)}:= \bm D^{(l)(j)} - M^{-1}(\bm\gamma^{(j)}+\bm H^{(j)})$.
\end{lemma}
Proofs of above lemmas are deferred to later sub-sections. 

By combining the results from Lemma~\ref{lemma:proof-variance-of-expectation} and Lemma~\ref{lemma:proof-expectation-of-variance}, the total variance of the difference-in-means estimator is given by
$$\mathrm{Var}_{\bm Z, \bm \pi}[\hat\tau]=V_0 + V_1\mu_{2c}+V_2\mu_{2c}^2 + V_3\mu_{3c}+V_4(\mu_{4c}-\mu_{2c}^2),$$
where 
\begin{align*}
    V_0&=\frac{1}{n_tn_c}\sum_{j=1}^MN_j\mathbb S[\tilde{\bm W}^{(j)}] + \frac{1}{2N^2}\sum_{j=1}^M\sum_{l=1}^MN_jN_l\mathbb S_\times [\bm D_{jl}]\\
    V_1&=\frac{N^2M}{n_t^2n_c^2}\mathbb S[\tilde{\bm W}^+] - \frac{N^2}{n_t^2n_c^2}\sum_{j=1}^MN_j\mathbb S[\tilde{\bm W}^{(j)}]\\
    & + \frac{1}{n_tn_c}\sum_{j=1}^M\sum_{l=1}^MN_j\mathbb S[\tilde{\bm  D}^{(l)(j)}]
    -\frac{1}{n_tn_c}\sum_{j=1}^M\sum_{l=1}^MN_jN_l\mathbb S_\times [\bm D_{jl}]\\
    &+ \frac{2N(n_c-n_t)}{n_t^2n_c^2}\sum_{j=1}^MN_j\mathbb S[\tilde{\bm W}^{(j)}, \tilde{\bm D}^{(j)(j)}]  +\frac{N^2(n_c-n_t)^2}{2n_t^2n_c^2}\sum_{j=1}^MN_j^2\mathbb S_\times[\bm D_{jj}]\\
    V_2&=\frac{N^2}{2n_t^2n_c^2}\sum_{j}\mathbb S[\{D^{(jl)}:l\neq j\}]
    - \frac{N^2}{n_t^2n_c^2}\sum_{j=1}^M\sum_{l=1}^MN_j\mathbb S[\tilde{\bm D}^{(l)(j)}]
    +\frac{N^2}{2n_t^2n_c^2}\sum_{j=1}^M\sum_{l=1}^MN_jN_l\mathbb S_\times[\bm D_{jl}]\\
    V_3&=\frac{N^2M}{n_t^2n_c^2}\mathbb S[\tilde{\bm W}^+, \bm D^+] -\frac{2N^2}{n_t^2n_c^2}\sum_{j=1}^MN_j\mathbb S[\tilde{\bm W}^{(j)},\tilde{\bm D}^{(j)(j)}]\\
    &+\frac{N(n_c-n_t)}{n_t^2n_c^2}\sum_{j=1}^MN_j\mathbb S[\tilde{\bm D}^{(j)(j)}]-\frac{N(n_c-n_t)}{n_t^2n_c^2}\sum_{j=1}^MN_j^2\mathbb S_\times[\bm D_{jj}]\\
    V_4&=\frac{N^2M}{4n_t^2n_c^2}\mathbb S[\bm D^+]-\frac{N^2}{n_t^2n_c^2}\sum_{j=1}^MN_j\mathbb S[\tilde{\bm D}^{(j)(j)}] + \frac{N^2}{2n_t^2n_c^2}\sum_{j=1}^MN_j^2\mathbb S_\times[\bm D_{jj}],
\end{align*}
where $\tilde W_i := W_i + \mu \gamma_i$.

\subsection{Proof of Corollary~\ref{cor:var-simplified}}
We start from the explicit forms for the coefficients given in Section~\ref{sec:proof:variance-dim} and do not assume the block-fixed effect assumption yet. 
With Assumptions~\ref{assump:boundedness} and \ref{assump:dense}, the sum of symmetrized interference coefficient $D_{ik}$ can be bounded above by 
\begin{align*}
  \sum_{j=1}^M\sum_{i,k\in\mathcal C_j}D_{ik}^2&\leqslant \sum_{i=1}^N\sum_{k=1}^ND_{ik}^2=\sum_{i,k}\left(\frac{\gamma_i}{|\mathcal N_i|}+\frac{\gamma_k}{|\mathcal N_k|}\right)^2\mathbb I\{i\sim k\}\\ 
  &\leqslant 2\frac{C}{cN/M}\sum_{i,k}\left(\frac{\gamma_i}{|\mathcal N_i|}+\frac{\gamma_k}{|\mathcal N_k|}\right)\mathbb I\{i\sim k\}\leqslant \frac{4C^2}{c}M,
\end{align*}
where $c$ and $C$ are the constants in Assumptions~\ref{assump:variance-sufficient}.
Therefore, the terms involving summations of $\mathbb S_\times[\bm D_{jl}]$ in $V_0,\dots, V_4$ is of order $O(M/N^2)$, which is negligible.\\
Next, we simplify $\mathbb S[\tilde{\bm D}^{(l)(j)}]$. Recall that
$$D_i^{(l)} =\sum_{k\in\mathcal C_l}D_{ik} = \gamma_i\frac{|\mathcal N_i\cap\mathcal C_l|}{|\mathcal N_i|} + \sum_{k\in\mathcal C_l}\frac{\gamma_k}{|\mathcal N_k|}\mathbb I\{i\sim k\}.$$
From Assumption~\ref{assump:edge-prob} and \ref{assump:network-unconfoundedness}, we have
$$\frac{|\mathcal N_i\cap\mathcal C_l|}{|\mathcal N_i|}=\frac{|\mathcal N_i\cap\mathcal C_l|/N_l}{\sum_{l}|\mathcal N_i\cap\mathcal C_l|/N_l}=\frac{p_{jl}}{\sum_{l}p_{jl}}\left(1+O\left(\sqrt{\frac{\log NM}{N_l}}\right)\right).$$
and
$$\sum_{k\in\mathcal C_l}\frac{\gamma_k}{|\mathcal N_k|}\mathbb I\{i\sim k\}=\frac{q_{lj}\gamma^{(l)}}{N_l}\left(1 + O\left(\sqrt{\frac{\log(NM)}{N_l}}\right)\right)$$
Then, we have
\begin{align*}
    \mathbb S\left[\tilde{\bm D}^{(l)(j)}\right]&=(q_{jl}-M^{-1})^2\mathbb S[\bm\gamma^{(j)}] + (q_{jl}+q_{lj}+M^{-1}) O\left(\sqrt{\frac{\log NM}{N_j}}\right),
\end{align*}
where $q_{jl}:= p_{jl}/\sum_l p_{jl}$.\\ 
Noticing that $N^{-2}N_j\sum_{jl}q_{jl}=N^{-1}$, the right hand side of above formula is negligible.\\ 
Simplify, for $\mathbb S[\tilde{\bm W}^{(j)},\tilde{\bm D}^{(j)(j)}]$, we get
$$\mathbb S[\tilde{\bm W}^{(j)},\tilde{\bm D}^{(j)(j)}]=(q_{jj}-M^{-1})\mathbb S[\tilde{\bm W}^{(j)},\bm\gamma^{(j)}] + (q_{jj} + M^{-1})O\left(\sqrt{\frac{\log NM}{N_j}}\right).$$

By plugging above simplifications, we have the reduced coefficients as follows (high order terms are not displayed).
\begin{align*}
    V_0&=\frac{1}{n_tn_c}\sum_{j=1}^MN_j\mathbb S[\tilde {\bm W}^{(j)}] \\
    V_1&=\frac{N^2M}{n_t^2n_c^2}\mathbb S[\tilde {\bm W}^+] - \frac{N^2}{n_t^2n_c^2}\sum_{j=1}^MN_j\mathbb S[\tilde {\bm W}^{(j)}]\\
    &+ \frac{2N(n_c-n_t)}{n_t^2n_c^2}\sum_{j=1}^M\tilde q_{jj}N_j\mathbb S[\tilde{\bm W}^{(j)}, \bm\gamma^{(j)}]+ \frac{1}{n_tn_c}\sum_{j=1}^M\sum_{l=1}^M\tilde q_{jl}^2N_j\mathbb S[\bm\gamma^{(j)}]\\
    V_2&=\frac{N^2M}{2n_t^2n_c^2}\sum_{j=1}^M\mathbb S[\{q_{lj}\gamma^{(l)} + q_{jl}\gamma^{(j)}:l\neq j\}]
    - \frac{N^2}{n_t^2n_c^2}\sum_{j=1}^M\sum_{l=1}^M\tilde q_{jl}^2N_j\mathbb S[\bm \gamma^{(j)}]\\ 
    V_3&=\frac{2N^2M}{n_t^2n_c^2}\mathbb S[\tilde{\bm W}^+, q_{jj}\gamma^{(j)}]-\frac{2N^2}{n_t^2n_c^2}\sum_{j=1}^M\tilde q_{jj}^2N_j\mathbb S[\tilde{\bm W}^{(j)}, \bm\gamma^{(j)}]\\
    &+\frac{N(n_c-n_t)}{n_t^2n_c^2}\sum_{j=1}^M\tilde q_{jj}^2N_j\mathbb S[\bm\gamma^{(j)}]\\
    V_4&=\frac{N^2M}{n_t^2n_c^2}\mathbb S[q_{jj}\gamma^{(j)}]-\frac{N^2}{n_t^2n_c^2}\sum_{j=1}^M\tilde q_{jj}^2N_j\mathbb S[\bm \gamma^{(j)}],
\end{align*}
Furthermore, if we assume the interference effect is block-fixed, all $\mathbb S[\bm\gamma^{(j)}]$ and $\mathbb S[\bm\gamma^{(j)}, \tilde{\bm W}^{(j)}]$ vanish, which gives Coroallary~\ref{cor:var-simplified}.

\subsection{Proof of Proposition~\ref{prop:optimizatin-closed-form}}
Within the distribution family $\mathcal F$, we have the objective function $\mathrm{Var}_{\bm Z}[\hat\tau] = V_0 + V_1\mu_{2c} + V_2\mu_{4c} + V_4(\mu_4c-\mu_2c^2)$, where the third moment disappears. Suppose $n_t/N \leqslant 1/2$ and define $\delta_j = (\pi_j-n_t/N)^2$ for $j=1,\dots, M$. The support for $\delta_j$ is $[0, n_t^2/N^2]$. Observe that
\begin{align*}
  \mu_{2c} &=\frac{1}{M}\sum_{j=1}^M\left(\pi_j - \frac{n_t}{N}\right)^2 = \frac{1}{M}\sum_{j=1}^M\delta_j = \overline{\delta},\\ 
  \mu_{4c} &=\frac{1}{M}\sum_{j=1}^M\left(\pi_j - \frac{n_t}{N}\right)^4 = \frac{1}{M}\sum_{j=1}^M\delta_j^2 = \overline{\delta^2}.
\end{align*}
Noticing that $\mathrm{Var}_{\bm Z}[\hat\tau]$ is linear in $\mu_{4c}$ when $\mu_{2c}$ is fixed, we consider optimizing $\mu_{4c}$ first. Depending on the sign of $V_4$, it suffices to provide the lower and upper bound for $\mu_{4c}-\mu_{2c}^2$. On the one hand, by Jensen's inequality, we have $\mu_{4c}-\mu_{2c}^2=\overline{\delta^2}-\overline{\delta}^2\geqslant 0$, attained when $\delta_j = \mu_{2c}$ for all $j$. On the other hand, observe that
$$\overline{\delta^2}-\overline{\delta}^2\leqslant \frac{N^2}{n_t^2}\overline{\delta}-\overline{\delta}^2,$$
where the upper bound is attained when $\delta_j\in\{0, n_t^2/N^2\}$ for all $j$. \\ 
Specifically, if $V_4 > 0$, the optimal allocation of $\bm\pi$ is that 
$$\frac{1}{M}\sum_{j=1}^M\mathbb I\{\pi_j = \frac{n_t}{N} - \sqrt{\mu_{2c}}\}=\frac{1}{M}\sum_{j=1}^M\mathbb I\{\pi_j = \frac{n_t}{N} + \sqrt{\mu_{2c}}\} = \frac{1}{2}.$$
In this case, $\mu_{4c}-\mu_{2c} = 0$. \\ 
If $V_4 < 0$, the optimal allocation of $\bm\pi$ is that
$$\frac{1}{M}\sum_{j=1}^M\mathbb I\{\pi_j = 0\}=\frac{1}{M}\sum_{j=1}^M\mathbb I\{\pi_j = \frac{2n_t}{N}\}=\frac{\mu_{2c}N^2}{2n_t^2},\quad \frac{1}{M}\sum_{j=1}^M\mathbb I\{\pi_j = \frac{n_t}{N}\}= 1 - \frac{\mu_{2c}N^2}{n_t^2}.$$
In this case, $\mu_{4c}-\mu_{2c}^2=\mu_{2c}n_t^2/N^2-\mu_{2c}^2$. \\ 
When $V_4 = 0$, either is optimal. The rest is to optimize $\mu_{2c}$ as in a quadratic function. 

\subsection{Proof of Examples~\ref{example:interference1}, \ref{example:interference2} and \ref{example:interference3}}
We proof Example \ref{example:interference1} here and leave Examples \ref{example:interference2} and \ref{example:interference3} for the readers as they follow the same procedure. 
When the clustering is perfect and the interference is block-fixed, we adopt the result from Corollary~\ref{cor:var-simplified} such that the coefficients are 
  \begin{align*}
    V_0&=\frac{N}{n_tn_c}\sigma^2 + \frac{1}{n_tn_c}\sum_{j=1}^MN_j\mathbb S^{(j)}\left[\alpha_i + \frac{n_c}{N}\beta_i\right]\\
    V_1&=\frac{N^2M}{n_t^2n_c^2}\mathbb S\left[\alpha^{(j)} + \frac{n_c}{N}\beta^{(j)} + \frac{n_t}{N}\gamma^{(j)}\right] - \frac{N^2}{n_t^2n_c^2}\sum_{j=1}^MN_j\mathbb S^{(j)}\left[\alpha_i + \frac{n_c}{N}\beta_i\right]\\
    V_2&=0\\
    V_3&=\frac{2N^2M}{n_t^2n_c^2}\mathbb S\left[\alpha^{(j)} + \frac{n_c}{N}\beta^{(j)} + \frac{n_t}{N}\gamma^{(j)}, \gamma^{(j)}\right]\\
    V_4&=\frac{N^2M}{n_t^2n_c^2}\mathbb S[\gamma^{(j)}].
\end{align*}
In this case, we have $V_4 > 0$ and it falls to the first case in Proposition~\ref{prop:optimizatin-closed-form}. Since $V_2=0$ the quadratic optimization of $\mu_{2c}$ is simply a linear optimization. When $V_1 > 0$, $\mu_{2c}$ needs to be minimized. The correspnding case is all $\pi_j$ are at $n_t/N$, resulting zero variance and zero four moment. When $V_1 < 0$, $\mu_{2c}$ needs to be maximized at $n_t/N$, corresponding to a two-point distribution for $\bm\pi$ with half points at $0$ and the other half at $2n_t/N$.

\subsection{Proof of Lemma~\ref{lemma:proof-variance-of-expectation}}
We first rewrite the conditional expectation of $\hat\tau$ in \eqref{eq:proof-dim-conditional-expectation} in terms of the symmetrized interference coefficients $D_{ik}$ such that
$$\mathbb E_{\bm Z}[\hat\tau\mid\bm\pi]=-\frac{N\bar\alpha}{n_c}+ \frac{N}{n_tn_c}\sum_{j=1}^MU^{(j)}\pi_j + \frac{N}{2n_tn_c}\sum_{j, l}\pi_j\pi_lD^{(jl)}+O(M/N),$$
where 
$$U^{(j)}:=W^{(j)} - \frac{n_t}{N}H^{(j)}=\sum_{i\in\mathcal C_j}\left(W_i - \frac{n_t}{N}H_i\right)$$
The conditional expectation above is now a quadratic function of $\bm\pi$. Its variance is given by
\begin{align}
    &\mathrm{Var}[\mathbb E_{\bm Z}[\hat\tau\mid\bm\pi]]\nonumber\\
    =&\frac{N^2}{n_t^2n_c^2}\left\{\sum_{j,l}U^{(j)}U^{(l)}\mathrm{Cov}[\pi_j,\pi_l] + \sum_{j,l,j'}D^{(jl)}U^{(j')}\mathrm{Cov}[\pi_j\pi_l,\pi_{j'}]+\frac{1}{4}\sum_{j,l,j',l'}D^{(jl)}D^{(j'l')}\mathrm{Cov}[\pi_j\pi_l,\pi_{j'}\pi_{l'}]\right\}\label{eq:proof-variance-of-expectation-raw}
\end{align}
For $\mathrm{Cov}[\pi_j,\pi_l]$, we have
$$
\mathrm{Cov}[\pi_j, \pi_l] = \begin{cases}
        \mu_{2c} & \text{if } j=l,\\
        -\frac{1}{M-1}\mu_{2c}&\text{if } j\neq l,
    \end{cases}
$$
where, in this proof, we use $\mu$ to denote $\overline{\bm\pi}$ and use $\mu_{kc}$ to denote $k$-th central moment of $\bm\pi$.

The first term in \eqref{eq:proof-variance-of-expectation-raw} is now
\begin{align*}
    \sum_{j,l}U^{(j)}U^{(l)}\mathrm{Cov}[\pi_j, \pi_l]=\sum_j \left[U^{(j)}\right]^2\frac{M}{M-1}\mu_{2c}-\frac{\mu_{2c}}{M-1}\left[\sum_j U^{(j)}\right]^2=\mu_{2c}M\mathbb S[U^{(j)}],
\end{align*}
where $\mathbb S[U^{(j)}]$ denotes the sample variance of $\{U^{(j)}\}_{j=1}^M$.\\
For $j\neq j'\neq l$, we define the following excessive covariance. 
\begin{align*}
    V^{111}&:= \mathrm{Cov}[\pi_j\pi_l, \pi_{j'}]\\
    V^{21}_{21}&:= \mathrm{Cov}[\pi_j^2, \pi_{j'}] - V^{111}\\
    V^{21}_{111}&:= \mathrm{Cov}[\pi_j\pi_l, \pi_j] - V^{111}\\
    V^{3}&:= \mathrm{Cov}[\pi_j^2,\pi_j] - V^{111}-V^{21}_{21} - 2V^{21}_{111}.
\end{align*}
The above covariance are expressed in accordance with the inclusion-exclusion principle. The excessive covariance in addition to the preceding cases are presented. Therefore, the second term in \eqref{eq:proof-variance-of-expectation-raw} is now
\begin{align*}
    &\sum_{j,l,j'}D^{(jl)}U^{(j')}\mathrm{Cov}[\pi_j\pi_l,\pi_{j'}]\\
    =&\sum_{j,l,j'}D^{(jl)}U^{(j')}V^{111}+ \sum_{j,j'}D^{(jj)}U^{(j')}V^{21}_{21} + \sum_{j,l}\left(D^{(jl)}U^{(j)}+D^{(lj)}U^{(j)}\right)V^{21}_{111} + \sum_{j=1}^MD^{(jj)}U^{(j)}V^3\\
    =&V^{111}\left(\sum_{j,l}D^{(jl)}\right)\left(\sum_j U^{(j)}\right) + V^{21}_{21}\left(\sum_j D^{(jj)}\right)\left(\sum_j U^{(j)}\right)\\
    &+ 2V^{21}_{111}\sum_j\left(\sum_l D^{(jl)}\right)U^{(j)} +V^3\sum_j D^{(jj)}U^{(j)}\\
    =&2V^{21}_{111}M\mathbb S\left[\bm\gamma^++\bm H^+, \bm U^+\right] + V^3M\mathbb S\left[\bm D^+, \bm U^+\right]\\
    =&\left(2\mu\mu_{2c}+\frac{2}{M-2}\mu_{3c}\right)M\mathbb S[\bm U^{(j)}, \bm\gamma^{+}+ \bm H^{+}] + \frac{M}{M-2}\mu_{3c}M\mathbb S[\bm U^{+}, \bm D^{+}]\\
    \approx&2\mu\mu_{2c}M\mathbb S[\bm U^+, \bm\gamma^{+}+ \bm H^{+}]+\mu_{3c}M\mathbb S[\bm U^{+}, \bm D^{+}]
\end{align*}

Similarly, for $j\neq j'\neq l\neq l'$, we have the following excess covariance.
\begin{align*}
    V^{1111}&:=\mathrm{Cov}[\pi_j\pi_l,\pi_{j'}\pi_{l'}]\\
    V^{211}_{211}&:=\mathrm{Cov}[\pi_j^2,\pi_{j'}\pi_{l'}]-V^{1111}\\
    V^{211}_{1111}&:=\mathrm{Cov}[\pi_j\pi_l,\pi_{j}\pi_{l'}]-V^{1111}\\
    V^{22}_{22}&:=\mathrm{Cov}[\pi_j^2,\pi_{j'}^2] - V^{1111}-2V^{211}_{211}\\
    V^{22}_{1111}&:=\mathrm{Cov}[\pi_j\pi_l,\pi_{j}\pi_{l}]-V^{1111}-2V^{211}_{1111}\\
    V^{31}&:=\mathrm{Cov}[\pi_j^2,\pi_{j}\pi_{l'}]-V^{1111}-V^{211}_{211}-2V^{211}_{1111}\\
    V^4&:=\mathrm{Cov}[\pi_j^2,\pi_{j}^2]-V^{1111}-2V^{211}_{211}-4V^{211}_{1111}-V^{22}_{22}-2V^{22}_{1111}-4V^{31}.
\end{align*}
Therefore, by the inclusion-exclusion principle, the third term in \eqref{eq:proof-variance-of-expectation-raw} is
\begin{align*}
    &\sum_{j,l,j',l'}D^{(jl)}D^{(j'l')}\mathrm{Cov}[\pi_j\pi_l,\pi_{j'}\pi_{l'}]\\
    =&V^{1111}\sum_{j,l,j',l'}D^{(jl)}D^{(j'l')}
    +V^{211}_{211}\sum_{j,j',l'}\left(D^{(jj)}D^{(j'l')}+D^{(j'l')}D^{(jj)}\right)\\ 
    &+V^{211}_{1111}\sum_{j,l,l'}\left(D^{(jl)}D^{(jl')}+D^{(jl)}D^{(l'j)}+D^{(jl)}D^{(j'l)}+D^{(jl)}D^{(lj')}\right)\\
    &+V^{22}_{22}\sum_{j,j'}D^{(jj)}D^{(j'j')}
    +V^{22}_{1111}\sum_{j,l}\left(D^{(jl)}D^{(jl)}+D^{(jl)}D^{(lj)}\right)\\
    &+V^{31}\sum_{j,l}\left(D^{(jj)}D^{(jl)}+D^{(jj)}D^{(lj)}+D^{(jl)}D^{(jj)}+D^{(lj)}D^{(jj)}\right)
    +V^4\sum_{j}D^{(jj)}D^{(jj)}\\
    =&V^{1111}\left(\sum_{jl}D^{(jl)}\right)^2 +2V^{211}_{211}\left(\sum_j D^{(jj)}\right)\left(
    \sum_{j,l}D^{(jl)}\right)+ 4V^{211}_{1111}\sum_j\left(\sum_l D^{(jl)}\right)^2\\
    &+V^{22}_{22}\left(\sum_j D^{(jj)}\right)^2+2V^{22}_{1111}\sum_{j,l}\left(D^{(jl)}\right)^2 + 4V^{31}\sum_j D^{(jj)}\sum_l D^{(jl)} + V^4\sum_{j}\left(D^{(jj)}\right)^2\\
    =&-M(M-1)\left\{V^{1111}\mathbb S\left[\bm\gamma^++\bm H^+\right]-2V^{211}_{211}\mathbb S\left[\bm D^+, \bm\gamma^++\bm H^+\right]-V^{22}_{22}\mathbb S\left[\bm D^+\right]\right\}\\
    &+[4V^{211}_{1111}+MV^{1111}]\sum_j\left(\sum_l D^{(jl)}\right)^2
    +[4V^{31} + 2MV^{211}_{211}]\sum_j D^{(jj)}\sum_l D^{(jl)} \\
    &+ [V^4+MV^{22}_{22}]\sum_{j}\left(D^{(jj)}\right)^2
    + 2V^{22}_{1111}\sum_{j,l}\left(D^{(jl)}\right)^2\\
    =&-M(M-1)\left\{V^{1111}\mathbb S\left[\bm\gamma^++\bm H^+\right]-2V^{211}_{211}\mathbb S\left[\bm D^+, \bm\gamma^++\bm H^+\right]-V^{22}_{22}\mathbb S\left[\bm D^+\right]\right\}\\
    &+\frac{2V^{22}_{1111}}{M-1}\left\{-\sum_j\left(\sum_l D^{(jl)}\right)^2
    +2\sum_j D^{(jj)}\sum_l D^{(jl)} -M\sum_{j}\left(D^{(jj)}\right)^2
    +(M-1)\sum_{j,l}\left(D^{(jl)}\right)^2\right\}\\
    =&-M(M-1)\left\{V^{1111}\mathbb S\left[\bm\gamma^++\bm H^+\right]-2V^{211}_{211}\mathbb S\left[\bm D^+, \bm\gamma^++\bm H^+\right]-V^{22}_{22}\mathbb S\left[\bm D^+\right]\right\}\\
    &+2V^{22}_{1111}\sum_j\left\{\sum_l \left(D^{(jl)}-D^{(jj)}\right)^2 - \frac{1}{M-1}\left(\sum_lD^{(jl)} - MD^{(jj)}\right)^2\right\}\\
    =&-M(M-1)\left\{V^{1111}\mathbb S\left[\bm\gamma^++\bm H^+\right]-2V^{211}_{211}\mathbb S\left[\bm D^+, \bm\gamma^++\bm H^+\right]-V^{22}_{22}\mathbb S\left[\bm D^+\right]\right\}\\
    &+2V^{22}_{1111}(M-1)\sum_j\mathbb S[\{D^{(jl)}|l\neq j\}]\\
    =&\left(4\mu^2\mu_{2c} - \frac{8}{M-2}\mu\mu_{3c} - \frac{2(M^2+M-3)}{(M-1)(M-2)(M-3)}\mu_{2c}^2 +\frac{6}{(M-2)(M-3)}\mu_{4c}\right)M\mathbb S[\bm\gamma^{+}+\bm H^+]\\
    &+\left(4\frac{M}{M-2}\mu\mu_{3c} + \frac{4M(2M-3)}{(M-1)(M-2)(M-3)}\mu_{2c}^2 -\frac{4M}{(M-2)(M-3)}\mu_{4c}\right)M\mathbb S[\bm D^+, \bm\gamma^++\bm H^+]\\
    &+\left(-\frac{M^2(M-3)}{(M-1)(M-2)(M-3)}\mu_{2c}^2 + \frac{M(M-1)}{(M-2)(M-3)}\mu_{4c}\right)M\mathbb S[\bm D^+]\\
    &+\left(\frac{2M(M^2-3M+3)}{(M-1)^2(M-3)}\mu_{2c}^2 - \frac{2M}{(M-1)(M-3)}\mu_{4c}\right)M\sum_{j=1}^M\mathbb S[\{D^{(jl)}|l\neq j\}]\\
    \approx & 4\mu^2\mu_{2c}M\mathbb S\left[\bm\gamma^++\bm H^+\right]+4\mu\mu_{3c}M\mathbb S\left[\bm D^+, \bm\gamma^++\bm H^+\right]+(\mu_{4c}-\mu_{2c}^2)M\mathbb S\left[\bm D^+\right]\\
    &+2\mu_{2c}^2M\sum_{j=1}^M\mathbb S[\{D^{(jl)}|l\neq j\}].
\end{align*}

Finally, by substituting the results back to \eqref{eq:proof-variance-of-expectation-raw}, we have
\begin{align*}
    &\mathrm{Var}_{\bm\pi}[\mathbb E_{\bm Z}[\hat\tau\mid \bm\pi]] \\
    =&\frac{N^2M}{n_t^2n_c^2}\Bigg\{ \mu_{2c}\mathbb S[\bm W^++\mu\bm\gamma^+] + \mu_{3c}\mathbb S\left[\bm D^+, \bm W^++\mu\bm\gamma^+\right] +\frac{1}{4}(\mu_{4c}-\mu_{2c}^2)\mathbb S\left[\bm D^+\right]\\
    &\quad +\frac{1}{2}\mu_{2c}^2\sum_{j=1}^M\mathbb S[\{D^{(jl)}|l\neq j\}]\Bigg\}.
\end{align*}

\subsection{Proof of Lemma~\ref{lemma:proof-expectation-of-variance}}

We start from the difference-in-means estimator
$$\hat\tau(Z) = -\frac{N\bar\alpha}{N_c} + \frac{N}{n_tn_c}\sum_{i=1}^N U_iZ_i + \frac{N}{2n_tn_c}\sum_{i=1}^N\sum_{k=1}^ND_{ik}Z_iZ_k,$$
where $U_i = W_i - \frac{n_t}{N}H_i$.
Its conditional variance is therefore,
\begin{multline*}
\mathrm{Var}[\hat\tau(Z)\mid\bm\pi]=\frac{N^2}{n_t^2n_c^2}\Bigg\{\sum_{i, i'}U_iU_{i'}\mathrm{Cov}[Z_i, Z_{i'}\mid\bm\pi] + \sum_{i,k,i'}D_{ik}U_{i'}\mathrm{Cov}[Z_iZ_k, Z_{i'}\mid\bm\pi]\\
+\frac{1}{4}\sum_{i,k,i',k'}D_{ik}D_{i'k'}\mathrm{Cov}[Z_iZ_k, Z_{i'}Z_{k'}\mid\bm\pi]\Bigg\}
\end{multline*}
We now calculate the three terms in above formula.\\

\noindent\textbf{Term 1}\\
Since for different cluster memberships of $i$ and $i'$, we have 
\begin{align*}
	\mathrm{Cov}[Z_i, Z_i\mid\bm\pi] &= \pi_j(1-\pi_j) &&\text{if }i\in\mathcal C_j\\
	\mathrm{Cov}[Z_i, Z_{i'}\mid\bm\pi]  &= -\frac{\pi_j(1-\pi_j)}{N_j-1} &&\text{if }i\neq i'\in\mathcal C_j\\
	\mathrm{Cvo}[Z_i, Z_{i'}\mid\bm\pi] &= 0 &&\text{if }C_i\neq C_{i'}.
\end{align*}
then 
\begin{align*}
	S_1:=&\sum_{i=1}^N\sum_{i'=1}^NU_iU_{i'}\mathrm{Cov}[Z_i, Z_{i'}\mid \bm\pi]\\
	=&\sum_{j=1}^N\sum_{i\in\mathcal C_j}U_i^2\pi_j(1-\pi_j) - \sum_{j=1^N}\sum_{i\neq i'\in\mathcal C_j}U_iU_{i'}\frac{\pi_j(1-\pi_j)}{N_j-1}\\
	=&\sum_{j=1}^N\pi_j(1-\pi_j)\left(\sum_{i\in\mathcal C_j}U_i^2-\frac{1}{N_j-1}\sum_{i\neq i'\in\mathcal C_j}U_iU_{i'}\right)\\
	=&\sum_{j=1}^NN_j\pi_j(1-\pi_j)\mathbb S[\bm U^{(j)}]\\
\end{align*}

\noindent\textbf{Term 2}\\
We consider the following cases for covaraince $\mathrm{Cov}[Z_iZ_k, Z_{i'}\mid\bm\pi]$.\\
\noindent\textbf{-Case 2-a}\\
When $i=i'\neq k\in\mathcal C_j$, we have 
$$\mathrm{Cov}[Z_iZ_k, Z_i\mid\bm\pi]=\pi_j(1-\pi_j)\left(\pi_j-\frac{1-\pi_j}{N_j-1}\right).$$
The subtotal is
\begin{align*}
	S_{2a}:=&\sum_{j=1}^M\sum_{i\neq k\in\mathcal C_j}(D_{ik}U_i + D_{ki}U_i)\pi_j(1-\pi_j)\left(\pi_j-\frac{1-\pi_j}{N_j-1}\right)\\
	=&2\sum_{j=1}^M\sum_{i\neq k\in\mathcal C_j}D_{ik}U_i\pi_j(1-\pi_j)\left(\pi_j-\frac{1-\pi_j}{N_j-1}\right)\\
	=&2\sum_{j=1}^M\sum_{i\in\mathcal C_j}D_{i}^{(j)}U_i\pi_j(1-\pi_j)\left(\pi_j-\frac{1-\pi_j}{N_j-1}\right).
\end{align*}

\noindent\textbf{-Case 2-b}\\
When $i=i'\in\mathcal C_j$ and $k\in\mathcal C_l, l\neq j$, we have 
$$\mathrm{Cov}[Z_iZ_k, Z_i\mid\bm\pi]=\pi_j(1-\pi_j)\pi_l.$$
The subtotal is
\begin{align*}
	S_{2b}:=&\sum_{j=1}^M\sum_{i\in\mathcal C_j}\sum_{l\neq j}\sum_{k\in\mathcal C_l}(D_{ik}U_i + D_{ki}U_i)\pi_j(1-\pi_j)\pi_l\\
	=&2\sum_{j=1}^M\sum_{i\in\mathcal C_j}\sum_{l\neq j}D_{i}^{(l)}U_i\pi_j(1-\pi_j)\pi_l\\
	=&2\sum_{j=1}^M\sum_{i\in\mathcal C_j}\pi_j(1-\pi_j)U_i(G_i - \pi_jD_i^{(j)}),
\end{align*}
where $G_i:= \sum_{l=1}^M\pi_lD_i^{(l)}$.

\noindent\textbf{-Case 2-c}\\
When $i\neq i'\neq k\in\mathcal C_j$, we have
\begin{align*}
\mathrm{Cov}[Z_iZ_k, Z_{i'}\mid\bm\pi]&=\pi_j\left(\pi_j - \frac{1-\pi_j}{N_j-1}\right)\left(\pi_j - \frac{2(1-\pi_j)}{N_j-2}\right)-\pi_j^2\left(\pi_j - \frac{1-\pi_j}{N_j-1}\right)\\
&=-2\frac{\pi_j(1-\pi_j)}{N_j-2}\left(\pi_j - \frac{1-\pi_j}{N_j-1}\right).
\end{align*}
The subtotal is
\begin{align*}
	S_{2c}:=&-2\sum_{j=1}^M\sum_{i\neq i'\neq k\in\mathcal C_j}D_{ik}U_{i'}\frac{\pi_j(1-\pi_j)}{N_j-2}\left(\pi_j - \frac{1-\pi_j}{N_j-1}\right)\\
	=&-2\sum_{j=1}^M\frac{\pi_j(1-\pi_j)}{N_j-2}\left(\pi_j - \frac{1-\pi_j}{N_j-1}\right)\sum_{i\neq i'\neq k\in\mathcal C_j}D_{ik}U_{i'}\\
	=&-2\sum_{j=1}^M\frac{\pi_j(1-\pi_j)}{N_j-2}\left(\pi_j - \frac{1-\pi_j}{N_j-1}\right)\sum_{i\neq i'\in\mathcal C_j}(D_i^{(j)}U_{i'}-D_{ii'}U_{i'})\\
	=&-2\sum_{j=1}^M\frac{\pi_j(1-\pi_j)}{N_j-2}\left(\pi_j - \frac{1-\pi_j}{N_j-1}\right)\left(\sum_{i\neq i'\in\mathcal C_j}D_i^{(j)}U_{i'}-\sum_{i\in\mathcal C_j}D_{i}^{(j)}U_{i}\right).
\end{align*}

\noindent\textbf{-Case 2-d}\\
When $i\neq i'\in\mathcal C_j$ and $k\in\mathcal C_l, l\neq j$, we have
$$\mathrm{Cov}[Z_iZ_k, Z_{i'}\mid\bm\pi]=-\frac{\pi_j(1-\pi_j)}{N_j-1}\pi_l.$$
The subtotal is
\begin{align*}
	S_{2d}:=&-\sum_{j=1}^M\sum_{i\neq i'\in\mathcal C_j}\sum_{l\neq j}\sum_{k\in\mathcal C_l}(D_{ik}U_{i'} + D_{ki}U_{i'})\frac{\pi_j(1-\pi_j)}{N_j-1}\pi_l\\
	=&-2\sum_{j=1}^M\sum_{i\neq i'\in\mathcal C_j}\sum_{l\neq j}D_{i}^{(l)}U_{i'}\frac{\pi_j(1-\pi_j)}{N_j-1}\pi_l\\
	=&-2\sum_{j=1}^M\sum_{i\neq i'\in\mathcal C_j}U_{i'}\frac{\pi_j(1-\pi_j)}{N_j-1}(G_i - \pi_jD_i^{(j)})\\
	=&-2\sum_{j=1}^M\frac{\pi_j(1-\pi_j)}{N_j-1}\sum_{i\neq i'\in\mathcal C_j}(G_iU_{i'} - \pi_j D_i^{(j)}U_{i'}).
\end{align*}

By summing up $S_{2a}$ to $S_{2d}$ we have
\begin{align*}
	S_2 :=&\sum_{i,k,i'}D_{ik}U_{i'}\mathrm{Cov}[Z_iZ_k, Z_{i'}\mid\pi]\\
	=&S_{2a}+S_{2b}+S_{2c}+S_{2d}\\
	=&2\sum_{j=1}^M\pi_j(1-\pi_j)\left(\sum_{i\in\mathcal C_j}G_iU_i - \frac{1}{N_j-1}\sum_{i\neq i'\in\mathcal C_j}G_iU_{i'}\right)\\
	&\qquad +2\sum_{j=1}^M\pi_j(1-\pi_j)\frac{2\pi_j-1}{N_j-2}\left(\sum_{i\in\mathcal C_j}D_i^{(j)}U_i - \frac{1}{N_j-1}\sum_{i\neq i'\in\mathcal C_j}D_i^{(j)}U_{i'}\right)\\
	=&2\sum_{j=1}^M\pi_j(1-\pi_j)N_j\left(\mathbb S[\bm G^{(j)}, \bm U^{(j)}]+\frac{2\pi_j-1}{N_j-2}\mathbb S[\bm D^{(j)(j)}, \bm U^{(j)}]\right),
\end{align*}
where $\bm D^{(j)(j)}:= (D^{(j)}_i)_{i\in\mathcal C_j}$.

\noindent\textbf{Term 3}\\
We consider the following cases for covariance $\mathrm{Cov}[Z_iZ_k, Z_{i'}Z_{k'}\mid\bm\pi]$.\\

\noindent\textbf{-Case 3-a}\\
When $i=i'\in\mathcal C_j$ and $k=k'\in\mathcal C_j$ but $i\neq k$, we have
$$\mathrm{Cov}[Z_iZ_k, Z_iZ_k\mid\bm\pi] = \pi_j\left(\pi_j-\frac{1-\pi_j}{N_j-1}\right)\left[1-\pi_j\left(\pi_j-\frac{1-\pi_j}{N_j-1}\right)\right].$$
The subtotal is
\begin{align*}
	S_{3a}:=&\sum_{j=1}^M\sum_{i\neq k\in\mathcal C_j}(D_{ik}D_{ik}+D_{ik}D_{ki})\pi_j\left(\pi_j-\frac{1-\pi_j}{N_j-1}\right)\left[1-\pi_j\left(\pi_j-\frac{1-\pi_j}{N_j-1}\right)\right]\\
	=&2\sum_{j=1}^M\pi_j\left(\pi_j-\frac{1-\pi_j}{N_j-1}\right)\left[1-\pi_j\left(\pi_j-\frac{1-\pi_j}{N_j-1}\right)\right]\sum_{i,k\in\mathcal C_j}D_{ik}^2.
\end{align*}

\noindent\textbf{-Case 3-b}\\
When $i=i'\in\mathcal C_j$ and $k=k'\in\mathcal C_l$ but $j\neq l$, we have
$$\mathrm{Cov}[Z_iZ_k, Z_iZ_k\mid\bm\pi] = \pi_j\pi_l(1-\pi_j\pi_l).$$
The subtotal is
\begin{align*}
	S_{3b}:=&\sum_{j=1}^M\sum_{l\neq j}\sum_{i\in\mathcal C_j}\sum_{k\in\mathcal C_l}(D_{ik}D_{ki} + D_{ik}D_{ki})\pi_j\pi_l(1-\pi_j\pi_l)\\
	=&2\sum_{j=1}^M\sum_{l\neq j}\pi_j\pi_l(1-\pi_j\pi_l)\sum_{i\in\mathcal C_j}\sum_{k\in\mathcal C_l}D_{ik}^2\\
	=&2\sum_{j=1}^M\sum_{l=1}^M\pi_j\pi_l(1-\pi_j\pi_l)\sum_{i\in\mathcal C_j}\sum_{k\in\mathcal C_l}D_{ik}^2 - 2\sum_{j=1}^M\pi_j^2(1-\pi_j^2)\sum_{i,k\in\mathcal C_j}D_{ik}^2.
\end{align*}

\noindent\textbf{-Case 3-c}\\
When $i=i'\in\mathcal C_j$ and $k\neq k'\in\mathcal C_j$, we have
$$\mathrm{Cov}[Z_iZ_k, Z_iZ_{k'}\mid\bm\pi] = \pi_j(1-\pi_j)\left(\pi_j - \frac{1-\pi_j}{N_j-1}\right)\left(\frac{N_j}{N_j-1}\pi_j - \frac{2}{N_j-2}\right).$$
The subtotal is
\begin{align*}
	S_{3c}:=&\sum_{j=1}^M\sum_{i\neq k\neq k'\in\mathcal C_j}(D_{ik}D_{ik'}+D_{ik}D_{k'i}+D_{ki}D_{ik'}+D_{ki}D_{k'i})\pi_j(1-\pi_j)\\
	&\times \left(\pi_j - \frac{1-\pi_j}{N_j-1}\right)\left(\frac{N_j}{N_j-1}\pi_j - \frac{2}{N_j-2}\right)\\
	=&4\sum_{j=1}^M\pi_j(1-\pi_j)\left(\pi_j - \frac{1-\pi_j}{N_j-1}\right)\left(\frac{N_j}{N_j-1}\pi_j - \frac{2}{N_j-2}\right)\sum_{i\neq k\neq k'\in\mathcal C_j}D_{ik}D_{ik'}\\
	=&4\sum_{j=1}^M\pi_j(1-\pi_j)\left(\pi_j - \frac{1-\pi_j}{N_j-1}\right)\left(\frac{N_j}{N_j-1}\pi_j - \frac{2}{N_j-2}\right)\left(\sum_{i\in\mathcal C_j}\left[D_{i}^{(j)}\right]^2 - \sum_{i,k\in\mathcal C_j}D_{ik}^2\right).
\end{align*}

\noindent\textbf{-Case 3-d}\\
When $i=i'\in\mathcal C_j$, $k\in\mathcal C_j$ and $k'\in\mathcal C_l, l\neq j$, we have
$$\mathrm{Cov}[Z_iZ_k, Z_iZ_{k'}\mid\bm\pi] = \pi_j(1-\pi_j)\left(\pi_j - \frac{1-\pi_j}{N_j-1}\right)\pi_l.$$
The subtotal is
\begin{align*}
	S_{3d}:=&8\sum_{j=1}^M\sum_{l\neq j}\sum_{i\neq k\in\mathcal C_j}\sum_{k'\in\mathcal C_l}D_{ik}D_{ik'}\pi_j(1-\pi_j)\left(\pi_j - \frac{1-\pi_j}{N_j-1}\right)\pi_l\\
	=&8\sum_{j=1}^M\pi_j(1-\pi_j)\left(\pi_j - \frac{1-\pi_j}{N_j-1}\right)\sum_{i\neq k\in\mathcal C_j}\sum_{l\neq j}\sum_{k'\in\mathcal C_l}\pi_lD_{ik}D_{ik'}\\
	=&8\sum_{j=1}^M\pi_j(1-\pi_j)\left(\pi_j - \frac{1-\pi_j}{N_j-1}\right)\sum_{i\neq k\in\mathcal C_j}D_{ik}G_i - \pi_jD_{ik}D_i^{(j)}\\
	=&8\sum_{j=1}^M\pi_j(1-\pi_j)\left(\pi_j - \frac{1-\pi_j}{N_j-1}\right)\sum_{i\in\mathcal C_j}D_{i}^{(j)}G_i - \pi_j\left[D_i^{(j)}\right]^2.
\end{align*}

\noindent\textbf{-Case 3-e}\\
When $i=i'\in\mathcal C_j$, $k\neq k'\in\mathcal C_l, l\neq j$, we have
$$\mathrm{Cov}[Z_iZ_k, Z_iZ_{k'}\mid\bm\pi] = \pi_j\pi_l\left(\pi_l(1-\pi_j)-\frac{1-\pi_l}{N_l-1}\right).$$
The subtotal is
\begin{align*}
	S_{3e}:=&4\sum_{j=1}^M\sum_{i\in\mathcal C_j}\sum_{l\neq j}\sum_{k\neq k'\in\mathcal C_l}D_{ik}D_{ik'}\pi_j\pi_l\left(\pi_l(1-\pi_j)-\frac{1-\pi_l}{N_l-1}\right)\\
	=&4\sum_{j=1}^M\sum_{i\in\mathcal C_j}\sum_{l\neq j}\pi_j\pi_l\left(\pi_l(1-\pi_j)-\frac{1-\pi_l}{N_l-1}\right)\left(\left[D_i^{(l)}\right]^2-\sum_{k\in\mathcal C_l}D_{ik}^2\right)\\
	=&4\sum_{j=1}^M\sum_{l=1}^M\sum_{i\in\mathcal C_j}\pi_j\pi_l\left(\pi_l(1-\pi_j)-\frac{1-\pi_l}{N_l-1}\right)\left(\left[D_i^{(l)}\right]^2-\sum_{k\in\mathcal C_l}D_{ik}^2\right)\\
	&\qquad - 4\sum_{j=1}^M\pi_j^2(1-\pi_j)\left(\pi_j-\frac{1}{N_j-1}\right)\left(\sum_{i\in\mathcal C_j}\left[D_i^{(j)}\right]^2-\sum_{i,k\in\mathcal C_j}D_{ik}^2\right).
\end{align*}

\noindent\textbf{-Case 3-f}\\
When $i=i'\in\mathcal C_j$, $k\in\mathcal C_l$, $k'\in\mathcal C_{l'}$ and $j\neq l\neq l'$, we have
$$\mathrm{Cov}[Z_iZ_k, Z_iZ_{k'}\mid\bm\pi] = \pi_j(1-\pi_j)\pi_l\pi_{l'}.$$
The subtotal is
\begin{align*}
	S_{3f}:=&4\sum_{j=1}^M\sum_{l\neq l'\neq j}\sum_{i\in\mathcal C_j}\sum_{k\in\mathcal C_l}\sum_{k'\in\mathcal C_{l'}}D_{ik}D_{ik'}\pi_j(1-\pi_j)\pi_l\pi_{l'}\\
	=&4\sum_{j=1}^M\sum_{l\neq l'\neq j}\sum_{i\in\mathcal C_j}D_{i}^{(l)}D_{i}^{(l')}\pi_j(1-\pi_j)\pi_l\pi_{l'}\\
	=&4\sum_{j=1}^M\pi_j(1-\pi_j)\sum_{i\in\mathcal C_j}\left((G_i - \pi_jD_i^{(j)})^2 - \sum_{l\neq j}\pi_l^2\left[D_i^{(l)}\right]^2\right)\\
	=&4\sum_{j=1}^M\pi_j(1-\pi_j)\sum_{i\in\mathcal C_j}\left(G_i^2 - 2\pi_jG_iD_i^{(j)} + 2\pi_j^2\left[D_i^{(j)}\right]^2 - \sum_{l=1}^M\pi_l^2\left[D_i^{(l)}\right]^2\right).\\
\end{align*}

\noindent\textbf{-Case 3-g}\\
When $i\neq i'\neq k\neq k'\in\mathcal C_j$, we have
\begin{align*}
\mathrm{Cov}[Z_iZ_k, Z_{i'}Z_{k'}\mid\bm\pi] =& -\pi_j(1-\pi_j)\left(\pi_j-\frac{1-\pi_j}{N_j-1}\right)\\
&\times \left(\frac{4N_j^2-12N_j+6}{(N_j-1)(N_j-2)(N_j-3)}\pi_j - \frac{6(1-\pi_j)}{(N_j-2)(N_j-3)}\right)
\end{align*}
The subtotal is
\begin{align*}
	S_{3g}:=&-\sum_{j=1}^M\sum_{i\neq i'\neq k\neq k'\in\mathcal C_j} D_{ik}D_{i'k'}\pi_j(1-\pi_j)\left(\pi_j-\frac{1-\pi_j}{N_j-1}\right)\\
	&\times\left(\frac{4N_j^2-12N_j+6}{(N_j-1)(N_j-2)(N_j-3)}\pi_j - \frac{6(1-\pi_j)}{(N_j-2)(N_j-3)}\right)\\
	=&-\sum_{j=1}^M\pi_j(1-\pi_j)\left(\pi_j-\frac{1-\pi_j}{N_j-1}\right)\left(\frac{4N_j^2-12N_j+6}{(N_j-1)(N_j-2)(N_j-3)}\pi_j - \frac{6(1-\pi_j)}{(N_j-2)(N_j-3)}\right)\\
	&\qquad\times\left(\left[D^{(jj)}\right]^2-4\sum_{i\in\mathcal C_j}\left[D_i^{(j)}\right]^2+2\sum_{i,k\in\mathcal C_j}D_{ik}^2\right).
\end{align*}

\noindent\textbf{-Case 3-h}\\
When $i\neq i'\neq k'\in\mathcal C_j$, $k\in\mathcal C_l, l\neq j$, we have
$$\mathrm{Cov}[Z_iZ_k, Z_{i'}Z_{k'}\mid\bm\pi] = -2\frac{\pi_j(1-\pi_j)}{N_j-2}\left(\pi_j-\frac{1-\pi_j}{N_j-1}\right)\pi_l.$$
The subtotal is
\begin{align*}
	S_{3h}:=&-8\sum_{j=1}^M\sum_{l\neq j}\sum_{i\neq i'\neq k'\in\mathcal C_j}\sum_{k\in\mathcal C_l}D_{ik}D_{i'k'}\frac{\pi_j(1-\pi_j)}{N_j-2}\left(\pi_j-\frac{1-\pi_j}{N_j-1}\right)\pi_l\\
	=&-8\sum_{j=1}^M\frac{\pi_j(1-\pi_j)}{N_j-2}\left(\pi_j-\frac{1-\pi_j}{N_j-1}\right)\sum_{i\neq i'\neq k'\in\mathcal C_j}D_{i'k'}(G_i-\pi_jD_i^{(j)})\\
	=&-8\sum_{j=1}^M\frac{\pi_j(1-\pi_j)}{N_j-2}\left(\pi_j-\frac{1-\pi_j}{N_j-1}\right)\sum_{i\neq i'\in\mathcal C_j}(D_{i'}^{(j)}-D_{i'i})(G_i-\pi_jD_i^{(j)})\\
	=&-8\sum_{j=1}^M\frac{\pi_j(1-\pi_j)}{N_j-2}\left(\pi_j-\frac{1-\pi_j}{N_j-1}\right)\\
	&\times\left(\sum_{i\neq i'\in\mathcal C_j}\left(G_iD_{i'}^{(j)}-\pi_jD_i^{(j)}D_{i'}^{(j)}\right)-\sum_{i\in\mathcal C_j}\left(G_iD_i^{(j)} - \pi_j\left[D_i^{(j)}\right]^2\right)\right).\\
\end{align*}

\noindent\textbf{-Case 3-i}\\
When $i\neq i'\in\mathcal C_j$, $k\neq k'\in\mathcal C_l, l\neq j$, we have
$$\mathrm{Cov}[Z_iZ_k, Z_{i'}Z_{k'}\mid\bm\pi] = -\pi_j\pi_l\left(\frac{\pi_l(1-\pi_j)}{N_j-1}+\frac{\pi_j(1-\pi_l)}{N_l-1}-\frac{(1-\pi_j)(1-\pi_l)}{(N_j-1)(N_l-1)}\right).$$
The subtotal is
\begin{align*}
	S_{3i}:=&-2\sum_{j=1}^M\sum_{l\neq j}\sum_{i\neq i'\in\mathcal C_j}\sum_{k\neq k'\in\mathcal C_l}D_{ik}D_{i'k'}\pi_j\pi_l\left(\frac{\pi_l(1-\pi_j)}{N_j-1}+\frac{\pi_j(1-\pi_l)}{N_l-1}-\frac{(1-\pi_j)(1-\pi_l)}{(N_j-1)(N_l-1)}\right)\\
	=&-2\sum_{j=1}^M\sum_{l\neq j}\pi_j\pi_l\left(\frac{\pi_l(1-\pi_j)}{N_j-1}+\frac{\pi_j(1-\pi_l)}{N_l-1}-\frac{(1-\pi_j)(1-\pi_l)}{(N_j-1)(N_l-1)}\right)\\
	&\qquad\times \left(\left[D^{(jl)}\right]^2 - \sum_{i\in\mathcal C_j}\left[D_i^{(l)}\right]^2 - \sum_{k\in\mathcal C_l}\left[D_k^{(j)}\right]^2+\sum_{i\in\mathcal C_j}\sum_{k\in\mathcal C_l}D_{ik}^2\right)\\
	=&-2\sum_{j=1}^M\sum_{l=1}^M\pi_j\pi_l\left(\frac{\pi_l(1-\pi_j)}{N_j-1}+\frac{\pi_j(1-\pi_l)}{N_l-1}-\frac{(1-\pi_j)(1-\pi_l)}{(N_j-1)(N_l-1)}\right)\\
	&\qquad\times \left(\left[D^{(jl)}\right]^2 - \sum_{i\in\mathcal C_j}\left[D_i^{(l)}\right]^2 - \sum_{k\in\mathcal C_l}\left[D_k^{(j)}\right]^2+\sum_{i\in\mathcal C_j}\sum_{k\in\mathcal C_l}D_{ik}^2\right)\\
	&+2\sum_{j=1}^M\pi_j^2\left(2\frac{\pi_l(1-\pi_j)}{N_j-1}-\frac{(1-\pi_j)^2}{(N_j-1)^2}\right)\left(\left[D^{(jj)}\right]^2 - 2\sum_{i\in\mathcal C_j}\left[D_i^{(j)}\right]^2 +\sum_{i,k\in\mathcal C_j}D_{ik}^2\right).\\
\end{align*}

\noindent\textbf{-Case 3-j}\\
When $i\neq i'\in\mathcal C_j$, $k\in\mathcal C_l$, $k'\in\mathcal C_{l'}$ and $j\neq l\neq l'$, we have
$$\mathrm{Cov}[Z_iZ_k, Z_{i'}Z_{k'}\mid\bm\pi] = -\frac{\pi_j(1-\pi_j)}{N_j-1}\pi_l\pi_{l'}.$$
The subtotal is
\begin{align*}
	S_{3j}:=&-4\sum_{j=1}^M\sum_{l\neq l'\neq j}\sum_{i\neq i'\in\mathcal C_j}\sum_{k\in\mathcal C_l}\sum_{k'\in\mathcal C_{l'}}D_{ik}D_{i'k'}\frac{\pi_j(1-\pi_j)}{N_j-1}\pi_l\pi_{l'}\\
	=&-4\sum_{j=1}^M\frac{\pi_j(1-\pi_j)}{N_j-1}\sum_{i\neq i'\in\mathcal C_j}\left(G_iG_{i'}-2\pi_jG_iD_{i'}^{(j)} + 2\pi_j^2D_i^{(j)}D_{i'}^{(j)}-\sum_{l=1}^M\pi_l^2D_i^{(l)}D_{i'}^{(l)}\right).
\end{align*}

By summing up $S_{3a}$ to $S_{3j}$, we have
\begin{align*}
	S_3 :=&\sum_{i,k,i',k'}D_{ik}D_{i'k'}\mathrm{Cov}[Z_iZ_k, Z_{i'}Z_{k'}\mid\bm\pi]\\
		 =&-\sum_{j=1}^M\left(\frac{2\pi_j(1-\pi_j)(1-2\pi_j)^2}{N_j-1}+O(N_j^{-2})\right)\left[\sum_{i,k\in\mathcal C_j}D_{ik}^2-N_j^{-1}\sum_{i\in\mathcal C_j}\left[D_i^{(j)}\right]^2\right]\\
		 &+\sum_{j=1}^M\sum_{l=1}^M\left(2\pi_j\pi_l(1-\pi_j)(1-\pi_l)\right)\frac{N_j}{N_j-1}\frac{N_l}{N_l-1}\left[\sum_{i\in\mathcal C_j}\sum_{k\in\mathcal C_l}D_{ik}^2-N_l^{-1}\sum_{i\in\mathcal C_j}[D_i^{(l)}]^2\right]\\
		 &-\sum_{j=1}^M\sum_{l=1}^M\left(2\frac{\pi_j\pi_l(1-\pi_j)(1-\pi_l)}{N_l-1}\right)\frac{N_j}{N_j-1}\left[\sum_{i\in\mathcal C_j}[D_{i}^{(l)}]^2-N_j^{-1}[D^{(jl)}]^2\right]\\
		 &+\sum_{j=1}^M\frac{8\pi_j(1-\pi_j)(2\pi_j-1)}{N_j-2}\left[\sum_{i\in\mathcal C_j}G_iD_i^{(j)} - (N_j-1)^{-1}\sum_{i\neq i'\in\mathcal C_j}G_iD_{i'}^{(j)}\right]\\
		 &+\sum_{j=1}^M4\pi_j(1-\pi_j)\left[\sum_{i\in\mathcal C_j}G_i^2 - (N_j-1)^{-1}\sum_{i\neq i'\in\mathcal C_j}G_iG_{i'}\right]\\
		 =&-\sum_{j=1}^M\left[2\pi_j(1-\pi_j)(1-2\pi_j)^2 + O(N_j^{-1})\right]\sum_{k\in\mathcal C_j}\mathbb S[\bm D^{k(j)}]\\
		 &+\sum_{j=1}^M\sum_{l=1}^M2\pi_j\pi_l(1-\pi_j)(1-\pi_l)N_jN_l \mathbb S_{\times}[\bm D_{jl}]\\
		 &+\sum_{j=1}^M8\pi_j(1-\pi_j)(2\pi_j-1)\frac{N_j}{N_j-2}\mathbb S[\bm G^{(j)}, \bm D^{(j)(j)}]\\
		 &+\sum_{j=1}^M4\pi_j(1-\pi_j)N_j\mathbb S[\bm G^{(j)}]\\
		 =&\sum_{j=1}^M4\pi_j(1-\pi_j)N_j\mathbb S\left[\bm G^{(j)}+ \frac{2\pi_j-1}{N_j-2}\bm D^{(j)(j)}\right]\\ 
		 &\qquad +\sum_{j=1}^M\sum_{l=1}^M2\pi_j\pi_l(1-\pi_j)(1-\pi_l)N_jN_l\mathbb S_\times[\bm D_{jl}]
\end{align*}
where $\bm D^{i(l)}:=(D_{ik})_{k\in\mathcal C_l}$, $\bm D_{jl}=[D_{ik}]_{i\in\mathcal C_j, k\in\mathcal C_l}$ and $\mathbb S_\times$ is the mean squared interaction defined by
$$\mathbb S_\times [\bm X] = \frac{1}{(m-1)(n-1)}\sum_{i=1}^m\sum_{j=1}^n \left(X_{ij} - m^{-1}\sum_{i=1}^mX_{ij} - n^{-1}\sum_{j=1}^nX_{ij} + (mn)^{-1}\sum_{ij}X_{ij}\right)^2$$
for any $m\times n$ matrix $\bm X$.\\

Finally, we have the conditional variance of the difference-in-means estimator 
\begin{align}
	\mathrm{Var}[\hat\tau\mid \bm \pi]=&S_1 + S_2 + S_3/4\nonumber\\
	=&\sum_{j=1}^MN_j\pi_j(1-\pi_j)\mathbb S\left[\bm G^{(j)}+\bm U^{(j)}+ \frac{2\pi_j-1}{N_j-2}\bm D^{(j)(j)}\right]\\ &+\frac{1}{2}\sum_{j=1}^M\sum_{l=1}^MN_jN_l\pi_j\pi_l(1-\pi_j)(1-\pi_l)\mathbb S_\times [\bm D_{jl}]\nonumber\\
	\approx & \sum_{j=1}^MN_j\pi_j(1-\pi_j)\mathbb S\left[\bm G^{(j)}+\bm U^{(j)}\right] +\frac{1}{2}\sum_{j=1}^M\sum_{l=1}^MN_jN_l\pi_j\pi_l(1-\pi_j)(1-\pi_l)\mathbb S_\times [\bm D_{jl}]\label{eq:proof-conditional-variance}
\end{align}
with an relative error of order $O(N_j^{-1})$.

Next, we derive the expectation of \eqref{eq:proof-conditional-variance} under the permutation of $\bm\pi$. Notice that $G_i$ depends on $\{\pi_l\}_{l=1}^M$ such that
\begin{align*}
    \mathbb S[\bm U^{(j)} + \bm G^{(j)}]&= \mathbb S\left[\bm U^{(j)} + \sum_l \pi_l \bm D^{(l)(j)}\right]\\
    &=\mathbb S[\bm U^{(j)}] + 2\sum_{l=1}^M\pi_l\mathbb S[\bm U^{(j)}, \bm D^{(l)(j)}] + \sum_{l, m} \pi_l\pi_{m}\mathbb S[\bm D^{(l)(j)}, D^{(m)(j)}].
\end{align*}
Hence the first term in \eqref{eq:proof-conditional-variance} is
\begin{align*}
    &\sum_{j=1}^MN_j\pi_j(1-\pi_j)\mathbb S[\bm U^{(j)}+\bm G^{(j)}]\\
    =&\sum_{j=1}^MN_j\pi_j(1-\pi_j)\mathbb S[\bm U^{(j)}] + 2\sum_{j=1}^M\sum_{l=1}^M\pi_j(1-\pi_j)\pi_l N_j\mathbb S[\bm U^{(j)}, \bm D^{(l)(j)}]\\
    &+ \sum_{j=1}^M\sum_{l=1}^M\sum_{m=1}^MN_j\pi_j(1-\pi_j)\pi_l\pi_{m}\mathbb S[\bm D^{(l)(j)}, \bm D^{(m)(j)}].
\end{align*}
The expectation of above equation involves higher moments of $\bm\pi$. Hence we denote the expectation terms as follows. For $j\neq l\neq m$,
\begin{align*}
    \mathbb E[\pi_j(1-\pi_j)]&= E_j\\
    \mathbb E[\pi_j(1-\pi_j)\pi_l]&= E_{jl}\\
    \mathbb E[\pi_j^2(1-\pi_j)]&= E_{jj} + E_{jl}\\
    \mathbb E[\pi_j(1-\pi_j)\pi_l\pi_{m}]&=E_{jlm}\\
    \mathbb E[\pi_j(1-\pi_j)\pi_j\pi_l]&=E_{jjl} + E_{jlm}\\
    \mathbb E[\pi_j(1-\pi_j)\pi_l^2] &= E_{jll} + E_{jlm}\\
    \mathbb E[\pi_j^3(1-\pi_j)]&=E_{jjj} + E_{jlm} + 2E_{jjl} + E_{jll}\\
    \mathbb E[\pi_j(1-\pi_j)\pi_l(1-\pi_l)]&=E_{jl0}\\
    \mathbb E[\pi_j^2(1-\pi_j)^2]&=E_{jl1}.
\end{align*}
Here $E$-terms denotes the excessive expectation compared to a more general case. 

Then we have
\begin{align*}
    &\mathbb E\left[\sum_{j=1}^MN_j\pi_j(1-\pi_j)\mathbb S[\bm U^{(j)}+\bm G^{(j)}]\right]\\
    =&\sum_{j=1}^MN_j\pi_j(1-\pi_j)\mathbb S[\bm U^{(j)}] + 2\sum_{j=1}^M\sum_{l=1}^M\pi_j(1-\pi_j)\pi_l N_j\mathbb S[\bm U^{(j)}, \bm D^{(l)(j)}]\\
    &+ \sum_{j=1}^M\sum_{l=1}^M\sum_{m=1}^MN_j\pi_j(1-\pi_j)\pi_l\pi_{m}\mathbb S[\bm D^{(l)(j)}, \bm D^{(m)(j)}]\\
    =&\sum_j N_j E_j\mathbb S[\bm U^{(j)}] + 2\sum_{j,l} N_j E_{jl}\mathbb S[\bm U^{(j)}, \bm D^{(l)(j)}] + 2\sum_j N_j E_{jj}\mathbb S[\bm U^{(j)}, \bm D^{(j)(j)}]\\
    &+\sum_{j,l,m}N_jE_{jlm}\mathbb S[\bm D^{(l)(j)}, \bm D^{(m)(j)}] + 2\sum_{j,l}N_j E_{jjl}\mathbb S[\bm D^{(j)(j)}, \bm D^{(l)(j)}] + \sum_{j,l}N_jE_{jll}\mathbb S[\bm D^{(l)(j)}]\\
    &+\sum_j N_j E_{jjj}\mathbb S[\bm D^{(j)(j)}]\\
    =&\sum_j N_j E_j\mathbb S[\bm U^{(j)}] + 2\sum_{j} N_j E_{jl}\mathbb S[\bm U^{(j)}, \bm \gamma^{(j)} + \bm H^{(j)}] + 2\sum_j N_j E_{jj}\mathbb S[\bm U^{(j)}, \bm D^{(j)(j)}]\\
    &+\sum_{j}N_jE_{jlm}\mathbb S[\bm\gamma^{(j)}+\bm H^{(j)}] + 2\sum_{j}N_j E_{jjl}\mathbb S[\bm D^{(j)(j)}, \bm\gamma^{(j)}+\bm H^{(j)}] + \sum_{j,l}N_jE_{jll}\mathbb S[\bm D^{(l)(j)}]\\
    &+\sum_j N_j E_{jjj}\mathbb S[\bm D^{(j)(j)}]\\
    =&\sum_j N_j E_j\mathbb S[\bm U^{(j)}] + 2\sum_{j} N_j \left(E_{jl}+\frac{1}{M}E_{jj}\right)\mathbb S[\bm U^{(j)}, \bm \gamma^{(j)} + \bm H^{(j)}]\\
    &+ 2\sum_j N_j E_{jj}\mathbb S[\bm U^{(j)}, \bm D^{(j)(j)} - (\bm\gamma^{(j)}+\bm H^{(j)})/M]\\
    &+\sum_{j}N_j\left(E_{jlm}+\frac{1}{M}E_{jll} + \frac{2}{M}E_{jjl} + \frac{1}{M^2}E_{jjj}\right)\mathbb S[\bm\gamma^{(j)}+\bm H^{(j)}]\\
    &+ 2\sum_{j}N_j E_{jjl}\mathbb S[\bm D^{(j)(j)} - (\bm\gamma^{(j)}+\bm H^{(j)})/M, \bm\gamma^{(j)}+\bm H^{(j)}]\\
    &+ \sum_{j}N_jE_{jll}\left(\sum_l\mathbb S[\bm D^{(l)(j)}]-\frac{1}{M}\mathbb S[\bm\gamma^{(j)}+\bm H^{(j)}]\right)\\
    &+\sum_j N_j E_{jjj}\left(\mathbb S[\bm D^{(j)(j)}]-\mathbb S[(\bm\gamma^{(j)}+\bm H^{(j)})/M]\right)\\
    =& \sum_j N_j E_j \mathbb S[\bm U^{(j)} + \mu\bm \gamma^{(j)} +\mu \bm H^{(j)}]\\
    &+2\sum_j N_j E_{jj}\mathbb S[\bm U^{(j)} + \mu\bm \gamma^{(j)} +\mu \bm H^{(j)}, \bm D^{(j)(j)} - (\bm\gamma^{(j)}+\bm H^{(j)})/M]\\
    &- 2\sum_{j}N_j \frac{E_{jjj}}{M}\mathbb S[\bm D^{(j)(j)} - (\bm\gamma^{(j)}+\bm H^{(j)})/M, \bm\gamma^{(j)}+\bm H^{(j)}]\\
    &+ \sum_{j}N_jE_{jll}\left(\sum_l\mathbb S[\bm D^{(l)(j)}]-\frac{1}{M}\mathbb S[\bm\gamma^{(j)}+\bm H^{(j)}]\right)\\
    &+\sum_j N_j E_{jjj}\left(\mathbb S[\bm D^{(j)(j)}]-\mathbb S[(\bm\gamma^{(j)}+\bm H^{(j)})/M]\right)\\
    =&\sum_j N_j E_j\mathbb S[\bm U^{(j)}+\mu\bm\gamma^{(j)} + \mu \bm H^{(j)}] + 2\sum_j N_j E_{jj}\mathbb S[\bm U^{(j)} + \mu\bm\gamma^{(j)} + \mu\bm H^{(j)}, \tilde{\bm D}^{(j)(j)}]\\
    &+\sum_j N_j E_{jjj}\mathbb S[\tilde{\bm D}^{(j)(j)}] + \sum_j N_j E_{jll}\sum_l \mathbb S[\tilde{\bm D}^{(l)(j)}]\\
    =&\sum_j N_j \left[\mu(1-\mu)  - \mu_{2c}\right] \mathbb S[\bm W^{(j)} + \mu\bm \gamma^{(j)}]\\
    &+2\sum_j N_j\left[(1-2\mu)\mu_{2c}-\mu_{3c}+O(M^{-1}]\right)\mathbb S[\bm W^{(j)} + \mu\bm \gamma^{(j)}, \tilde{\bm D}^{(j)(j)}]\\
    & + \sum_j N_j \left[(1-2\mu)\mu_{3c}+\mu_{2c}^2-\mu_{4c}+O(M^{-1})\right]\mathbb S[\tilde{\bm D}^{(j)(j)}]\\
    &+\sum_jN_j\left[\mu(1-\mu)\mu_{2c}-\mu_{2c}^2+O(M^{-1})\right]\sum_l\mathbb S[\tilde{\bm D}^{(l)(j)}],
\end{align*}
where we define
$$\tilde{\bm D}^{(l)(j)} = \bm D^{(l)(j)} - M^{-1}\left(\bm \gamma^{(j)} + \bm H^{(j)}\right).$$

For the second term in \eqref{eq:proof-conditional-variance}, we have
\begin{align*}
    &\mathbb E\left[\sum_{j=1}^M\sum_{l=1}^MN_jN_l\pi_j(1-\pi_j)\pi_l(1-\pi_l)\mathbb S_\times[\bm D_{jl}\right]\\
    =&\sum_{j,l}N_jN_lE_{jl0}\mathbb S_\times[\bm D_{jl}]+\sum_{j}N_j^2E_{jl1}\mathbb S_\times[\bm D_{jj}]\\
    =&\left[(\mu(1-\mu) - \mu_{2c})^2 + O(M^{-1})\right]\sum_{j=1}^M\sum_{l=1}^MN_jN_l\mathbb S_\times[\bm D_{jl}]\\
    &+\left[(1-2\mu)\mu_{2c}-2(1-2\mu)\mu_{3c}+\mu_{4c}-\mu_{2c}^2\right]\sum_{j=1}^MN_j^2\mathbb S_\times[\bm D_{jj}]
\end{align*}

By combining terms together, we have
\begin{align*}
    &\mathbb E_{\bm\pi}[\mathrm{Var}_{\bm Z}[\hat\tau\mid \bm\pi]]\\
    =&\frac{N^2}{n_t^2n_c^2}\Bigg\{\left[\mu(1-\mu)-\mu_{2c}\right]\sum_{j=1}^MN_j\mathbb S[\bm W^{(j)}+\mu\bm\gamma^{(j)}]\\
    &+ 2\left[(1-2\mu)\mu_{2c} -\mu_{3c}\right]\sum_{j=1}^MN_j\mathbb S[\bm W^{(j)}+\mu\bm\gamma^{(j)}, \bm {\tilde{D}}^{(j)(j)}]\\
    &+\left[\mu(1-\mu)\mu_{2c}-\mu_{2c}^2\right]\sum_{j=1}^M\sum_{l=1}^MN_j\mathbb S[\tilde{\bm D}^{(l)(j)}]\\
    &+\left[\mu_{2c}^2-\mu_{4c}+(1-2\mu)\mu_{3c}\right]\sum_{j=1}^MN_j\mathbb S[\tilde{\bm D}^{(j)(j)}]\\
    &+ \frac{1}{2}\left[\mu(1-\mu) - \mu_{2c}\right]^2\sum_{j=1}^M\sum_{l=1}^MN_jN_l\mathbb S_\times[\bm D_{jl}]\\ &+\frac{1}{2}\left[(1-2\mu)\mu_{2c}-2(1-2\mu)\mu_{3c}+\mu_{4c}-\mu_{2c}^2\right]]\sum_{j=1}^MN_j^2\mathbb S_\times[\bm D_{jj}]\Bigg\}.
\end{align*}
The result is up to a relative error of $O(M^{-1})$.

\section{Analysis on Random Graph Extension}
\subsection{Proof of Proposition~\ref{prop:random-graph-assumption}}
For Assumption~\ref{assump:dense}, let $i$ be an unit in $\mathcal C_j$. Then
$|\mathcal N_i| = \sum_{l=1}^M\sum_{k\in\mathcal C_j}\mathbb I\{k\in\mathcal N_i\}$, where for $k\in\mathcal C_l$, each $\mathbb I\{k\in\mathcal N_i\}$ is a Bernoulli random variable with probability $A_{jl}$. 
By Bernstein's inequality, we have for $0\leqslant t\leqslant \mathbb E[|\mathcal N_i|]=\sum_{l=1}^MN_lA_{jl}$,
\begin{align}
  P[|\mathcal N_i| \leqslant \mathbb E[|\mathcal N_i|] - t]&\leqslant \exp\left\{-\frac{\frac{1}{2}t^2}{\sum_{l=1}^M\sum_{k\in\mathcal C_l}A_{jl}(1-A_{jl}) + \frac{1}{3}\left(A_{jl}\vee (1-A_{jl})\right)t}\right\}\notag\\
  &\leqslant \exp\left\{-\frac{\frac{1}{2}t^2}{\sum_{l=1}^MN_lA_{jl} + \frac{1}{3}\sum_{l=1}^MN_lA_{jl}}\right\}\notag\\
  &\leqslant \exp\left\{-\frac{1}{3}\frac{t^2}{\sum_{l=1}^MN_lA_{jl}}\right\}.\label{eq:proof-bernstein}
\end{align}
If we plug in $t=\mathbb E[|\mathcal N_i|]-\epsilon_2N_j=N_j\left[\sum_{l=1}^MA_{jl} - \epsilon_2\right]$, we have
\[
  P\left[|\mathcal N_i| \leqslant \epsilon_2 \frac{N}{M}\right]\leqslant \exp\left\{-\frac{N}{3M}\frac{\left(\sum_{l=1}^MA_{jl} - \epsilon_2\right)^2}{\sum_{l=1}^MA_{jl}}\right\}.
\]
Take the union bound over all units, we have
\[
  P\left[\min_i\ |\mathcal N_i| \leqslant \epsilon_2 \frac{N}{M}\right]\leqslant N\exp\left\{-\frac{N}{3M}\frac{\left(\min_j\ \sum_{l=1}^MA_{jl} - \epsilon_2\right)^2}{\min_j\ \sum_{l=1}^MA_{jl}}\right\}.
\]
Therefore, one can choose $\epsilon_2 \leqslant \underline{A}- \sqrt{\frac{3M}{N}\underline{A}\log \frac{N}{\delta}}$ with $\underline{A}=\min_j\ \sum_l A_{jl}$ to control the probability that Assumption~\ref{assump:dense} is violated below $\delta$. If $\frac{M}{N}\log N\rightarrow 0$, one can choose $\epsilon_2=\underline{A} - \sqrt{-\underline{A}\log\ \delta}$.\\ 
For Assumption~\ref{assump:edge-prob}, $A_{jl}$ is a natural edge forming probability for $p_{jl}$. Similar to \eqref{eq:proof-bernstein}, we have the probability for the edges between unit $i$ and cluster $l$ to violate Assumption~\ref{assump:edge-prob} as
\begin{align*}
  P[||\mathcal N_i\cap\mathcal C_j| - A_{jl}N_l| \geqslant \epsilon_3\sqrt{A_{jl}N_l\log NM}] &\leqslant \exp\left\{-\frac{1}{3}\epsilon_3^2\log (NM)\right\}=(NM)^{-\frac{\epsilon_3^2}{3}},
\end{align*}
which gives the union bound
\begin{align*}
&P\left[\bigcup_{i,l}\ \left\{||\mathcal N_i\cap\mathcal C_j| - A_{jl}N_l| \geqslant \epsilon_3\sqrt{A_{jl}N_l\log (NM)}\right\}\right]\\ 
&\leqslant \sum_{i, l}P\left[||\mathcal N_i\cap\mathcal C_j| - A_{jl}N_l| \geqslant \epsilon_3\sqrt{A_{jl}N_l\log (NM)}\right]\\ 
&\leqslant (NM)^{1-\epsilon_3^2/3}.
\end{align*}
Hence, as long as one choose $\epsilon_3^2\geqslant 3 - \frac{3\log \delta}{\log(NM)}$, Assumption~\ref{assump:edge-prob} is satisfied with probability at least $1-\delta$. One such choice that does not depend on $N, M$ is $\epsilon_3 = \sqrt{3(1-\log\delta)}$.

Assumption~\ref{assump:network-unconfoundedness} can be done in a very similar way by observing that $\|[f(\alpha_i, \beta_i, \gamma_i)-\bar f_j]\mathbb I\{k\in\mathcal N_i\}\|_\infty\leqslant \|[f(\alpha_i, \beta_i, \gamma_i)-\bar f_j]\|_\infty \leqslant 2\|f\|_\infty$, where $\bar f_j = N_j^{-1}=\sum_{i\in\mathcal C_j}f(\alpha_i, \beta_i, \gamma_i)$ is the averaged function value in cluster $\mathcal C_j$.
Again, by Bernstein's inequality, we have
\[
    P\left[\left|\sum_{i\in\mathcal C_j}\ [f(\alpha_i, \beta_i, \gamma_i)-\bar f_j]\mathbb I\{k\in\mathcal N_i\}\right| \geqslant \epsilon_f\sqrt{N_j\log(NM)}\right]\leqslant (NM)^{-\frac{1}{3}\left(\frac{\epsilon_f}{2\|f\|_\infty}\right)^2},
\]
which yields a union bound for the probability that Assumption~\ref{assump:network-unconfoundedness} is not satisfied as $(NM)^{1-\frac{1}{3}\left(\frac{\epsilon_f}{2\|f\|_\infty}\right)^2}$. As long as $\epsilon_f\geqslant 2\|f\|_\infty\sqrt{3\left(1-\frac{\log \delta}{\log(NM)}\right)}$, Assumption~\ref{assump:network-unconfoundedness} is satisfied with probability at least $1-\delta/3$. One can choose $\epsilon = 2\|f\|_\infty \sqrt{3(1-\log\delta)}$ to remove dependence on $N, M$.


\section{Analysis on the Stratified Estimator}

\subsection{Presentation of Initial Results}

Recall the definition of the stratified estimator $\hat\tau^s$.
Let $N_j$ be the total number of units in cluster $\calC_j$
and $n_j$ be the number of units assigned to treatment in that cluster. Let
$\bm{\lambda} \in \mathbb{R}^M_+$ be a vector of positive coefficients, usually
chosen to sum to $1$, we define the stratified estimator as
\begin{equation*}
  \hat \tau^s \defeq \sum_{j =1}^M \lambda_j \hat \tau_j = \sum_{j =1}^M
  \lambda_j \left(\sum_{i \in \calC_j} \frac{Z_i}{n_j} Y_i(\Z) - \sum_{i \in
  \calC_j} \frac{1 - Z_i}{N_j - n_j} Y_i(\Z) \right)
\end{equation*}

In general, we recommend choosing $\lambda_j = \frac{N_j}{N}$, as evidenced by
the following result on the bias of the stratified estimator under the stable
treatment value assumption:

\begin{proposition}
  \label{prop:bias_sutva_stratified}
  Assume that the standard unit treatment value assumption holds. The
  conditional expectation of the stratified estimator is
  \begin{equation}
    \forall \bm{\pi},~\E_\Z[\hat \tau^s | \bm{\pi} ] = \E_\Z[\hat \tau^s] =
    \sum_{j=1}^N \frac{\lambda_j}{N_j} \sum_{i \in \calC_j} Y_i(1) - Y_i(0)
  \end{equation}
  If $\lambda_j = \frac{N_j}{N}$, then the stratified estimator is unbiased for
  the total treatment effect, conditionally on the assignment of treatment
  proportions to clusters. The same holds true in expectation over a
  randomized saturation assignment.
\end{proposition}

A proof can be found in the following subsections. Similarly, the variance has an easily interpretable closed-form under the
standard unit treatment value assumption. Recall $S_{tj} \defeq \mathbb S(\Y^{(j)}(\mathbf{1}))$,  $S_{cj} \defeq \mathbb S(\Y^{(j)}(\mathbf{0}))$, and $S_{tcj} \defeq \mathbb S (\Y^{(j)}(\mathbf{1}) - \Y^{(j)}(\mathbf{0}))$.

\begin{proposition}
  \label{prop:variance_sutva_stratified}
  Assume that the stable unit value assumption holds. 
  The variance of the stratified estimator under a randomized saturation design
  is
  \begin{equation}
    \label{eq:stratified_estimator_variance_sutva}
    \var_\Z[\hat \tau^s] = \sum_{j=1}^M \lambda_j^2 \frac{N}{N_j}
    \left(\frac{S_{tj}}{\pi^\dagger N} + \frac{S_{cj}}{(1  - \pi)^\dagger N} -
    \frac{S_{tcj}}{N} \right)
  \end{equation}
  where $\pi^\dagger \defeq \left(\frac{1}{M} \sum_{j=1}^M \pi_j^{-1}
  \right)^{-1}$ is the harmonic mean of $\pi$ and $(1 - \pi)^\dagger$ is the
  harmonic mean of $1 - \pi$. Constrained to maintain $\bar{\pi} =
  \frac{n_t}{N}$, the stratified completely randomized design with $\bm{\pi} =
  \left(\frac{n_t}{N}\right)_M$ minimizes the variance of the stratified
  estimator in Eq.~\ref{eq:stratified_estimator_variance_sutva}, which is then
  equal to $\var_\Z \left[\hat \tau^s \right] =  \sum_{j = 1}^M \lambda_j^2 \frac{N}{N_j}
  \left(\frac{S_{tj}}{n_t} + \frac{S_{cj}}{n_c} - \frac{S_{tcj}}{N} \right)$.
\end{proposition}

A proof can be found in the following subsections. In
contrast to the variance of the difference-in-means estimator $\var[\hat \tau]$,
which is linear in the variance of the treatment-proportions vector
$\var[\bm{\pi}]$ (cf.  Prop~\ref{prop:variance_sutva}), the variance of the
stratified estimator $\var[\hat \tau^s]$ depends on the variance of the
treatment-proportions vector only through the inverse of the harmonic mean of
$\bm{\pi}$, as stated in Proposition~\ref{prop:variance_sutva_stratified}. Since
any mean-preserving spread decreases the harmonic mean~\citep{mitchell200488},
when holding $n_t$ constant, any increase in the variance of the
treatment-proportions vector increases the variance of the stratified estimator
under the stable unit treatment value assumption.

We can also express the bias of the
stratified estimator under the linear interference model of
Equation~\ref{eq:dirg:ch3}.  Recall that $\rho_\calC = \frac{1}{N} \sum_{i=1}^N
\frac{|\N_i \cap \calC_j|}{|\N_i|}$ is the proportion of a unit's neighborhood
that also belongs to its cluster, averaged over all units.
\begin{proposition}
  \label{prop:bias_linear_interference_stratified}
  Under the linear model of interference in Eq.~\ref{eq:dirg:ch3}, the
  expectation of the stratified estimator conditioned on the assignment of
  clusters to treatment-proportions is:
  \begin{equation*}
    \E_\Z[\hat \tau^s | \bm{\pi}] = \E_\Z[\hat \tau^s] = \bar{\beta}
    + \frac{1}{N} \sum_{j=1}^M \sum_{i\in \calC_j} \gamma_i \frac{|\N_i \cap
    \calC_j|}{|\N_i|}
  \end{equation*}
  If the interference effects are constant, then the formula becomes $\E_\Z[\hat \tau^s] = \bar{\beta} + \rho_\calC \gamma$.
\end{proposition}
A proof can be found in
the following subsections. In conclusion, the
bias of the stratified estimator does not depend on the treatment-proportions
vector $\bm{\pi}$ under SUTVA or under
the suggested linear model of interference. Its variance, when SUTVA can be assumed, decreases with $\var[\bm{\pi}]$.

\subsection{Proof of Proposition~\ref{prop:bias_sutva_stratified}}
\label{proof:prop:bias_sutva_stratified}

The stratified estimator is given by:
\begin{align*}
  \hat{\tau}^s & = \sum_{j = 1}^M \lambda_j \hat \tau(j) \\
  & = \sum_{j = 1}^M \lambda_j  \sum_{i = 1}^{N_j} \left( Z_i Y_i(1)
  + (1 - Z_i) Y_i(0)\right) \frac{(-1)^{1 - Z_i}}{n_j^{Z_i} (N_j - n_j)^{1 -
  Z_i}}
\end{align*}

The expectation of the stratified difference-in-means estimator conditioned on
the proportion of units assigned to treatment is given by:
\begin{align*}
  \E_\Z \left[\hat \tau^s \middle| \bm{\pi} \right] &= \sum_{j =1}^M
  \frac{\lambda_j}{N_j} \sum_{i \in \calC_j} Y_i(1) - Y_i(0)
\end{align*}

If $\lambda_j = \frac{N_j}{N}$, then the stratified estimator is unbiased for
the total treatment effect conditioned on the assignment of treatment
proportions to clusters. Same in expectation over that assignment.

\subsection{Proof of Proposition~\ref{prop:variance_sutva_stratified}}
\label{proof:prop:variance_sutva_stratified}

According to Eve's law, we must compute two terms. The first term is equal to
$0$.
\begin{equation*}
  \var_{\bm{\pi}}\left[ \E_\Z [\hat \tau | \bm{\pi} \right] = \var_{\bm{\pi}}
  \left[ \sum_{j = 1}^M \frac{\lambda_j}{N_j} \sum_{i \in \calC_j} Y_i(1) - Y_i(0)
  \right]  = 0
\end{equation*}
such that $\var_\Z \left[\hat \tau^s \right] = \E_{\bm{\pi}}\left[ \var_\Z \left[
  \hat \tau^s | \bm{\pi} \right] \right]$. We compute this remaining term:
\begin{align*}
  \var_\Z[\hat \tau^s | \bm{\pi}] &= \sum_{j =1}^M \lambda_j^2 \var_{\Z} \left[ \hat
  \tau(j) | \pi_j \right] \\
  & = \sum_{j = 1}^M \lambda_j^2 \left(\frac{S_{tj}}{n_j} + \frac{S_{cj}}{N_j
  - n_j} - \frac{S_{tcj}}{N_j}  \right)
\end{align*}
\begin{align*}
  \E_{\bm{\pi}} \left[ \var_\Z[\hat \tau^s | \bm{\pi}] \right] &= \sum_{j = 1}^M
  \lambda_j^2 \left(S_{tj} \E_{\bm{\pi}} \left[\frac{1}{n_j}\right] + S_{cj}
  \E_{\bm{\pi}} \left [\frac{1}{N_j - n_j} \right] - \frac{S_{tcj}}{N_j}
  \right) \\
  & = \sum_{j = 1}^M \frac{\lambda_j^2}{N_j} \left(S_{tj} \E_{\bm{\pi}}
  \left[\frac{1}{\pi_j}\right] + S_{cj} \E_{\bm{\pi}} \left [\frac{1}{1 -
  \pi_j} \right] - S_{tcj} \right)
\end{align*}
Let $\bm{\pi}^\dagger = \left(\frac{1}{M} \sum_{j =1}^M \frac{1}{\pi_j}
\right)^{-1}$ be the harmonic mean of $\pi$.  If we use $\lambda_j =
\frac{N_j}{N}$, then the above formula becomes:
\begin{equation*}
  \var_\Z \left[\hat \tau^s \right] = \sum_{j = 1}^M \frac{N_j}{N}
   \left( \frac{S_{tj}}{N \bm{\pi}^\dagger}
   + \frac{S_{cj}}{N \bm{(1 - \pi)}^\dagger} - \frac{S_{tcj}}{N} \right)
\end{equation*}
Since any mean-preserving spread~\citep{mitchell200488} of $\bm{\pi}$ will
decrease the harmonic mean, the optimal randomized saturation design is one with
the lowest variance for $\bm{\pi}$, i.e. $\bm{\pi} =
\left(\frac{n_t}{N}\right)_M$.

\subsection{Proof of Proposition~\ref{prop:bias_linear_interference_stratified}}
\label{proof:prop:bias_linear_interference_stratified}

Recall that $\hat \tau^s$ is the stratified estimator. We seek to understand its
expectation under the linear interference model in Equation~\ref{eq:dirg:ch3}.

\begin{align*}
  \E_\Z \left[\hat \tau^s|\bm{\pi} \right] &= \E_\Z\left[\sum_{j = 1}^M
  \lambda_j \sum_{i \in \calC_j} \left(\alpha_i + \beta_i Z_i +
  \frac{\gamma_i}{|\N_i|} \sum_{k \in \N_i} Z_k \right)\frac{(-1)^{1 -
  Z_i}}{n_j^{Z_i} (N_j - n_j)^{1 - Z_i}} \right] \\
  & = \sum_{j = 1}^M \lambda_j \sum_{i \in \calC_j} \beta_i \frac{n_j}{N_j}
  \frac{1}{n_j} + \frac{\gamma_i}{|\N_i|} \sum_{l \neq j} \sum_{k \in \calC_l \cap
  \N_i}
  \frac{n_l}{N_l} \left(\frac{n_j}{N_j} \frac{1}{n_j} - \frac{N_j - n_j}{N_j}
  \frac{1}{N_j - n_j} \right)\\
  & \qquad + \frac{\gamma_i}{|\N_i|} \sum_{k \in \calC_j \cap \N_i} \frac{n_j}{N_j}
  \frac{1}{n_j}
  \\
  & = \sum_{j =1}^M \lambda_j \overline{\bm{\beta}_j} + \frac{\lambda_j}{N_j}
  \sum_{i \in \calC_j}  \gamma_i \frac{|\N_i \cap \calC_j|}{|\N_i|}
\end{align*}

If $\lambda_j = \frac{N_j}{N}$, the previous formula simplifies to:
\begin{equation*}
  \E_\Z \left[\hat \tau^s|\bm{\pi} \right] = \bar{\bm{\beta}} + \frac{1}{N}
  \sum_{j =1}^M \sum_{i \in \calC_j} \gamma_i \frac{|\N_i \cap \calC_j|}{|\N_i|}
\end{equation*}
If the interference effects are constant, then the formula becomes:
\begin{equation*}
  \E_\Z \left[\hat \tau^s|\bm{\pi} \right] = \bar{\bm{\beta}} + \gamma
  \rho_\calC
\end{equation*}

\section{Results on Optimal Deterministic Saturation Design}

\subsection{Proof of Example~\ref{ex:sutva_bias_optimized}}
\label{proof:ex:sutva_bias_optimized}

Let $Y_j^+ \defeq \sum_{i \in \calC_j} Y_i$ be the cluster-level outcomes.
Recall the definition of $f$ and the difference-in-means estimator $\hat \tau$
under the stable unit treatment value assumption:

\begin{align*}
  f(\bm{\pi}, \calC, \Theta)  & = \left| TTE - \E_\Z\left[\hat \tau|
  \bm{\pi}\right] \right| \\
  & = \left|\frac{1}{N} \sum_{i=1}^N \left( Y_i(1) - Y_i(0) \right) - \left(
  \frac{1}{n_t} \sum_{j=1}^M \pi_j \sum_{i \in \calC_j} Y_i(1) - \frac{1}{n_c}
  \sum_{j=1}^M (1 - \pi_j) \sum_{i \in \calC_j} Y_i(0) \right)  \right| \\
  &= \left| \sum_{j =1}^M \left(\frac{\pi_j}{n_t} - \frac{1}{N} \right) Y_j^+(1) -
  \left( \frac{1 - \pi_j}{n_c} - \frac{1}{N} \right) Y_j^+(0) \right|
\end{align*}
It is easy to see that for $\bm{\pi}^* = \left(\frac{n_t}{N}\right)_M$, we
have $f(\bm{\pi}^*, \calC, \Theta) = 0$, such that
\begin{equation*}
\left(\frac{n_t}{N}\right)_M \in \arg \min_{\bm{\pi} \in \calS} f(\bm{\pi},
  \calC, \Theta)
\end{equation*}

%
\subsection{Proof of Example~\ref{ex:sutva_mse_optimized}}
\label{proof:ex:sutva_mse_optimized}

Let $f$ be the mean-squared error of the difference-in-means
estimator $\hat \tau$ under the stable unit treatment value assumption:

\begin{equation*}
  f : (\bm{\pi}, \calC, \{\Y(1), \Y(0)\}) \mapsto   \left( TTE - \E_\Z \left[ \hat
  \tau \middle| \bm{\pi} \right]  \right)^2 + \var_{\Z}\left[\hat \tau |
  \bm{\pi}\right]
\end{equation*}
From Proposition~\ref{prop:bias_sutva} and the proof of Proposition~\ref{prop:variance_sutva}, we have
\begin{align*}
    \mathbb E_{\bm Z}[\hat\tau\mid \bm\pi] - TTE &= \frac{N}{n_tn_c}\sum_{j=1}^M(\pi_j - \overline{\pi})W^{(j)}=\frac{N}{n_tn_c}\bm\pi^T\tilde{\bm W}^+\\
    \mathrm{Var}_{\bm Z}[\hat\tau\mid\bm\pi] &=\frac{N^2}{n_t^2n_c^2}\sum_{j=1}^M\pi_j(1-\pi_j)N_j\mathbb S[\bm W^{(j)}]=\frac{N^2}{n_t^2n_c^2}\left(\bm\pi^T\bm S^+ - \bm \pi^T\cal S^+\bm\pi\right).
\end{align*}
Therefore, the conditional MSE is given by
\[
    f(\bm{\pi}, \calC, \{\Y(1), \Y(0)\})=
    \frac{N^2}{n_t^2n_c^2}\left[\bm\pi^T\left(\tilde{\bm W}^+[\tilde{\bm W}^+]^T - \cal S^+\right)\bm\pi + \bm\pi^T\bm S^+\right].
\]
Let 
$$\tilde f(\bm\pi) = \bm\pi^T\left(\tilde{\bm W}^+[\tilde{\bm W}^+]^T - \cal S^+\right)\bm\pi + \bm\pi^T\bm S^+.$$
It is sufficient to minimize $\tilde f$ with respect to the constraints in \eqref{eq:example-optimize-sutva-constr}. Now let us show that the optimal proportion vector $\bm\pi$ must lie on the boundary of $\Omega_\pi$.

Suppose $\bm\pi^*$ is the optimal proportion vector and is an interior point of $\Omega_\pi$. Then there exist $\delta > 0$ such that the $\ell_2$ open ball in the linear subspace $\{\bm\pi: \bm 1^T\bm\pi= M\overline{\pi}\}$ centered at $\bm\pi^*$ with radius $\delta$ is subset of $\Omega_\pi$:
\[
    B_\delta(\bm\pi^*):=\{\bm\pi: \bm 1^T\bm\pi = M\overline{\pi},\  \|\bm\pi-\bm\pi^*\|_2<\delta\}\subset\Omega_\pi.
\]
Next, choose a unit vector $\bm u \in\mathbb R^M$ such that $\bm u^T\bm 1 = 0$ and $\bm u^T\tilde{\bm W}^+=0$, which is always possible. Consider the set $D_\delta(\bm\pi^*)=\{\bm\pi^* + \tau \bm u: |\tau|\leqslant\frac{\delta}{2}\}$. One can easily verify that $D_\delta(\bm\pi^*)\subset B_\delta(\bm\pi^*)\subset \Omega_\pi$ and hence is feasible. Notice that
$$g(\tau):=\tilde{f}(\bm\pi^* + \tau\bm u)=\tilde{f}(\bm\pi^*) - \tau^2\bm u^T{\cal S}^+ \bm u + \tau \bm u^T\left(\bm S^+ -2\cal S^+\bm\pi^*\right)$$
is a concave quadratic function of $\tau$, whose minimum should be attained at either $\tau=-\frac{\delta}{2}$ or $\tau=\frac{\delta}{2}$. This contradicts the assumption that $\bm\pi^*$ (with $\tau=0$) is optimal.

\end{document}